\documentclass[10pt,twocolsize,letterpaper,notitlepage]{iopart}

\usepackage[%
    style=nature-jjs,
    sorting=none,
    backend=biber,
    natbib=true,
    autocite=inline,
    maxcitenames=3,
    maxbibnames=15,
    backref=false,
    refsection=none]{biblatex}
\usepackage{amssymb}
    
\usepackage{graphicx}
\usepackage{titletoc}
\usepackage{hyperref}
\usepackage{comment}
\hypersetup{
    unicode=true, 
    plainpages=false,
    colorlinks=true,
    linkcolor=blue,
    citecolor=blue,
    filecolor=blue,
    urlcolor=blue
}

\usepackage{etoolbox}

\makeatletter
\newcommand{\mainmatter}{%
  \setcounter{footnote}{0}%
  \patchcmd{\@makefntext}{\fnsymbol}{\alph}{}{}%
  \patchcmd{\@thefnmark}{\fnsymbol}{\alph}{}{}%
  \def\@makefnmark{\textsuperscript{\alph{footnote}}}%
}
\makeatother

\newcommand\text[1]{{\rm{#1}}}
\newcommand{\eqref}[1]{(\ref{#1})}

\renewcommand{\Re}{\mathop{\mathrm{Re}}}
\renewcommand{\Im}{\mathop{\mathrm{Im}}}
\newcommand{\sign}{\mathop{\mathrm{sign}}}
\newcommand{\ket}[1]{\left|#1\right\rangle}
\newcommand{\bra}[1]{\left\langle#1\right|}
\newcommand{\braket}[2]{\left\langle#1|#2\right\rangle}
\newcommand{\sbraket}[1]{\left\langle#1\right\rangle}
\newcommand{\rect}{\mathop{\mathrm{rect}}}
\newcommand{\sinc}{\mathop{\mathrm{sinc}}}

\newcommand{\intinf}{\int_{-\infty}^{\infty}}

\newcommand{\etabol}{\eta_\text{OL}}

\newcommand{\XinG}{X_\text{in}}
\newcommand{\YinG}{Y_\text{in}}

\newcommand{\XLin}{X_\text{L,in}}
\newcommand{\YLin}{Y_\text{L,in}}
\newcommand{\YLv}{Y_\text{L,v}}
\newcommand{\YLout}{Y_\text{L,out}}

\newcommand{\XMout}{X_\text{M,out}}
\newcommand{\YMout}{Y_\text{M,out}}

\newcommand{\Qin}{Q_\text{in}}
\newcommand{\Pin}{P_\text{in}}

\newcommand{\XL}{X_\text{L}}
\newcommand{\YL}{Y_\text{L}}

\newcommand{\XLloss}{X_{\text{L},\ell}}
\newcommand{\YLloss}{Y_{\text{L},\ell}}

\newcommand{\XMin}{X_\text{M,in}}
\newcommand{\YMin}{Y_\text{M,in}}
\newcommand{\XMloss}{X_{\text{M},\ell}}
\newcommand{\YMloss}{Y_{\text{M},\ell}}

\newcommand{\CL}{C_\text{L}}
\newcommand{\kL}{\kappa_\text{L}}
\newcommand{\gL}{g_\text{L}}
\newcommand{\CM}{C_\text{M}}
\newcommand{\kM}{\kappa_\text{M}}
\newcommand{\gM}{g_\text{M}}

\newcommand{\ain}{a_\text{in}}
\newcommand{\bin}{b_\text{in}}

\newcommand{\cin}{c_\text{in}}
\newcommand{\cout}{c_\text{out}}
\newcommand{\bopt}{b_\text{opt}}
\newcommand{\bdrv}{b_\text{drv}}
\newcommand{\al}{a_{\ell}}
\newcommand{\cl}{c_{\ell}}
\newcommand{\alT}{a_{\ell}^{T}}
\newcommand{\ainT}{a_{\text{in}}^{T}}

\newcommand{\ainN}[1]{a_{#1,\text{in}}}
\newcommand{\aoutN}[1]{a_{#1,\text{out}}}
\newcommand{\pumpinN}[1]{\alpha_{#1,\text{in}}}
\newcommand{\pumpoutN}[1]{\alpha_{#1,\text{out}}}
\newcommand{\pumpN}[1]{\alpha_{#1}}

\newcommand{\gLN}[1]{g_{\text{L}#1}}
\newcommand{\kLN}[1]{\kappa_{\text{L}#1}}
\newcommand{\XLN}[1]{X_{\text{L}#1}}
\newcommand{\YLN}[1]{Y_{\text{L}#1}}
\newcommand{\CLN}[1]{C_{\text{L}#1}}

\newcommand{\XLinN}[1]{X_{\text{L}#1,\text{in}}}
\newcommand{\YLinN}[1]{Y_{\text{L}#1,\text{in}}}

\newcommand{\XLoutN}[1]{X_{\text{L}#1,\text{out}}}
\newcommand{\YLoutN}[1]{Y_{\text{L}#1,\text{out}}}

\newcommand{\Hom}{H_\text{OM}}

\newcommand{\Hb}{H_{b}}
\newcommand{\Hem}{H_\text{EM}}

\newcommand{\chib}{\chi_{b}}
\newcommand{\chic}{\chi_{c}}
\newcommand{\chibld}{\chi_{b,\text{EM}}}
\newcommand{\chicld}{\chi_{c,\text{EM}}}
\newcommand{\chix}{\chi_\leftrightarrow}

\newcommand{\effd}{\eta_\text{d}}
\newcommand{\effL}{\eta_\text{L}}
\newcommand{\effM}{\eta_\text{M}}
\newcommand{\effbin}{\eta_{b}}
\newcommand{\effbopt}{\eta_\text{OM}}
\newcommand{\effbol}{\eta_\text{OL}}

\newcommand{\tac}{t_{ac}}
\newcommand{\tinf}{t_{\infty}}
\newcommand{\tcc}{t_{cc}}
\newcommand{\tbc}{t_{bc}}
\newcommand{\tvc}{t_{\text{v}c}}
\newcommand{\tclc}{t_{\cl c}}
\newcommand{\talc}{t_{\al c}}
\newcommand{\taa}{t_{aa}}
\newcommand{\tba}{t_{ba}}
\newcommand{\tca}{t_{ca}}

\newcommand{\Tac}{T_{ac}}
\newcommand{\Tinf}{T_{\infty}}
\newcommand{\Tcc}{T_{cc}}
\newcommand{\Tbc}{T_{bc}}
\newcommand{\Tvc}{T_{\text{v}c}}
\newcommand{\Tclc}{T_{\cl c}}
\newcommand{\Talc}{T_{\al c}}
\newcommand{\Taa}{T_{aa}}
\newcommand{\Tba}{T_{ba}}
\newcommand{\Tca}{T_{ca}}

\newcommand{\lwfact}{\mu}  
\newcommand{\signY}{\sigma}
\newcommand{\hineff}{h'}

\newcommand{\SBpSymb}{+}

\newcommand{\SBp}[1]{#1^{\SBpSymb}}

\newcommand{\pXLinT}{X_{\text{L,in}}^{T,\SBpSymb}}

\newcommand{\pXLlossT}{X_{\text{L},\ell}^{T,\SBpSymb}}

\newcommand{\pXLjT}{X_{\text{L},j}^{T,\SBpSymb}}

\newcommand{\pYLinT}{Y_{\text{L,in}}^{T,\SBpSymb}}

\newcommand{\pYLlossT}{Y_{\text{L},\ell}^{T,\SBpSymb}}

\newcommand{\pYLjT}{Y_{\text{L},j}^{T,\SBpSymb}}

\newcommand{\spectSymb}{\mathcal{S}}
\newcommand{\SXMout}{\spectSymb[\XMout]}
\newcommand{\SYMout}{\spectSymb[\YMout]}
\newcommand{\SpXLin}{\spectSymb[\pXLinT]}
\newcommand{\SpXLl}{\spectSymb[\pXLlossT]}
\newcommand{\SpQin}{\spectSymb[\SBp{\Qin}]}
\newcommand{\SXMin}{\spectSymb[\XMin]}
\newcommand{\SXMl}{\spectSymb[\XMloss]}
\newcommand{\SYLout}{\spectSymb[\YLout]}
\newcommand{\bSYLout}{\bar{\spectSymb}[\YLout]}
\newcommand{\SYv}{\spectSymb[\YLv]}

\newcommand{\SalTXMout}{\spectSymb[\XMout|\alT]}
\newcommand{\SbinXMout}{\spectSymb[\XMout|\bin]}
\newcommand{\ScinXMout}{\spectSymb[\XMout|\cin]}
\newcommand{\SclXMout}{\spectSymb[\XMout|\cl]}
\newcommand{\SYLvXMout}{\spectSymb[\XMout|\YLv]}
\newcommand{\SopXMout}{\spectSymb[\XMout|\mathcal{O}]}

\newcommand{\faa}{\phi_{aa}}

\newcommand{\qzpf}{q_{\text{zpf}}}
\newcommand{\Fclass}{\tilde{F}}

\newcommand{\asb}{\alpha_{\text{sb}}}

\newcommand{\CA}{\mathcal{C}_A}
\newcommand{\CB}{\mathcal{C}_B}
\newcommand{\CC}{\mathcal{C}_C}
\newcommand{\CD}{\mathcal{C}_D}
\newcommand{\nxi}{\bar{n}_\xi}
\newcommand{\zxi}{\bar{z}_\xi}
\newcommand{\zsqxi}{(\zxi)^2}
\newcommand{\zsqxicj}{(\zxi^*)^2}

\newcommand{\nth}{\bar{n}}
\newcommand{\sigdet}{\omega}

\newcommand{\ainTN}[1]{a_{#1,\text{in}}^{T}}
\newcommand{\aoutRN}[1]{a_{#1,\text{out}}^{R}}
\newcommand{\ainRN}[1]{a_{#1,\text{in}}^{R}}
\newcommand{\XLv}{X_\text{L,v}}

\newcommand{\bSXoutCFB}[1][]{\bar{\spectSymb}[\XLoutN{2}^{\SBpSymb} #1]}
\newcommand{\bSYoutCFB}[1][]{\bar{\spectSymb}[\YLoutN{2}^{\SBpSymb} #1]}
\newcommand{\bSXMout}[1][]{\bar{\spectSymb}[\XMout #1]}
\newcommand{\bSYMout}[1][]{\bar{\spectSymb}[\YMout #1]}

\newcommand{\VXadd}{V_X^\text{add}}
\newcommand{\VYadd}{V_Y^\text{add}}
\newcommand{\qtw}{\mathcal{W}_T}

\newcommand{\ourtitle}{Mechanically mediated optical--microwave quantum state transfer by feedback}

\begin{document}
\startcontents[sections]
\jjstwocolsetup
\twocolumn[
\title{\ourtitle}
\author{Max P. Foreman$^{1,2}$\footnotemark[1], Jesse J. Slim$^{1,2,3}$\footnotemark[1], and Warwick P. Bowen$^{1,2,3}$}
\address{$^1$ School of Mathematics and Physics, The University of Queensland, St. Lucia, Queensland 4072, Australia}
\address{$^2$ ARC Centre of Excellence in Biotechnology, The University of Queensland, St. Lucia, Queensland 4072, Australia}
\address{$^3$ ARC Centre of Excellence for Engineered Quantum Systems, The University of Queensland, St. Lucia, Queensland 4072, Australia}
\ead{w.bowen@uq.edu.au}


\begin{abstract}
State transfer between light and microwaves is a key challenge in quantum networks.
Promising transducers use a mechanical intermediary that couples to both fields via radiation pressure. Such electro-optomechanical devices have achieved high efficiencies, yet require resolved-sideband cavities, and generally compromise in scalability and noise performance. Here, we relax this constraint by extending the protocol of Navarathna \textit{et al.} that transfers optical quantum information onto a mechanical resonator using a broadband, sideband-unresolved cavity and feedback. 
Combining this with parametric mechanical-to-microwave conversion, we show that continuous optical-to-microwave quantum state transfer is possible using measurement-based feedback, while all-optical coherent feedback enables bidirectional transfer.
To assess the transfer, we introduce the \textit{quantum transfer witness} \(\qtw\), which---though similar to the input-referred added noise---also identifies whether a channel is capable of both preserving Gaussian entanglement and outperforming classical transduction schemes. Finally, we show that quantum-compatible noise performance is within reach of current experimental capabilities. 
Our results unlock a new design space for electro-optomechanical transducers and strengthens their candidacy as scalable quantum links between distant nodes. 
\end{abstract}

]
\footnotetext[1]{M.P.F. and J.J.S. contributed equally to this work.}
\markboth{\ourtitle}{\ourtitle}
Interconversion of quantum information between travelling optical modes and the microwave domain is a key capability for a future quantum internet~\cite{Quantum_Internet, Han_microwave-optical-review_2021,
lauk_perspectives-on-transduction_2020}.
Important applications include entanglement sharing~\cite{delaney_qubit-readout-via-electro-optic-transducer_2022,
sahu_entangling-MW-with-light-electro-optic_2023,
zhong_microwave-optical-entanglement-blue-sideband-pulses_2024,
meesala_optical-microwave-entanglement-using-blue-sideband-transducer_2024,
meesala_optical-microwave-non-classical-using-blue-sideband_2024,mirhosseini_qubit-to-optical-photon_2020}, distributed quantum sensing~\cite{lachance-quirion_magon-detection-using-superconducting-qubit_2020, degen_quantum-sensing_2017}, secure communication~\cite{zhang_quantum-comm-100km-fibre_2022,takesue_quantum-teleportation-100km-fibre_2015}, accessing long-lived quantum memory \cite{Han_microwave-optical-review_2021, shandilya_optomech-interface-spin-q-memory_2021}, and on-demand generation and efficient detection of optical and microwave photons \cite{Han_microwave-optical-review_2021,lauk_perspectives-on-transduction_2020}. Even within a single node, optical-microwave transduction offers a benefit: multiplexed quantum signals can enter the cryostat through fibres, rather than thermally conductive metal leads \cite{Han_microwave-optical-review_2021}. 

To date, highest conversion efficiencies have been achieved with cold Rydberg atoms (58-82\%)~\cite{tu_cold-atom-off-res-scattering-record-efficiency_2022, kumar_Rb_bulk_transducer_2023}. These approaches have limited scalability due to large components, and are often constrained to pulsed operation~\cite{tu_cold-atom-off-res-scattering-record-efficiency_2022}. 
Alternatively, optical and microwave degrees of freedom can interact directly in bulk~\cite{hease_electro-optic_convertor_2020,sahu_electro-optic-effect_2022, sahu_entangling-MW-with-light-electro-optic_2023} and on-chip~\cite{fan_electro-optic-convertor_2018,
fu_electro-optic_2021,
holzgrafe_electo-optics-LiN_2020} nonlinear elements. 
Direct electro-optic transducers achieve quantum-compatible noise performance \cite{hease_electro-optic_convertor_2020,
sahu_electro-optic-effect_2022, sahu_entangling-MW-with-light-electro-optic_2023}. 
However, high conversion efficiencies require high-power optical pump pulses \cite{sahu_electro-optic-effect_2022}, which are generally incompatible with co-localised on-chip superconducting elements \cite{fu_electro-optic_2021, lauk_perspectives-on-transduction_2020}. Alternative 
designs leverage a mechanical intermediary, 
helping to minimise pump-induced heating and quasiparticle poisoning~\cite{Benevides2024quasiparticle} by spatially separating microwave and optical components~\cite{mirhosseini_qubit-to-optical-photon_2020}. The mediating mechanical resonator either couples via radiation pressure to both electromagnetic fields (electro-optomechanics)~\cite{andrews_bidirectional_membrane_2014,higginbotham_harnessing-electro-optic-correlations_2018,arnold_electro-opto-mechanical_2020,delaney_qubit-readout-via-electro-optic-transducer_2022,brubaker_optomechanical-ground-state-cooling-electro-optic-transducer_2022} or substitutes the parametric electromechanical interaction with a resonant piezoelectric coupling (piezo-optomechanics)~\cite{mirhosseini_qubit-to-optical-photon_2020,
meesala_optical-microwave-non-classical-using-blue-sideband_2024,
meesala_optical-microwave-entanglement-using-blue-sideband-transducer_2024,
jiang_piezo-opto-mechanical_2020,
jiang_piezo-opto-mechanical_2023,
forsch_piezo-opto-mechanics_2020,
han_piezo-optomechanics_2020,
weaver_piezo-opto-mechanics_2024_Groblacher,
optica_painter_piezo-optomechanical-design_2023}. 

Electro-optomechanical transducers have demonstrated competitive transmission efficiencies (47\%) exceeding their piezo-optomechanical and electro-optic counterparts~\cite{brubaker_optomechanical-ground-state-cooling-electro-optic-transducer_2022}.
These devices rely on optical and microwave cavities to boost the interaction and typically operate in the resolved-sideband regime, where the cavity decay is slower than the mechanical period. While this is readily attained in electromechanical systems (for example, \cite{teufel_electromechanical_sideband_cooling_2011}), practical limits on the optical quality factor necessitate high-frequency mechanical resonators that generally exhibit only moderate radiation pressure interaction. As such, these devices generally trade off in noise performance and scalability, motivating the search for new transducer designs.

Recently, Navarathna et al.~\cite{Warwick_Paper} proposed an optical-to-mechanical transduction scheme that operates continuously in the unresolved-sideband limit. 
This protocol combines radiation pressure and feedback of a homodyne measurement to transfer the upper mechanical sideband's amplitude and phase, respectively. Feedback cooling also suppresses mechanical noise, 
enabling faithful transfer of quantum states.
In fact, Navarathna \textit{et al.} show that feedback cooling itself can be viewed as a transfer of optical vacuum onto mechanical motion. 

Here, we extend the protocol in Ref.~\cite{Warwick_Paper} with subsequent transfer from the mechanical mode to a sideband-resolved microwave cavity, which then couples out into a transmission line. 
After modelling the transducer's dynamics including key loss sources, we show that measurement-based feedback permits unidirectional state transfer, with high fidelity and quantum-compatible noise performance for realistic, state-of-the-art system parameters.
We then introduce the \textit{quantum transfer witness} \(\qtw\)---inspired by the gain-normalised conditional variance product \cite{bowen_M_2003}---as a useful single-valued figure of merit for comparing transducers. Though similar to the input-referred added noise \cite{Han_microwave-optical-review_2021}, we show that it unlocks new insights. Namely, \(\qtw < 1\) marks both performance beyond a classical detection-reconstruction scheme, and is a necessary and sufficient condition for the transfer to preserve perfect input Gaussian entanglement. Using $\qtw$, we identify optical coupling inefficiency as the most detrimental loss source for our transducer.
Finally, we investigate the content of the feedback signal. We identify that optical noise in the cavity output is replaced by microwave noise, so that the classical measurement photocurrent does not reveal the input state. Motivated by this observation, we show that upgrading to all-optical coherent feedback~\cite{ernzer_coherent-feedback_2023} permits bidirectional transfer with similar performance. By enabling faithful quantum state conversion in sideband-unresolved (`bad-cavity') optomechanical systems, our findings, together with Ref.~\cite{Warwick_Paper}, substantially broaden the design space for optical-microwave transducers. We expect this to inspire compact, higher-performance devices that further establish electro-optomechanical converters as viable quantum interconnects.

\begin{figure*}
    \centering
    \includegraphics{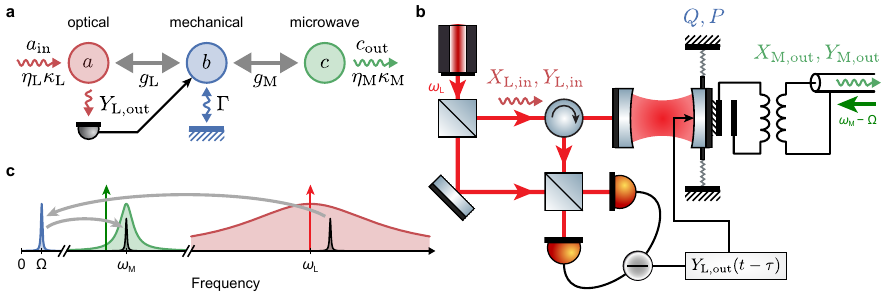}
    \caption{\textbf{Optical-to-microwave quantum transducer.} \textbf{(a)} Transduction from an optical input channel $a_\text{in} = (\XLin + \rmi \YLin)/\sqrt{2}$ to a microwave output channel $c_\text{out} =  (\XMout + \rmi \YMout)/\sqrt{2}$ is facilitated by cascading an optical ($a$, input coupling rate $\effL\kL$), a mechanical ($b$) and a microwave ($c$, output coupling rate $\effM\kM$) resonance. Modes interact with optomechanical and electromechanical coupling rates $g_\text{L}$ and $g_\text{M}$, respectively. The mechanical resonator couples to the thermal environment with rate $\Gamma$. Optical output quadrature $\YLout$ is detected and fed back onto $b$. \textbf{(b)} Schematic opto-electromechanical transfer system. A resonant optical pump is coupled into an optomechanical cavity, imprinting the amplitude quadrature $\XLin$ onto $b$ through radiation pressure. Phase quadrature $\YLin$ contributes to output quadrature $\YLout$, which is measured through homodyne detection, delayed by time $\tau$ and fed back onto the momentum of $b$. Mechanical position $Q$ and momentum $P$ are transferred to $c$ by a microwave pump tuned to the red electromechanical sideband. Finally, microwave output quadratures $\XMout$ and $\YMout$ carry the transferred quantum state out. \textbf{(c)} Lab-frame frequency diagram of the transfer process. The optical resonance at $\omega_\text{L}$ has linewidth $\kL$ larger than the mechanical frequency $\Omega$. The upper optomechanical sideband (grey) is transferred to the mechanical resonator by a resonant pump and feedback. The microwave resonance at $\omega_\text{M}$ is sideband-resolved ($\kM \ll \Omega$), so that a red-detuned microwave pump (at $\omega_\text{M} - \Omega$) transfers the mechanical state to the microwave resonator.}
    \label{fig:1}
\end{figure*}

\section{Set-up}
In our set-up, we consider a mechanical resonator with frequency $\Omega$, energy decay rate $\Gamma \ll \Omega$ and hiqh quality factor $\mathcal{Q} = \Omega/\Gamma$. As sketched in Figs.~\ref{fig:1}a and~\ref{fig:1}b, it is coupled dispersively to both an optical and a microwave cavity (decay rates $\kL$ and $\kM$, resonance frequencies $\omega_\text{L} \gg \omega_\text{M} \gg \Omega$). For electromechanical systems, the resolved sideband condition $\kM \ll \Omega$ is routinely achieved \cite{teufel_electromechanical_sideband_cooling_2011}. In this regime, state transfer from mechanical motion to the microwave mode is implemented in the usual manner by driving the microwave cavity with a control tone red-detuned from resonance by the mechanical frequency \cite{Warwick_Textbook}. However, for the same mechanical frequency, the condition $\kL \ll \Omega$ is far more stringent -- given that
typical optical resonance frequencies are four orders of magnitude greater than that of microwave cavities. 
Our transducer therefore relies on a low-quality optical mode that does not resolve the mechanical sidebands $\kL \gg \Omega$. 
We first propose a one-directional state transfer protocol from light, through the mechanical resonator, and onto the microwave field. In the final section, we will turn to conversion in the opposite direction. 

To transfer an optical input state onto the mechanical resonator, we follow the protocol in~\cite{Warwick_Paper}. By applying a strong, resonant drive to the optical cavity, the linearised radiation pressure interaction imprints the amplitude quadrature $\XLin$ of the optical input field directly onto the mechanical motion. In contrast, the phase quadrature $\YLin$ does not affect the resonator directly. It is, however, encoded in the cavity output phase quadrature $\YLout$, which is detected by a homodyne measurement, filtered and subsequently fed back onto the resonator's momentum, as shown in Fig.~\ref{fig:1}b. After appropriately choosing the feedback filter function and gain, the upper (blue) mechanical sideband is transferred onto the resonator while mechanical noise is suppressed~\cite{Warwick_Paper}. This downconversion from light to motion completes our continuous optical-to-microwave transduction protocol, sketched in Fig.~\ref{fig:1}c in the frequency domain.

\section{Optical to mechanical conversion}
We first model the conversion from optical photons to phonons in the absence of the microwave cavity, closely following the analysis in \cite{Warwick_Paper}. We do, however, include an additional rotating wave approximation, detailed below, to simplify the results and make them more naturally compatible with a subsequent sideband-resolved mechanical-to-microwave transfer.

The linearised interaction Hamiltonian of the optomechanical subsystem, driven on resonance to a large coherent optical amplitude, is given by \cite{Warwick_Textbook}
\begin{equation}
    \Hom = \hbar \gL(a^\dagger + a)(b^\dagger + b) = 2\hbar \gL \XL Q. \label{eq:h-om}
\end{equation}
for the mechanical annihilation operator $b$, expressed in the lab frame and evolving under 
$\Hb = \hbar \Omega b^\dagger b$. The optical annihilation operator $a$ describes fluctuations on top of the strong carrier and is defined in a frame rotating at the optical resonance frequency $\omega_\text{L}$. The amplitude $\alpha$ of the carrier itself, taken to be real without loss of generality, is incorporated in the coherent-amplitude boosted optomechanical coupling rate $\gL = \alpha g_{\text{L},0}$, where $g_{\text{L},0}$ is the vacuum optomechanical coupling rate. This determines the optomechanical cooperativity $\CL = 4\gL^2/\Gamma \kL$. Alternatively, the interaction can be expressed using the quadratures $\XL = (a^\dagger + a)/\sqrt{2}$, $\YL = \rmi (a^\dagger - a) / \sqrt{2}$ of the cavity light field, and the dimensionless mechanical position $Q = (b^\dagger + b)/\sqrt{2}$ and momentum operators $P = \rmi(b^\dagger - b)/\sqrt{2}$. These satisfy commutation relations $[\XL, \YL] = [Q,P] = \rmi$ and exhibit ground-state fluctuations $\langle \XL^2 \rangle = \langle \YL^2 \rangle = 1/2$.

We model the open system using quantum Langevin equations that incorporate dissipation and driving. In the bad-cavity limit \(\kL \gg \Omega\), after adiabatic elimination of the optical dynamics, the intracavity quadrature $\XL(t) = 2\XLin(t)/\sqrt{\kL}$ is addressed instantaneously
by the amplitude quadrature $\XLin$ of the optical input field $\ain$. Thus, the interaction in Eq.~\eqref{eq:h-om} imprints only the quadrature $\XLin$ onto the mechanical momentum. As the mechanical resonator is mostly sensitive to driving around its resonance frequency, we expect the sidebands of $\XLin$ (and thus $\ain$) at $\pm\Omega$ to be transferred most efficiently. In the lab frame, beating between the carrier and such a detuned input field generates radiation pressure modulations that resonantly drive the resonator.

Next, we use the input-output relation $\mathcal{O}_\text{out} = \mathcal{O}_\text{in} - \sqrt{\gamma} \mathcal{O}$ \cite{Warwick_Textbook} to obtain the optical output phase quadrature
\begin{equation}
    \YLout = -\sqrt{\effd} \YLin + 2\sqrt{\effd \Gamma \CL} Q + \sqrt{1-\effd} \YLv. \label{eq:yout}
\end{equation}
Here, imperfect detection is expressed by the efficiency $\effd$ and corresponding vacuum noise $\YLv$ arising from detection loss~\cite{Warwick_Textbook, Warwick_Paper}. 
For simplicity, we do not yet consider any coupling inefficiencies---these will be included later.

In addition to the mechanical position $Q$, the output $\YLout$ importantly encodes the phase quadrature $\YLin$ of the input field. This observation motivates the protocol in~\cite{Warwick_Paper} to transfer the full quantum state of the optical input onto the mechanical resonator. Following \cite{Warwick_Paper}, we measure the output quadrature $\YLout$ by homodyne detection and continuously feed it back onto the oscillator's momentum via a force proportional to the photocurrent $\YLout \circledast f(t)$ filtered by a real, causal function $f(t)$. This allows us to imprint the phase quadrature $\YLin$ on the resonator. As with $\XLin$, we expect the frequency components of $\YLin$ at $\pm\Omega$ to drive the resonator most efficiently. Again, this can be understood as a beating --- in this instance between the homodyne local oscillator and the reflected input sideband --- that is picked up by the detector and fed back onto the resonator.

We define the feedback force as
\begin{equation}
    F_\text{fb}(t) = \frac{\hbar \Gamma G}{2} \left( f(t) \circledast \frac{\YLout(t)}{2\sqrt{\effd \Gamma \CL}} \right)
\end{equation}
where the photocurrent --- representing a measurement result that is fed back as a classical parameter --- has been rescaled so that $\YLout(t) / 2\sqrt{\effd \Gamma \CL} = Q(t) + \cdots$ constitutes an estimate of oscillator position with added optical input terms. The strength of the feedback is quantified relative to the intrinsic friction by the dimensionless gain $G$. While we treat the feedback force generally, in experiments it can be applied by e.g. electrostatic actuation or optical forces \cite{rossi_feedback-by-optical-forces_2017, lee_feedback-by-actuation_2010}. 

In general, the filter function $f(t)$ can be chosen to optimise the feedback with minimal added noise, effectively isolating mechanical signal from photocurrent noise with maximal signal-to-noise ratio. However, to illustrate our protocol, we pick the particularly simple filter $f(t) = -\delta(t-\tau)$ that delays the measured photocurrent by $\tau=\pi/2\Omega$, one-quarter of the mechanical cycle. A discussion for arbitrary filters can be found in~\cite{Warwick_Paper}. 



After combining the resonator's damped evolution with the radiation pressure and feedback forces, we obtain the quantum Langevin equation
\begin{eqnarray}
    \dot{b}(t) = &-\rmi \Omega b(t) - \frac{\Gamma}{2} b(t) + \sqrt{\Gamma} \bin(t) - \rmi \sqrt{2\Gamma \CL} \XLin(t) \nonumber \\
    &+ \frac{\rmi \Gamma G}{4} \frac{\YLout(t-\tau)}{\sqrt{2\effd\Gamma\CL}} \label{eq:eom-b-1}
\end{eqnarray}
for the mechanical annihilation operator.
Here, we have applied a rotating wave approximation (RWA) with respect to the mechanical bath, such that $\bin(t)$ represents white thermal noise that drives both mechanical operators symmetrically and satisfies $[\bin(t), \bin^\dagger(t')] = \delta(t-t')$ with variance $\langle \bin^\dagger \bin \rangle = \bar{n} \delta(t-t')$ \cite{Warwick_Textbook}. The phonon occupancy of the mechanical environment is denoted by $\bar{n}$.

The mechanical contribution to the delayed photocurrent $\YLout(t-\tau)$ is proportional to the delayed position operator $Q(t-\tau) = [b^\dagger(t-\tau) + b(t-\tau)] / \sqrt{2}$. As $Q(t-\tau) \approx -P(t)$ for a high-quality resonator, this provides velocity damping that suppresses mechanical fluctuations and cools the resonator.
While not essential~\cite{Warwick_Paper}, for simplicity we apply a second RWA to remove the counter-rotating term $b^\dagger(t-\tau)$. This choice neglects the retarded response of the mechanical position \(Q\) to the feedback forcing on the momentum quadrature \(P\), and effectively symmetrises the evolution of the two mechanical observables $Q$ and $P$. The resulting optical-to-mechanical transfer is more naturally compatible with subsequent sideband-resolved transfer onto the microwave mode. If the delayed response is retained, a different optical mode is best transferred onto \(Q\) versus \(P\). This introduces an additional noise contribution, referred to as mode mismatch noise in Ref.~\cite{Warwick_Paper}, that can ultimately constrain the optical-to-mechanical state transfer. However, provided that the resonator maintains a high quality factor \(\mathcal{Q}'\gg 1\) even with feedback cooling on, this contribution is negligible.
We will shortly see that the condition \(\mathcal{Q}'\gg 1\) is equivalent to \(\CL \ll \mathcal{Q}\), an important constraint that we carefully consider when characterising the transducer's noise performance in Section \ref{sec:state_of_the_art_params}. 

Proceeding, we assume the resonator to be weakly coupled to the cavity ($\gL \ll \Omega$) and to retain a high quality factor $\mathcal{Q}' \gg 1$ after feedback. We can then neglect back-action and damping within a single cycle and approximate $b(t-\tau) \approx e^{i\Omega \tau} b(t) = \rmi b(t)$ from the resonator's free evolution. 
The mechanical dynamics then simplify to
\begin{eqnarray}
    &\dot{b}(t) = -\rmi\Omega b(t) - \frac{\Gamma'}{2} b(t) + \sqrt{\Gamma}\bin(t) 
    - \rmi \sqrt{2\Gamma \CL} \times \label{eq:eom-b-2} \\
    &\quad\Bigg( \XLin(t) 
    + \frac{G}{8\CL} \YLin(t-\tau) 
    - \frac{G}{8\CL} \varepsilon \YLv(t-\tau)  \Bigg), \nonumber
\end{eqnarray}
where $\varepsilon = \sqrt{(1-\effd)/\effd}$ and the mechanical feedback term $-\Gamma G b(t) / 4$ is absorbed into the feedback-broadened linewidth $\Gamma' = (1+G/2)\Gamma$. We read off that the choice $G_\text{sym}=8\CL$ for the feedback gain symmetrises the transfer of the optical amplitude and phase quadratures, leading to state transfer without squeezing. This is consistent with the analysis in Ref. \cite{Warwick_Paper}. 

We continue our analysis in the frequency domain, using the Fourier transform $a(\omega) = \int_{-\infty}^\infty e^{\rmi\omega t} a(t)\,\mathrm{d}t$, noting that $[a^\dagger](\omega) = \int_{-\infty}^\infty e^{\rmi \omega t} a^\dagger(t)\,\mathrm{d}t = a(-\omega)^\dagger$. To avoid confusion, we will use the latter notation for conjugated operators.
From Eq. \eqref{eq:eom-b-2} we obtain the mechanical frequency response
\begin{eqnarray}
    b(\omega) = &\sqrt{\Gamma} \chib(\omega) \bigg[ \bin(\omega)
    - \rmi \sqrt{2\CL} \Big( \XLin(\omega) \nonumber \\ 
    &\quad\quad+ h  e^{\rmi \omega \tau} \YLin(\omega) - h \varepsilon e^{\rmi \omega \tau} \YLv(\omega) \Big) \bigg], \label{eq:eom-b-freq}
\end{eqnarray}
in which we define the gain ratio $h=G/G_\text{sym}=G/8\CL$ and the mechanical susceptibility 
\begin{equation}
    \chib(\omega) = \frac{1}{\Gamma'/2 - \rmi(\omega - \Omega)}.
\end{equation}
The susceptibility is sharply peaked around $+\Omega$ in the high-quality regime $\mathcal{Q}' = \Omega / \Gamma' \gg 1$, so that we can make the approximation $e^{\rmi \omega \tau} \approx e^{\rmi \Omega \tau} = \rmi$.
Formally, the same condition underlies the approximation $b(t-\tau) \approx \rmi b(t)$ that we made earlier, as well as both RWAs.

Expressed in terms of the input optical field annihilation operator $\ain$, the mechanical response then reads
\begin{eqnarray}
    b(\omega) = &\sqrt{\Gamma}\chib(\omega) \bigg[ \bin(\omega) - \sqrt{2\CL} h \varepsilon \YLv(\omega) \nonumber \\
    &- \rmi \sqrt{\CL} \left[(1+h)\ain(\omega) + (1-h)\ain(-\omega)^\dagger \right] 
     \bigg],
\end{eqnarray}
in which the susceptibility $\chib(\omega)$ only selects frequencies around $\omega \approx \Omega$.
For $h\neq1$, we see that a Bogoliubov mode comprising a superposition of positive and negative sidebands $\ain(\omega \approx \Omega)$ and $\ain(-\omega \approx -\Omega)^\dagger$ is transferred to the mechanical resonator. In contrast, for $h=1$, the transfer of the lower mechanical sideband is suppressed, as the radiation pressure modulations it induces are exactly cancelled by the feedback force it generates (see SI Section~\ref{sec:lower_sideband_cancel}). The upper mechanical sideband is then transferred without squeezing. For simplicity, we operate at $G=8\CL$ and set $h=1$ in the remainder of this analysis. 
The case $h \neq 1$ is covered in SI Section~\ref{sec:apx:optomechanical_subsystem_arbgain}.

In principle, the lower mechanical sideband can be coupled to the resonator as well, by setting $h=-1$. However, the corresponding negative feedback gain will then amplify fluctuations rather than suppress them. This is analogous to the parametric gain induced by a blue-detuned control tone in a resolved-sideband cavity, completing the parallel between sideband-pumped good-cavity and feedback-enabled bad-cavity processes \cite{ernzer_coherent-feedback_2023}.


At the optimal operating point $h=1$, the mechanical response reduces to
$b(\omega) = \chib(\omega)(\sqrt{\Gamma} \bin(\omega) + \sqrt{4\Gamma\CL} \bopt(\omega))$,
in which we identify the optical input term 
\begin{eqnarray}
    \bopt(\omega) = -\rmi \ain(\omega) - \varepsilon \YLv(\omega)/\sqrt{2}.
\end{eqnarray}
Moreover, by rewriting the mechanical response using the broadened linewidth $\Gamma' = (4\CL+1)\Gamma$ as 
\begin{eqnarray}
    b(\omega) &= \sqrt{\Gamma'} \chib(\omega) \Bigg[ \sqrt{\effbin} \bin(\omega) + \sqrt{\effbopt} \bopt(\omega) \Bigg], \label{eq:mech-eff-channels}
\end{eqnarray}
we recognise that $\bopt$ constitutes an effective mechanical input channel that couples to the resonator with efficiency 
\begin{eqnarray}
    \effbopt = 4\CL / (4\CL + 1). \label{eq:eff_bopt}
\end{eqnarray}
Importantly, this efficiency tends to unity as $\CL \to \infty$. In contrast, as the cooperativity is increased, the effective coupling efficiency
\begin{eqnarray}
    \effbin = 1/(4\CL + 1) = 1-\effbopt \label{eq:eff_bin}
\end{eqnarray}
of the mechanical bath tends to zero. As a result, mechanical fluctuations in the resonator are suppressed and replaced by optical input fluctuations, crucially allowing faithful transfer in the presence of environmental noise. This concludes our analysis of the optical to mechanical transfer process.

\section{Mechanical to microwave conversion}
We now turn to the mechanical to microwave conversion step. As shown in Fig.~\ref{fig:1}b, the mechanical resonator we consider is coupled dispersively to a narrow-linewidth microwave resonator. The resulting interaction is formally the same as the optomechanical interaction in the optical subsystem. However, to avoid confusion, when applied to the microwave subsystem we will refer to it as the electromechanical interaction.

As this microwave cavity resolves the electromechanical sidebands ($\kM \ll \Omega$), driving the microwave cavity with a strong, coherent control tone, red-detuned from the microwave resonance $\omega_\text{M}$ by the mechanical frequency $\Omega$, selects the interaction terms
\begin{equation}
    \Hem = - \hbar \gM \left( c^\dagger b e^{\rmi \Omega t} + c b^\dagger e^{-\rmi \Omega t} \right) \label{eq:em-interaction}
\end{equation}
in the linearised electromechanical Hamiltonian~\cite{Warwick_Textbook}. Here, the microwave field operator $c$ is represented in a frame rotating at the microwave resonance frequency and $\gM$ is the coherent-amplitude boosted electromechanical interaction rate. 

From the perspective of the microwave cavity, the mechanical resonator---under the influence of the fast-cavity optomechanical interaction and feedback---appears like a colder, more damped resonator that is mostly driven by an optical environment rather than mechanical fluctuations, as expressed by Eq.~\eqref{eq:mech-eff-channels}. 
Conveniently, this allows us to solve for the dynamics of the full opto-electro-mechanical system by feeding the effective mechanical input channels in Eq.~\eqref{eq:mech-eff-channels} into an independent solution for the electro-mechanical sub-system driven by a placeholder input \(\bdrv(t)\). We will now follow this approach.

To extract converted microwave photons, we couple the microwave cavity to a transmission line with rate $\kM$. In absence of the electromechanical interaction,
the microwave resonator then exhibits a susceptibility
$\chic(\omega) = (\kM/2 - \rmi \omega)^{-1}$
to the microwave field $\cin$ coming through the line. 
After turning on the electromechanical interaction in Eq.~\eqref{eq:em-interaction}, the system evolves as 
\begin{eqnarray}
    \dot{b} &= -\rmi \Omega b - \frac{\Gamma'}{2} b + \sqrt{\Gamma'} \bdrv + \rmi \gM e^{-\rmi \Omega t} c, \\
    \dot{c} &= -\frac{\kM}{2}c + \sqrt{\kM}\cin + \rmi \gM e^{\rmi \Omega t} b. \label{eq:em-time-evo} 
\end{eqnarray}
By applying the Fourier transform and rearranging, we obtain the mechanical and microwave frequency responses
\begin{eqnarray}
    b(\omega) &= \chibld(\omega) \sqrt{\Gamma'}\bdrv(\omega) \nonumber \\ 
    &\quad + \rmi  \chix(\omega - \Omega) \sqrt{\kM} \cin(\omega-\Omega), \label{eq:b_freq_response_perfect} \\
    c(\omega) &= \chicld(\omega) \sqrt{\kM} \cin(\omega) + \rmi \chix(\omega) \sqrt{\Gamma'} \bdrv(\omega), \label{eq:c_freq_response_perfect}
\end{eqnarray}
which depend on the electromechanically modified mechanical and microwave susceptibilities
\begin{eqnarray}
    \chibld(\omega) &= \frac{\chib(\omega)}{1+\gM^2 \chib(\omega)\chic(\omega-\Omega)}, \\
    \chicld(\omega) &= \frac{\chic(\omega)}{1+\gM^2 \chib(\omega+\Omega)\chic(\omega)}, \label{eq:chicld_defn}
\end{eqnarray}
respectively. These susceptibilities also appear in the description of electromechanically induced absorption and transmission \cite{Warwick_Textbook}. Moreover, the transduction of one mode's input onto the other is quantified by the cross-susceptibility
\begin{eqnarray}
    \chix(\omega) &= \gM \chib(\omega+\Omega)\chicld(\omega) = \gM \chic(\omega)\chibld(\omega+\Omega) \nonumber \\
    &= \frac{\gM\chib(\omega+\Omega) \chic(\omega)}{1+\gM^2 \chib(\omega+\Omega) \chic(\omega)}.
\end{eqnarray}

From Eq.~\eqref{eq:mech-eff-channels} in the previous section, we extract the effective (opto)mechanical drive term 
\begin{eqnarray}
    \bdrv(t) = \sqrt{\effbin} \bin(t) + \sqrt{\effbopt} \bopt(t). \label{eq:b_drv_defn}
\end{eqnarray}
After plugging this drive term into the cavity response Eq.~\eqref{eq:c_freq_response_perfect} and applying the input-output relation, we obtain
\begin{eqnarray}
    &\cout(\omega) = \left(1-\kM \, \chicld(\omega)\right) \cin(\omega) \\
    &\quad- \rmi \chix(\omega)\sqrt{\effbin\Gamma'\kM} \bin(\omega+\Omega) \nonumber \\
    &\quad- \chix(\omega)\sqrt{\effbopt\Gamma'\kM} \left[ \ain(\omega+\Omega) - \rmi \varepsilon \frac{\YLv(\omega+\Omega)}{\sqrt{2}} \right]. \nonumber
\end{eqnarray}
for the full transduction from the optical input field $\ain$ to the microwave output field $\cout$. We conveniently rewrite this as 
\begin{equation}
    \cout(\omega) = \tac(\omega) \ain(\omega+\Omega) + c_\text{noise}(\omega), \label{eq:c_out_with_c_noise}
\end{equation}
to recognise that the upper mechanical sideband of $\ain$ is transferred onto the microwave output field with frequency dependent transfer gain $\tac(\omega) = \sqrt{\effbopt}  \tinf(\omega)$, where 
\begin{eqnarray}
     \tinf(\omega) &= -\chix(\omega)\sqrt{\Gamma'\kM} \nonumber \\
     &= \frac{-2 \sqrt{\CM'}}{\CM' + (1 - 2\rmi\omega/\Gamma')(1-2\rmi\omega/\kM)} \label{eq:tinf}
\end{eqnarray}
is the transfer gain in the limit of infinite optomechanical cooperativity. Here, we have defined the electromechanical cooperativity $\CM' = 4\gM^2 / \Gamma' \kM$ using the feedback-broadened linewidth $\Gamma'$---the only mechanical linewidth relevant to the microwave cavity. This central result fully describes the efficiency of the optical-to-microwave transduction for a given signal frequency $\omega$ in terms of $\effbopt$, $\CM'$ and the linewidth ratio $\beta = \Gamma'/\kM$.

\begin{figure*}
    \centering
    \includegraphics{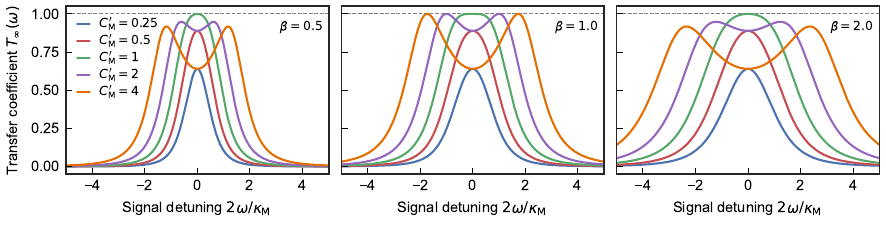}
    \caption{\textbf{Bandwidth of the transfer process.} Transmission $\Tinf(\omega)$ as a function of signal detuning $\omega$ from the upper mechanical sideband for different linewidth ratios $\beta = \Gamma'/\kM$ and electromechanical cooperativities $\CM' = 4\gM^2/\Gamma'\kM$. For $\CM' \leq 1$, the transmission exhibits a single peak at $\omega=0$. For $\CM' > 1$, when the microwave cavity and the feedback-broadened resonator are strongly coupled, the transmission spectrum splits into two peaks corresponding to hybridised modes. For any value of $\beta$, transfer can be achieved with unity efficiency by setting $\CM' = 1$. Moreover, for matched linewidths $\beta=1$, this can be achieved for any $\CM' \geq 1$. }
    \label{fig:transfer-coeff-freq}
\end{figure*}

To examine the bandwidth of the transfer process, in Fig.~\ref{fig:transfer-coeff-freq} we plot the ideal transmission coefficient $\Tinf(\omega) = |\tinf(\omega)|^2$ for different linewidth ratios and electromechanical cooperativities. For $\CM' \leq 1$, the transmission exhibits a single peak at zero signal detuning from the upper mechanical sideband ($\omega=0$). For any value of $\beta$, this peak reaches unity efficiency at $\CM' = 1$, when the impedances of the optomechanical and electromechanical transfer processes are matched. The width of the peak at $\CM' = 1$ is limited by the smaller of $\Gamma'$ and $\kM$: in the limit $\kM \ll \Gamma'$ ($\Gamma' \ll \kM$), $\Tinf(\omega)$ reduces to a Lorentzian function with a bandwidth (full width at half maximum, FWHM) of $2\kM$ ($2\Gamma'$).

In contrast, for $\CM' > 1$, the microwave cavity and the feedback-broadened resonator are strongly coupled. The transmission spectrum then splits into two peaks, corresponding to the two hybridised microwave-mechanical modes. Interestingly, for matched linewidths $\beta=1$, both peaks attain unity efficiency (Fig. \ref{fig:transfer-coeff-freq}, middle), allowing the simultaneous transmission of two frequency-multiplexed signals. In the limit $\CM' \gg 1$ with \(\beta=1\), the bandwidth (FWHM) of each peak tends to $\Gamma' = \kM$ while the mode splitting tends to \(2g_M\).

Finally, the case of simultaneously matched cooperativities $\CM' = 1$ and linewidths $\beta = 1$ is of particular interest (green curve, Fig. \ref{fig:transfer-coeff-freq}, middle). The transmission is then given by $\Tinf(\omega) = 1/(1+4\omega^4/\kM^4)$, which is approximated by a quartic function around the blue mechanical sideband where $\omega\approx0$. As such, it features a flatter top, with high efficiency over a broader range, than the parabolic profile of the Lorentzian lineshape obtained for either $\beta \ll 1$ or $\beta \gg 1$.

In all cases, for finite optomechanical cooperativity, the transmission coefficient $\Tac(\omega) = |\tac(\omega)|^2 = \effbopt \Tinf(\omega)$ is related to the ideal transmission by a reduction proportional to the optomechanical coupling efficiency Eq.~\eqref{eq:eff_bopt}. The optomechanical cooperativity $\CL$ also affects the bandwidth of the transfer process through the feedback gain $G=8\CL$, by tuning the mechanical linewidth $\Gamma'/\Gamma = G/2 + 1= 4\CL + 1$. 
In contrast, the microwave linewidth $\kM$ is typically fixed by device geometry, while the electromechanical cooperativity $\CM'$ can be tuned independently via the strength of the microwave pump. Ultimately, the optimal settings for these parameters depend on efficiency, bandwidth and frequency-multiplexing requirements, as well as on the intrinsic values of $\Gamma$ and $\kM$. 
Note that an upper limit to $G$ and $\CL$ is given by our protocol's need for a high-quality resonator $\mathcal{Q}' \gg 1$ even after feedback, which necessitates \(\Gamma' \ll \Omega\) and therefore \(\CL \ll \mathcal{Q}\). Given this, and the requirement of a sideband resolved microwave cavity \(\kM \ll \Omega\), the bandwidth of the transducer is ultimately constrained by the mechanical frequency \(\Omega\).

Next, we analyse the contributions to the noise term in Eq.~\eqref{eq:c_out_with_c_noise},
    $c_\text{noise}(\omega) = \tcc(\omega) \cin(\omega) + \tbc(\omega) \bin(\omega) + \tvc(\omega) \YLv(\omega + \Omega)$,
in which we recognise the input-to-microwave transfer gains
\begin{eqnarray}
    \tcc(\omega) &= 1-\kM\chicld(\omega), \quad
    \tbc(\omega) = \rmi \sqrt{\effbin} \tinf(\omega), \nonumber \\
    \tvc(\omega) &= -\rmi\varepsilon \sqrt{\effbopt / 2} \, \tinf(\omega).
\end{eqnarray}
Microwave noise in the input $\cin$ is reflected with efficiency $T_{cc}(\omega) = |\tcc(\omega)|^2 = 1 - T_\infty(\omega)$, while mechanical noise is imprinted on the microwave output with efficiency $\Tbc(\omega) = |\tbc(\omega)|^2 = \effbin T_\infty(\omega) = (1-\effbopt)T_\infty(\omega)$. Along with the light-to-microwave transduction, these processes ensure the correct level of quantum fluctuations in the microwave output, as $T_\text{ac}(\omega) + T_\text{bc}(\omega) + T_\text{cc}(\omega) = 1$. In addition, the optical detection loss introduces noise from the vacuum fluctuations in $\YLv$ with transfer coefficient $\Tvc(\omega) = \varepsilon^2 \Tac(\omega) / 2 =(1/\effd - 1) \effbopt \Tinf(\omega) / 2$, which increases with decreasing $\effd$.

\section{Coupling inefficiencies}
Before we assess the magnitude of noise added to the microwave output channel, we extend our model to include optical and microwave cavity coupling inefficiencies. In this section we point out the main differences to the case of perfect coupling considered above, while the full analysis is included in SI Section~\ref{sec:apx:system_dynamics}.

We introduce a loss port $\al$ for (vacuum) photons to leak in and out of the optical cavity with efficiency $1 - \effL$, so that the coupling efficiency of the signal input port $\ain$ is reduced to $\effL$. 
The mechanical contribution to the homodyne photocurrent $\YLout$ is then weaker by a factor $\sqrt{\effL}$, so that the feedback-broadened decay rate is reduced to $\Gamma' = (1 + \sqrt{\effL} G/2)\Gamma$.
Moreover, the input field phase quadrature now contributes to $\YLout = (2\effL - 1) \YLin + \cdots $ with a prefactor that changes sign $\sigma$ depending on whether the cavity is overcoupled ($\effL > 1/2$, $\sigma = +1$) or undercoupled ($\effL < 1/2$, $\sigma = -1$). 


The feedback gain that symmetrises both quadrature transfer strengths is changed to 
\begin{eqnarray}
    G_\text{sym} = 8 \CL \frac{\sqrt{\effL}}{|2\effL - 1|},
\end{eqnarray}
where the absolute value operator ensures that we select a positive, noise-suppressing gain. For $G=G_\text{sym}$, the feedback-broadened linewidth is given by 
\begin{eqnarray}
    \Gamma' = \left(1 + 4\CL \frac{\effL}{|2\effL - 1|} \right) \Gamma = \left( 1 + 4\lwfact\CL \right) \Gamma, \label{eq:Gamma'_inefficient}
\end{eqnarray}
where we define the factor $\lwfact = \effL / |2\effL - 1|$.
Positive gain imprints $\YLin$ with sign $\sigma$ onto the mechanical resonator. Consequently, the optical sideband that is transferred changes with each coupling condition. To designate that sideband, we introduce the notation
\begin{eqnarray}
    \ainT(\omega) &= \left[ \XLin(\omega) + \rmi \signY \YLin(\omega) \right] / \sqrt{2} \nonumber \\
    &= \cases{\ain(\omega) & $\effL > 1/2$ \\ [\ain(-\omega)]^\dagger & $\effL < 1/2$}.
\end{eqnarray}
When the cavity is overcoupled (undercoupled) the upper (lower) mechanical sideband is transferred at the optimal operating point $G=G_\text{sym}$. At critical coupling ($\effL = 1/2$), the input field is fully lost into the loss port. In that case, the reflected output field holds no information on $\YLin$ and transfer of the phase quadrature is ruled out.

In addition, the optomechanical interaction imprints fluctuations from the amplitude quadrature $\XLloss$ of $\al$ onto the mechanical resonator, while its phase quadrature $\YLloss$ feeds into the homodyne photocurrent $\YLout$ and is consequently transferred onto the resonator by feedback. Notably, the resonator does not respond symmetrically to both loss port quadratures at $G=G_\text{sym}$, so that it is susceptible to the Bogoliubov loss port mode
\begin{eqnarray}
    \alT(\omega) = \left[ \XLloss(\omega) / \sqrt{2\lwfact} + \rmi \sqrt{2\lwfact} \YLloss(\omega) \right]/\sqrt{2}, \label{eq:alT_def}
\end{eqnarray}
where the prefactors ensure that $[\alT(\omega), \alT(\omega')^\dagger] = \delta(\omega-\omega')$ is canonically normalised. The optical loss port field is then transferred with squeezing.

We now similarly introduce a microwave loss port $\cl$, so that the microwave input in Eq.~\eqref{eq:c_freq_response_perfect} is replaced by
    $\cin(\omega) \to \sqrt{\effM} \cin(\omega) + \sqrt{1-\effM} \cl(\omega)$
with coupling efficiency $\effM$. 
After proceeding identically to the analysis in the case of perfect couplings, we obtain the microwave output
\begin{eqnarray}
    \cout(\omega) = &\tac(\omega) \ainT(\omega+\Omega) + \talc(\omega) \alT(\omega + \Omega) \nonumber \\ 
    &+ \tbc(\omega) \bin(\omega + \Omega) + \tcc(\omega) \cin(\omega) \label{eq:cout_inefficient} \\ 
    &+ \tclc(\omega) \cl(\omega) + \tvc(\omega) \YLv(\omega+\Omega), \nonumber 
\end{eqnarray}
expressed succinctly in terms of the optical transfer modes $\ainT$ and $\alT$; the mechanical resonators' thermal environment $\bin$; vacuum loss $\cl$ and input $\cin$ modes of the microwave resonator; and vacuum noise introduced by imperfect homodyne detection \(\YLv\). The transfer gains for each channel are given by
\begin{eqnarray}
    &\tac(\omega) = \sqrt{\effM \effbopt} \tinf(\omega), 
    \hspace{2.1em} \tbc(\omega) = \rmi \sqrt{\effM \effbin} \tinf(\omega), \nonumber \\
    &\tcc(\omega) = 1 - \effM \kM \chicld(\omega),  
    \hspace{0.5em} \talc(\omega) = \sqrt{\effM \effbol} \tinf(\omega), \nonumber \\
    &\tclc(\omega) = -\sqrt{\effM(1-\effM)}\kM \chicld(\omega), \nonumber \\
    &\tvc(\omega) = -\rmi \sqrt{\effM\effbopt/2\,}\, (\lwfact\varepsilon /\effL)\, \tinf(\omega), \label{eq:tgains_inefficient}
\end{eqnarray}
in terms of the updated effective coupling efficiencies
\begin{eqnarray}
    \effbin &= \frac{1}{1+ 4\lwfact \CL} , \quad \effbopt = \frac{4\effL\CL}{1+4\lwfact\CL}, \nonumber \\
    \etabol &= \frac{4(\lwfact - \signY \effL)\CL}{1 + 4\lwfact\CL }.
\end{eqnarray}
The susceptibility $\chicld(\omega)$ and ideal gain $\tinf(\omega)$ are unchanged compared to Eqs. \eqref{eq:chicld_defn} and \eqref{eq:tinf}, but with updated \(\Gamma'\) according to Eq.~\eqref{eq:Gamma'_inefficient}. We verify in SI Section~\ref{sec:apx:microwave_coupling_ineff} that $\cout(\omega)$ is canonically normalised, except for the special case of $\effL = 1/2$ when the required feedback gain diverges. If the cavity is overcoupled, the efficiencies satisfy $\effbin + \effbopt + \effbol = 1$. However, if $\effL < 1/2$, the efficiencies sum to the larger-than-unity value $1+8\effL\CL/(4\lwfact\CL + 1)$---an observation we will return to shortly. 

After accounting for coupling inefficiencies, the optical-to-microwave transmission is finally given by $\Tac(\omega) = |\tac(\omega)|^2 = \effM \effbopt \Tinf(\omega)$, where $\Tinf(\omega) = |\tinf(\omega)|^2$ is still given by Eq.~\eqref{eq:tinf} and shown in Fig.~\ref{fig:transfer-coeff-freq}. However, $\Tinf(\omega)$ now depends on the updated definition of $\Gamma'$ in Eq.~\eqref{eq:Gamma'_inefficient}, both directly and through the electromechanical cooperativity $\CM' = 4\gM^2 / \Gamma'\kM$. The microwave coupling efficiency $\effM$ linearly reduces the transmission $\Tac(\omega)$, while the optical coupling efficiency $\effL$ impacts the transmission through $\effbopt = 4\effL\CL / (4\lwfact \CL + 1)$. In the high-cooperativity limit, the maximum transmission tends to 
\begin{eqnarray}
    \Tac^\text{max} &= \lim_{\CL \to \infty} \effM \effbopt =  \effM \effL / \lwfact \nonumber \\
    &= \cases{\effM |2\effL - 1| & $\effL > 0$ \\ 0 & $\effL = 0$}. \label{eq:Tac_max}
\end{eqnarray}

Surprisingly, for an undercoupled cavity with $\effL < 1/2$, $\Tac^\text{max}$ is maximal for small $\effL \to 0$. This counter-intuitive results stems from the fact that for such a cavity,
the mechanical resonator $b(\omega)$ is coupled to $\ain(-\omega)^\dagger$, so that the feedback in fact induces a two-mode squeezing (parametric gain) process characterised by conversion gains $T_{jc}$, rather than a state transfer. The input phase quadrature $\YLin$ is then reflected strongly off the undercoupled cavity and fed back with high gain $G \propto \CL$, providing the amplification that allows the total efficiency $\effbin + \effbopt + \effbol > 1$ to exceed unity.

\section{Added noise}
In addition to a high transfer efficiency $\Tac(\omega)$, faithful transfer of quantum correlations requires minimal added noise. We therefore calculate the noise contributions to the power spectral density (PSD) of the quadratures $\XMout = (\cout^\dagger + \cout) / \sqrt{2}$ and $\YMout = \rmi (\cout^\dagger - \cout) / \sqrt{2}$ of the microwave output field. Starting with the amplitude quadrature $\XMout$, its PSD is given by \cite{Warwick_Textbook}
\begin{eqnarray}
    \SXMout(\omega) = \int_{-\infty}^\infty \rmd \omega' \langle \XMout(\omega) \XMout(\omega') \rangle. \label{eq:def_SXM}
\end{eqnarray}
As we see from Eq.~\eqref{eq:cout_inefficient}, the frequency component of the optical and mechanical inputs at $+\Omega$ is transferred onto the microwave output at resonance.
We thus define the frequency-shifted (sideband) operators $\SBp{\mathcal{O}}(\omega) = \mathcal{O}(\omega + \Omega)$ for each of these inputs, along with the sideband quadratures 
$\pXLjT(\omega) = [a_{j}^{T,\SBpSymb}(-\omega)^\dagger + a_{j}^{T,\SBpSymb}(\omega)] / \sqrt{2}$, $\pYLjT(\omega) = \rmi[a_{j}^{T,\SBpSymb}(-\omega)^\dagger - a_{j}^{T,\SBpSymb}(\omega)] / \sqrt{2}$ and similarly $\SBp{\Qin}, \SBp{\Pin}$ for $\SBp{\bin}$.

Assuming all inputs are uncorrelated, we find the expression 
\begin{eqnarray}
    &\SXMout(\omega) = \Tac(\omega) \SpXLin(\omega) + \Tbc(\omega) \SpQin(\omega) \nonumber \\
    &+ \Tcc(\omega) \SXMin(\omega) + \Talc(\omega) \SpXLl(\omega) \\
    &+ \Tclc(\omega) \SXMl(\omega)  + \Tvc(\omega) \SYv(\omega) \nonumber
\end{eqnarray}
that relates $\SXMout(\omega)$ to the PSDs of the input channels
via the transmission coefficients $T_j(\omega) = |t_j(\omega)|^2$ corresponding to 
Eq.~\eqref{eq:tgains_inefficient}. At temperatures relevant for quantum information processing, the microwave environment constituted by $\cin$ and $\cl$ is in its ground state, so that $\SXMin(\omega) = \SXMl(\omega) = 1/2$. The same holds for the optical environment formed by $\al$. 
However, as this couples to the resonator via the rotating squeezed input $\alT(\omega+\Omega)$, the relevant spectral contribution is the average variance of the quadratures of $\alT(\omega)$ (see Eq.~\eqref{eq:alT_def}). This is larger than the vacuum noise in $\al$ and given by 
$\SpXLl(\omega) = (2\lwfact + 1/2\lwfact) / 4 \geq 1/2$ (SI Section~\ref{sec:apx:optical_loss_noise}). Finally, the fluctuations in the lower-frequency mechanical environment contain a thermal component $\bar{n}$ and satisfy $\SpQin(\omega) = \bar{n} + 1/2$, while the noise spectrum of the detection loss is given by $\SYv(\omega) = 1/2$ (SI Section~\ref{sec:apx:detection_noise}).

The total noise in the microwave output amplitude quadrature is thus given by
\begin{eqnarray}
    &\mathcal{S}^\text{noise}[\XMout](\omega) = \frac{1}{2}\Big[
    \overbrace{\effM \effbol \frac{1}{2} (2\lwfact + 1/2\lwfact) \Tinf(\omega)}^\text{optical} \nonumber \\
    &\quad+ \underbrace{\effM \effbin (2\bar{n} + 1) \Tinf(\omega)}_\text{mechanical} 
    + \underbrace{(1-\effM \Tinf(\omega))}_\text{microwave} \label{eq:output-noise-specific} \\
    &\quad+ \underbrace{\frac{\effM \effbopt}{(2\effL - 1)^2} \frac{1 - \effd}{2\effd} \Tinf(\omega) }_\text{detection}
    \Big], \nonumber
\end{eqnarray}
while the signal contribution equals $\mathcal{S}^\text{signal}[\XMout](\omega) = \effM \effbopt \Tinf(\omega) \, \SpXLin(\omega)$. Because our transfer model is quadrature-symmetric, Eq.~\eqref{eq:output-noise-specific} describes the noise $\mathcal{S}^\text{noise}[\YMout](\omega) = \mathcal{S}^\text{noise}[\XMout](\omega)$ in the orthogonal microwave output quadrature $\YMout$ as well.



For a monochromatic input mode, the noise variance added to the output channel is simply the noise spectrum evaluated at the signal detuning \(\sigdet\) from the upper optical sideband (SI Section~\ref{sec:apx:variance_narrowband_sigs})
\begin{eqnarray}
    \VXadd =\mathcal{S}^\text{noise}[\XMout](\sigdet)\, . \label{eq:V_PSD_relation}
\end{eqnarray}
Again, the quadrature-symmetry of the transfer model grants \(\VXadd = \VYadd\). We next introduce the notation 
\begin{eqnarray}
    \VXadd &= V_\text{opt} + V_\text{mw} + V_\text{mech} + V_\text{det} \, \label{eq:V_contributions}
\end{eqnarray}
to designate each contribution to the noise variance identified in Eq.~\eqref{eq:output-noise-specific}. We may further simplify our analysis by observing that the frequency and electomechanical cooperativity \(\CM'\) dependence of the noise spectrum is folded entirely into 
\(\Tinf(\omega)\). Fig.~\ref{fig:transfer-coeff-freq} identifies that unity \(\Tinf(\sigdet)\) can be readily achieved by tuning \(\CM'\) and choosing an 
appropriate\footnote{Explicitly, either: setting zero detuning \(\sigdet = 0\) from the blue optical sideband and tuning the microwave pump so that \(\CM'=1\); or, for \(\CM'\geq 1\) and matched linewidths \(\Gamma' = \kappa_M\), choosing a signal detuning \(\sigdet = \pm (\Gamma'\sqrt{\CM'-1})/2 \rightarrow \pm g_M\) (for \(\CM' \gg 1\)).\label{footnote:optimal_detunings}} 
signal detuning, \(\sigdet\). 
For this reason, we take \(T_\infty(\sigdet)=1\) in the proceeding calculations and refer to this condition as `matched transfer'.

\section{Transducer performance for state-of-the-art parameters}
\label{sec:state_of_the_art_params}

In the following section, we show that our protocol can achieve competitive performance for realistic system parameters, on par with the leading conversion efficiencies of cold atom scattering approaches \cite{tu_cold-atom-off-res-scattering-record-efficiency_2022, kumar_Rb_bulk_transducer_2023} while adding noise beneath the vacuum level. The latter is necessary for preserving the Wigner negativity of the input state \cite{leonhardt_measuring_1995, sahu_electro-optic-effect_2022} and unlocks a suite of quantum-compatible applications, for example, heralded entanglement of remote qubits \cite{mirhosseini_qubit-to-optical-photon_2020, zeuthen_transducer-figures-of-merit_2020}.

For definiteness, we consider a $\Omega/2\pi = 1$~MHz mechanical resonator with $\mathcal{Q} = 10^7$ and $\bar{n} = 10^3$ (bath temperature $T \sim 50$~mK), coupled to an optical cavity that can be operated in the back-action-dominated regime $\CL \gg \bar{n}$~\cite{andrews_bidirectional_membrane_2014,higginbotham_harnessing-electro-optic-correlations_2018,delaney_qubit-readout-via-electro-optic-transducer_2022,brubaker_optomechanical-ground-state-cooling-electro-optic-transducer_2022,arnold_electro-opto-mechanical_2020,Groblacher_feedback-cooling-ground-state_2019, rossi_measurement-based-mech-q-control_2018}. Crucially, as our protocol does not require optical sideband resolution, we take it to be highly overcoupled with $\effL = 0.95$. We assume a homodyne detection efficiency of $\effd = 0.85$~\cite{rossi_measurement-based-mech-q-control_2018, Groblacher_feedback-cooling-ground-state_2019}. In addition, the resonator couples to a sideband-resolved microwave resonator with electromechanical cooperativity $\CM$ on the order of $\CL$ (so that we can set $\CM' \geq 1$)~\cite{yuan_large-Cem-bulk-superconducting-cav_2015} and outcoupling efficiency $\effM = 0.98$~\cite{xu_radiative-cooling_2020, Han_microwave-optical-review_2021, wu_mech-supermode-piezo-transducer-params_2020}. While still challenging to attain in an integrated hybrid device, these parameters have been demonstrated in state-of-the-art optomechanical and electromechanical components (SI Section~\ref{sec:apx:state_ot_art_params}).

\begin{figure}[t]
    \centering
    \includegraphics[width=\linewidth]{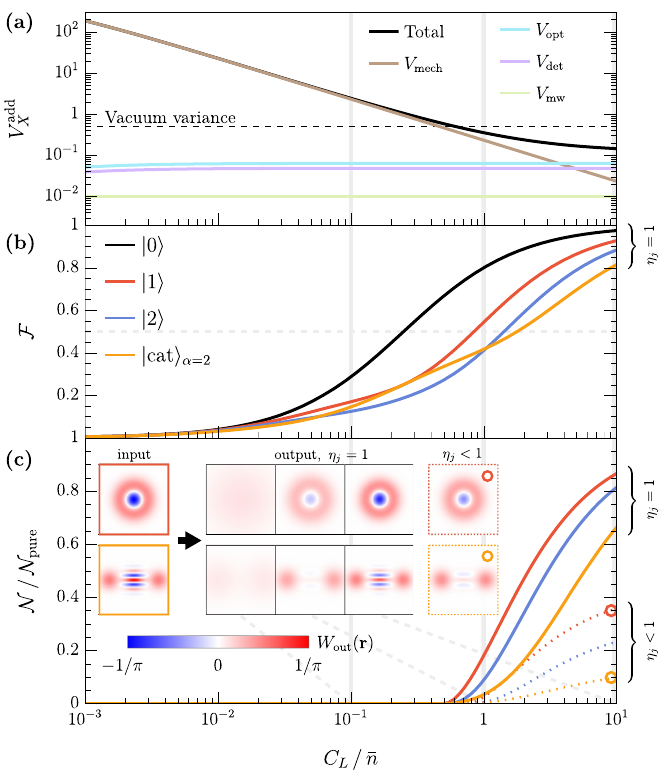}
    \caption{\textbf{Transducer performance with optomechanical cooperativity}. (a) Contributions to the noise variance in Eq.~\eqref{eq:V_contributions} added to the output microwave field amplitude as a function of \(\CL/\nth\). 
    Here, \((\eta_L, \eta_M, \eta_d) = (0.95,0.98, 0.85)\) and \(\nth = 10^3\). (b) Transfer fidelity \(\mathcal{F}\) (Eq.~\eqref{eq:fidelity_def}) for various pure input states and unit efficiencies. (c) Total negativity \(\mathcal{N}\) (Eq.~\eqref{eq:negativity_def}) of the output Wigner function relative to that of the pure input state, \(\mathcal{N}_\text{pure}\). Solid curves indicate unit efficiencies (\(\eta_L, \, \eta_M, \, \eta_d = 1\)), while dotted curves indicate the efficiency choices in (a). Insets: output Wigner function \(W_\text{out}(\mathbf{r})\) (Eq.~\eqref{eq:def_W_out}) in phase space \(\mathbf{r}=(X,Y)^T\) generated at the cooperativities identified by grey vertical lines (left-to-right: \(\CL/\nth=0.1,\, 1, \, 10\)). The pure input state is identified with a solid colour border. The transferred state at \(\CL/\nth = 10\) for the efficiencies in (a) is identified with a dotted border. 
    }
    \label{fig:CL_noise_analysis}
\end{figure}

For this choice of system parameters, Fig.~\ref{fig:CL_noise_analysis}a shows the total added noise (solid black) as a function of the optomechanical cooperativity ratio \(\CL/\nth\). 
In the high cooperativity limit \(C_L \gg 1\), the extraction efficiencies \(\eta_L,\eta_M\) (blue, green) and feedback detection loss \(\eta_d\) (purple) contribute to a noise floor independent of \(C_L\). 
In the same limit, the thermal mechanical noise (brown) tends to \(V_\text{mech} \approx (\eta_M /4\lwfact)(\nth/C_L)\), so that $V_\text{mech}$ drops below the vacuum level when \(C_L \gg \nth\), coinciding with the familiar condition for ground-state feedback cooling~\cite{Warwick_Paper, Warwick_Textbook}. For our realistic inefficiencies and \(\CL/\nth = 10\), the total added noise variance is a factor of \(3.4\) below the vacuum noise level (dashed black), dominated by the optical losses $\effL$ and $\effd$. 
The corresponding transmission efficiency \(\Tac \approx 0.88\) is on par with record-holding cold atom scattering transducers (82\%)~\cite{tu_cold-atom-off-res-scattering-record-efficiency_2022}. We note that moderate improvements to the cavity extraction efficiency (\(\eta_L \rightarrow 0.98\)) and mechanical quality (\(\mathcal{Q}\sim 10^8\) --- to permit \(\CL/\nth = 10^2\) with \(\mathcal{Q}' \gg 1\)), alongside a marked improvement in detection efficiency (\(\eta_d \rightarrow 0.96\)), allow the added noise to drop below $10$ times vacuum noise. These also yield a transmission efficiency \(\Tac \rightarrow 0.94\).

To model the degradation of a pure input state after transfer, we combine the total added noise variance and the Wigner function formalism. Given that the output microwave field in Eq.~\eqref{eq:cout_inefficient} is linear in each of its Gaussian noise sources, the Wigner function of the output state is given by the convolution (SI Section~\ref{sec:apx:transferred_wigner})
\begin{eqnarray}
    W_\text{out}(\mathbf{r}) = (W_\text{in}'\circledast \mathcal{G})(\mathbf{r}), \label{eq:def_W_out}
\end{eqnarray}
where \(W_\text{in}'(\mathbf{r})=W_\text{in}(\mathbf{r}/|\tac|)/|\tac|^2\) is the Wigner function of the input optical mode rescaled by the transfer gain, \(\mathcal{G}\) is a Gaussian noise kernel determined by the added noise variance, and \(\mathbf{r}=(X,Y)^T\) is a phase space coordinate~\cite{bennett_PhD_2017}. Since the added noise is quadrature-symmetric,
the convolution in Eq.~\eqref{eq:def_W_out} applies a symmetric Gaussian blur with variance \(\VXadd\) to the gain-modified input state, where  \(\VXadd = 1/2\) corresponds to convolving the input with a vacuum state. The transfer fidelity is then given by the overlap integral \cite{bennett_PhD_2017} 
\begin{eqnarray}
    \mathcal{F} &= 2\pi \int \rm{d}^2 \mathbf{r} \,\,  W_\text{in}(\mathbf{r}) W_\text{out}(\mathbf{r}). \label{eq:fidelity_def} 
\end{eqnarray}

In Fig.~\ref{fig:CL_noise_analysis}b, we present the transfer fidelity as a function of the cooperativity ratio \(\CL/\nth\) for input Fock states \(\ket{0},\ket{1}, \ket{2}\) and a cat state \(\ket{\text{cat}}_{\alpha=2}\) (SI Section~\ref{sec:apx:transferred_wigner}), for unit efficiencies so that $\Tac \to 1$. The noise is then entirely thermomechanical, so that an input vacuum state (black) surpasses the \(\mathcal{F}=1/2\) coherent state classical fidelity limit at \(\CL/\nth = 1/4\) (when \(\VXadd=1\)) and the \(\mathcal{F}=2/3\) coherent state no-cloning bound at \(\CL/\nth = 1/2\) (when \(\VXadd = 1/2\)). For the remaining non-classical states considered, near-unity transfer fidelities \(\mathcal{F}=0.93, \, 0.88, \, 0.81\) are obtained at \(\CL/\nth = 10\) for the single photon \(\ket1\) (red), two photon \(\ket2\) (blue) and cat state \(\ket{\text{cat}}_{\alpha=2}\) (orange), respectively. At the same cooperativity, transfer of the vacuum attains a fidelity \(\mathcal{F}=0.98\).


We then study the effect of losses. Remarkably, the transfer is highly robust to $\effd$---as reported in Ref.~\cite{Warwick_Paper}. Detection loss only contributes noise \(V_\text{det} \propto (1-\eta_d)/\eta_d\) without degrading the transfer efficiency. For \(\CL/\nth = 10\) and unit cavity extraction efficiencies, vacuum transfer can still attain \(\mathcal{F}=1/2\) with 80\% detection loss, and all non-classical states considered in Fig.~\ref{fig:CL_noise_analysis} attain \(\mathcal{F} > 1/2\) for \(\eta_d \gtrsim 0.7\) (SI Section~\ref{sec:apx:insensitivity_feedback_loss}). 

To characterise the transducer's sensitivity to cavity extraction efficiencies $\effL$ and $\effM$, we need to consider the resulting non-unity transfer gain $\Tac < 1$. In that case, fidelity is no longer considered an appropriate figure of merit~\cite{bowen_M_2003}. We will address this explicitly in the next section. However, another figure of merit that can still provide insight in the quality of transfer is the integrated volume of the negative part of the Wigner function \cite{kenfack_wigner-negativity_2004, hoff_measurement-induced-macroscopic-sups-pulsed_2016}
\begin{eqnarray}
    \mathcal{N} &= \bigg(\int \rm{d}^2\mathbf{r} \, |W_\text{out}(\mathbf{r})|\bigg) - 1, \label{eq:negativity_def}
\end{eqnarray}
relative to that of the pure input state, \(\mathcal{N}_\text{pure}\). Wigner negativity is a direct signature of the quantum nature of a state: its existence implies that the state cannot be described in terms of a classical phase space probability distribution \cite{kenfack_wigner-negativity_2004}. The ratio \(\mathcal{N}/\mathcal{N}_\text{pure}\) therefore directly indicates whether a non-unity gain channel maintains non-classical features of the input.

The dotted lines in Fig.~\ref{fig:CL_noise_analysis}c identify the transferred negativity \(\mathcal{N}/\mathcal{N}_\text{pure}\) for our chosen system parameters including losses, as used in Fig.~\ref{fig:CL_noise_analysis}a. We see that, at \(\CL/\nth=10\), appreciable negativity ratios \(\mathcal{N}/\mathcal{N}_\text{pure} = 0.36, \, 0.23, \, 0.10\) are maintained for the single photon, two photon and cat state inputs, respectively. For comparison, the corresponding \(\mathcal{N}/\mathcal{N}_\text{pure}\) in the absence of losses are shown as solid lines.
In general, Wigner negativity is fully erased (\(\mathcal{N}/\mathcal{N}_\text{pure}=0\)) when more than vacuum-equivalent noise is added, as observed in Fig.~\ref{fig:CL_noise_analysis}c for all states considered once \(\CL/\nth < 1/2\) and \(\VXadd > 1/2\). The loss of negativity is visualised in the insets to Fig.~\ref{fig:CL_noise_analysis}c, which display Wigner functions of the single photon \(\ket1\) and cat \(\ket{\text{cat}}_{\alpha=2}\) state for increasing cooperativity ratios. At low \(\CL/\nth \ll 1\), we recognize the destructive noise-induced Gaussian blur, while for \(\CL/\nth \gg 1\) features sharpen and negativity (blue) is recovered.

\section{A quantum transfer witness \(\qtw\) to quantify non-unity gain performance}
\label{sec:quantum_transfer_witness}

Extra care must be taken to characterise the transducer's sensitivity to the cavity extraction efficiencies \(\eta_L, \eta_M\) since the transfer gain depends linearly on each (see Eq.~\eqref{eq:Tac_max}). Importantly, the fidelity then becomes a somewhat ambiguous figure of merit, given its extreme sensitivity to that gain \cite{bowen_M_2003}. While the total Wigner negativity can complement fidelity, both are input-state-dependent metrics.
A popular input-independent figure of merit to compare transducer performance across platforms is the input-referred added noise, which normalises the noise at the output by the conversion efficiency \cite{Han_microwave-optical-review_2021, zeuthen_transducer-figures-of-merit_2020}. In this section, we introduce a figure of merit that is similar yet unlocks further insight into the transducer's capacity to outperform classical protocols and preserve Gaussian entanglement. 
 
The transmission efficiency and added noise are conventionally analysed together in a parametric transmission-noise (T-V) diagram. The ratio of the two -- the added noise referenced to the input channel -- condenses the information of the T-V diagram into a single-valued figure of merit. Similarly, we introduce the \textit{quantum transfer witness} \(\qtw\), based on the gain-normalised conditional variance product originally proposed in Ref.~\cite{bowen_M_2003},
\begin{eqnarray}
    \qtw &= \frac{\sqrt{4 \, \VXadd \, \VYadd}}{|g_X g_Y| + 1},\label{eq:def_quantum_transfer_witness}
\end{eqnarray}
defined for a general frequency-domain channel of the form \(X_\text{out} = g_X X_\text{in}+ X_\text{noise}\) (\(X\leftrightarrow Y\), \([X_\text{in}, Y_\text{in}]=i\)). Most generally, Eq.~\eqref{eq:def_quantum_transfer_witness} normalises the geometric mean of the output noise across the quadratures by a shifted transfer efficiency. For quadrature-symmetric transfer gains \(g_X = g_Y\) (\(|g_X|^2=T\)) and output noise \(\tilde{V}^\text{add} = 2\VXadd\) (such that it has unit ground-state variance), Eq.~\eqref{eq:def_quantum_transfer_witness} reduces to \(\qtw =  \tilde{V}^\text{add}/(T+1)\). Indeed, this is precisely the input-referred added noise with instead \(T\rightarrow T+1\). This subtle difference equips \(\qtw\) with two interesting properties. Firstly, a transducer leveraging only local operations and classical communication (LOCC) necessarily satisfies \(\qtw \geq 1\) (SI Section~\ref{sec:apx:witness_beyond_LOCC}) \cite{bowen_M_2003}. Secondly, we show that \(\qtw = \mathcal{I}\), where \(\mathcal{I}\) is the degree to which the Simon separability criterion~\cite{simon_separability_2000} is violated after one mode of a perfectly entangled two-mode Gaussian state encounters degradation of the above channel form (SI Section~\ref{sec:apx:witness_entanglement}). Therefore, \(\qtw<1\) is both a condensed signature for identifying advantage beyond LOCC and a necessary and sufficient condition for a transducer to maintain the inseparability of a perfectly entangled Gaussian input. That is, \(\qtw\) is a witness of the ability to transfer two-mode Gaussian entangled states. It is important to note that \(\qtw \leq 2 \VXadd\) for quadrature-symmetric transfer, and so \(\VXadd<1/2\) is indeed a stronger condition than \(\qtw<1\) -- i.e.,  preserving Gaussian entanglement is less demanding than preserving Wigner negativity.

\begin{figure}[t]
    \centering
    \includegraphics[width=\linewidth]{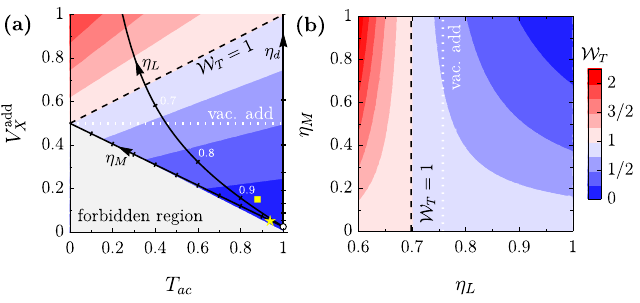}
    \caption{Quantum transfer witness \textbf{\(\qtw\) as a figure of merit for non-unity gain transfer}. (a) The signal transmission coefficient \(T_{ac}\) (Eq.~\eqref{eq:Tac_max}) and added noise variance \(\VXadd\) (Eq.~\eqref{eq:V_contributions}) traced as a function of each loss parameter \(\eta_L, \, \eta_M, \, \eta_d\) (black lines) from an initial starting point where \(\eta_L, \, \eta_M, \, \eta_d = 1\) and \(\CL/\nth=10\) (white circle). Black ticks along the parametric curves indicate \(10\%\) loss increments. In the blue region where \(\qtw<1\) (see Eq.~\eqref{eq:def_quantum_transfer_witness}), performance exceeds that of an LOCC transducer and the inseparability of a perfectly entangled Gaussian input is preserved. The grey forbidden region is identified from Heisenberg uncertainty requirements (SI Section~\ref{sec:apx:witness_beyond_LOCC}). Gold square: \((\eta_L, \eta_M, \eta_d, \CL/\nth) = (0.95, 0.98, 0.85, 10)\). Gold star: \((\eta_L, \eta_M, \eta_d, \CL/\nth) = (0.98,0.98,0.96, 100)\). (b) 
    The transducer's capacity to preserve entanglement (\(\qtw<1\)) is insensitive to microwave cavity external coupling losses \(\eta_M\), and relatively insensitive to that of the optical cavity \(\eta_L\). Here, \(C_L/n_\text{th}=10\) and \(\eta_d = 0.85\).
    }
    \label{fig:TV+W}
\end{figure}

We now apply this metric to our transducer. In Fig.~\ref{fig:TV+W}a, the solid black parametric curves trace how the added noise \(\VXadd\) and transfer efficiency \(T_{ac}\) change as a function of each inefficiency, \(\eta_L, \eta_M, \eta_d\), starting from a point of unit-efficiency (\(\eta_L, \eta_M, \eta_d = 1\)) and \(\CL/\nth = 10\) (white circle, bottom right). Ticks on each curve indicate 10\% loss increments (\(\eta_i = 1, 0.9, 0.8, ...\)). Overlaid are coloured contours that indicate the value of \(\qtw\) in the T-V parameter space (blue: \(\qtw<1\), red: \(\qtw>1\)). Since detection losses \(\eta_d\) add noise without compromising the transfer efficiency, the corresponding parametric curve is vertical from the starting point (see right border of Fig.~\ref{fig:TV+W}a). Again, the transfer is highly resilient to detection losses: only exceeding vacuum-equivalent added noise (white dotted line) for \(\eta_d < 1/3\) while the \(\qtw=1\) threshold (black dashed line) is breached for \(\eta_d  < 1/5\)  (for \(\eta_L,\eta_M = 1\), \(\CL/\nth \gg 1\)). Note that the tick-spacing along the parametric curve is non-linear, given that the detection noise scales as \(V_\text{det} \propto (1-\eta_d)/\eta_d\).

Conversely, microwave cavity losses \(\eta_M\) linearly degrade the transmission efficiency (see Eq.~\eqref{eq:Tac_max}) and all other noise contributions (i.e. \((\VXadd-V_\text{mw}) \propto \eta_M\)) by substituting the cavity output with microwave vacuum as \(\eta_M \rightarrow 0\). For this reason, the \(\eta_M\) parametric curve in Fig.~\ref{fig:TV+W}a is linear in the T-V parameter space with equal tick spacing. 
Correspondingly, the transfer is also resilient to fluctuations in the microwave out-coupling efficiency \(\eta_M\). 
Provided the microwave environment is in the ground state, additional microwave cavity losses cannot increase the added noise beyond the vacuum level. Consequently, \(\qtw < 1\) for all values of \(\eta_M\) with \(\eta_L,\eta_d = 1\) and \(\CL/\nth \gg 1\) (confer black dashed line, Fig.~\ref{fig:TV+W}a). This is indeed an established property of inseparability, whereby loss -- on its own -- cannot transform an inseparable state into a separable state~\cite{duan_inseparability_2000,bowen_CV_entanglement_2003,bowen_CV_entanglement_2004}. 

The non-linearity and far larger tick-spacing along the optical loss \(\eta_L\) parametric curve in Fig.~\ref{fig:TV+W}a illustrates that inefficient optical cavity extraction is most detrimental to the transfer performance. Such losses imprint the  squeezed optical vacuum in Eq.~\eqref{eq:alT_def} onto the output with variance~\(V_\text{opt}/\eta_M \rightarrow \lwfact/2\) for~\(\{C_L, \lwfact\} \gg 1\). Degradation to \(\eta_L\) also increases the feedback detection noise \(V_\text{det} \propto (2\eta_L -1)^{-1}\) (for \(C_L \gg 1\)) due to the compensating change in the feedback gain that enables transfer without squeezing.
The same compensating change in feedback gain broadens the mechanical linewidth (see Eq.~\eqref{eq:Gamma'_inefficient}), offering a generally minor boost to the thermal noise suppression \(V_\text{mech} \propto 1/\lwfact\) (for~\(\CL/\nth \gg 1\)). 
The high-cooperativity transmission coefficient \(T_{ac}^\text{max}\) in Eq.~\eqref{eq:Tac_max} is also twice as susceptible to optical in-coupling inefficiencies than that of the microwave cavity. Therefore, the optical loss parametric curve in Fig.~\ref{fig:TV+W}a breaches the vacuum noise threshold (white dotted line) for \(\eta_L \lesssim 0.72\) and the \(\qtw=1\) threshold (black dashed line) for \(\eta_L \lesssim 0.67\) (with \(\eta_M, \eta_d = 1\), \(\CL/\nth \gg 1\)).

We separately analyse the sensitivity of the quantum transfer witness \(\qtw\) to the cavity extraction efficiencies in Fig.~\ref{fig:TV+W}b. Here, we see that for \(\eta_d = 0.85\) \cite{rossi_measurement-based-mech-q-control_2018}, up to 30\% optical cavity extraction loss can be sustained before the inseparability of a perfectly entangled two-mode Gaussian state is lost after one mode is transduced (\(\qtw\geq 1\)). This threshold is markedly independent of the microwave extraction efficiency, as previously noted. 

Overall, our analysis suggests that optimising the optical cavity extraction efficiency is critical for effective transducer performance, since the feedback mechanism compounds the effect of optical losses. Fortunately, our protocol accommodates for this. The optomechanical system can operate deeply in the sideband-unresolved regime, so that the external coupling rate can, in principle, be made sufficiently large to permit overcoupling \(\eta_L \rightarrow 1\). Comparatively, the transfer is relativity insensitive to feedback detection and microwave cavity losses. Indeed, operation deeply within both the entanglement- and negativity- preserving regimes \(\qtw < 1\) and \(\VXadd < 1/2\), respectively, is readily attainable with our chosen realistic parameters (gold square, Fig.~\ref{fig:TV+W}a), while moderate improvements discussed unlock greater than \(90\%\) transfer efficiency and noise performance a factor of 10 below the vacuum-add threshold (gold star, Fig.~\ref{fig:TV+W}a).

\section{Using classical feedback to transfer a quantum state}

At first sight, the possibility of transferring quantum information using classical measurement and feedback may seem counter-intuitive. In quantum teleportation, similar transfer is made possible by making joint measurements of an input state that has been interfered with one half of an entangled ancilla. The entanglement acts to shroud the properties of the input, so that---in the limit of ideal entanglement---the photocurrents which carry the classical signal reveal no information about the input state. In this section, we show that a similar mechanism allows faithful transfer without state collapse in our scheme.

To do so, we quantify the content of the classical homodyne photocurrent by its symmetrised spectral density
    $\bSYLout(\omega) = [\SYLout(\omega) + \SYLout(-\omega)] / 2$. 
For simplicity, we assume perfect detection and cavity coupling efficiencies. After turning on the electromechanical interaction, the output phase quadrature can be expressed in terms of the optical, mechanical and microwave input fields as 
$\YLout(\omega) = \rmi [a_\text{out}(-\omega)^\dagger - a_\text{out}(\omega)]/\sqrt{2}$ with 
\begin{eqnarray}
    a_\text{out}(\omega) = &-\taa(\omega) \ain(\omega) + \rmi \tba(\omega) \bin(\omega) \nonumber \\
    &+ \tca(\omega) \cin(\omega-\Omega) \label{eq:homodyne_from_inputs}
\end{eqnarray}
using the input-to-light transfer gains (SI Section~\ref{sec:apx:homodyne_content})
\begin{eqnarray}
    \taa(\omega) &= 1 - \effbopt \Gamma' \chibld(\omega) \nonumber \\
    \tba(\omega) &= \sqrt{\effbopt \effbin} \Gamma' \chibld(\omega) \label{eq:input_to_light_gains} \\
    \tca(\omega) &= \sqrt{\effbopt} \tinf(\omega-\Omega). \nonumber
\end{eqnarray}

\begin{figure}
    \centering
    \includegraphics{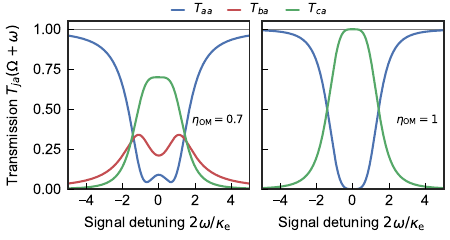}
    \caption{\textbf{Information content of the homodyne photocurrent.} Transmission coefficients $T_{ja}(\Omega + \omega)$ of the optical (blue), mechanical (red) and microwave (green) input fields to the optical output around the upper mechanical sideband. Here, $\CM' = 1$ and $\beta = 1$. For finite cooperativity $\CL$ and limited transfer efficiency $\effbopt = 0.7$ (left), the optical output is contains contributions from each channel. In the limit $\CL \to \infty$ where $\effbopt \to 1$ (right), optical and mechanical fluctuations are fully suppressed at $\omega = \Omega$, and replaced by noise from the microwave input channel. In all cases, $\Taa(\omega) + \Tba(\omega) + \Tca(\omega) = 1$.}
    \label{fig:homodyne_content}
\end{figure}


Around the frequency $-\Omega$ of the lower mechanical sideband, the photocurrent $\YLout(\omega)$ is fully dominated by the optical input, as $\taa(\omega) = 1$ and $\tba(\omega) = \tca(\omega) = 0$ for $\omega \approx -\Omega$. In contrast, at the upper mechanical sideband $\omega \approx +\Omega$, the light-to-light gain $\taa(\omega)$ is strongly suppressed, so that the homodyne measurement does not reveal the state of the input field $\ain(\Omega)$. Instead, optical fluctuations are replaced by microwave and mechanical noise, as we see from the transmission coefficients $T_j(\omega) = |t_j(\omega)|^2$ plotted in Fig.~\ref{fig:homodyne_content} around $\omega \approx \Omega$. 
Moreover, in limit of high cooperativity $\CL \to \infty$, where $\effbopt \to 1$, the mechanical contribution $\tba(\omega)$ is fully suppressed as the resonator is effectively decoupled from its mechanical environment ($\effbin \to 0$).

We will now show that these transfer gains indeed determine the information content of the homodyne photocurrent. The mechanical and microwave contributions to $\bSYLout(\omega)$ are readily calculated, since we expect the corresponding environments to be a thermal state or ground state, respectively. The optical environment instead embeds the arbitrary quantum state to be transferred, and therefore requires more care. A general optical input \(\ket\psi\) may be constructed as a coherent superposition of environmental excitations \(\ain(\omega)^\dagger\) weighted by a spectral mode shape \(\xi(\omega)\) \cite{Warwick_Textbook}. A single-photon pulse \( \ket{1_\xi} = Z^\dagger \ket0\) is then created by the operator \(Z^\dagger = \int_{-\infty}^\infty \rmd\omega\, \xi(\omega) \ain(\omega)^\dagger\), which inherits the usual ladder operator properties (SI Section~\ref{sec:apx:construct_optical_input}). We find that, for a general input state $\ket\psi$ constructed using \(Z\), the optical contribution to $\bSYLout(\omega)$ can be conveniently expressed in terms of the light-to-light transfer gain \(T_{aa}(\omega)\) and a quadrature \(B_\theta = \rmi (Z^\dagger e^{-\rmi \theta} - Z e^{\rmi \theta}) / \sqrt{2}\) of the pulse operator. The single-sided spectrum of the total photocurrent then reads
\begin{eqnarray}
    \spectSymb[\YLout](\omega) &= \frac{1}{2} + \frac{\bar n }{2} \Tba(\omega) \nonumber \\
    &+\Taa(\omega_p) x(\omega) \bra\psi B_\theta^2 - 1/2\ket\psi, \label{eq:homodyne_spectrum_main}
\end{eqnarray}
where $x(\omega)$ is proportional to the modeshape $\xi(\omega)$ and the quadrature angle $\theta$ sums the phases of the optical pulse and the reflection $\taa(\omega_p)$. 
Here, we have assumed that the input pulse \(\xi(\omega)\) (with centre frequency \(\omega_p\)) is spectrally much narrower than the bandwidth of the transfer gain \(T_{aa}(\omega)\).

From Eq.~\eqref{eq:homodyne_spectrum_main}, we identify that information about the input state, as encoded in the quadrature variance $\langle B_\theta^2 \rangle$, is imprinted on the spectrum of the classical photocurrent with intensity $\Taa(\omega_p)$. 
Importantly, as we increase the optomechanical cooperatitivity, the efficiency $\effbopt$ tends to one so that $\Taa(\omega_p)$ is minimised. 
Explicitly, in the limit $\effbopt \to 1$, the light-to-light transfer coefficient reduces to $\Taa(\omega_p) = 1 - \Tinf(\omega_p-\Omega)$. Then, by matching the pulse centre frequency $\omega_p = \Omega + \sigdet$ to a value close to the positive sideband where the light-to-microwave transfer efficiency $\Tinf(\sigdet) \approx 1$ is high, we find that $\Taa(\omega_p) \approx 0$---indicating that, indeed, the photocurrent does not reveal the input state.

\begin{figure}
    \centering
    \includegraphics[]{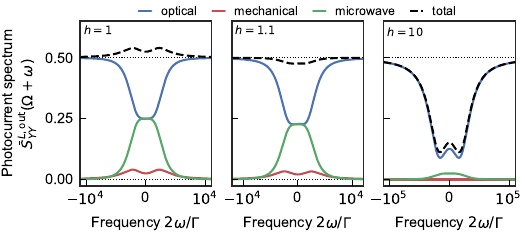}
    \caption{\textbf{Spectrum of the classical homodyne photocurrent and noise squashing.} Contributions from optical (blue), mechanical (red) and microwave (green) inputs to the classical photocurrent spectrum are plotted around the mechanical frequency $\Omega$, for increasing values of the feedback gain $h=G / 8\CL$. Here, $\CL = 500$, $\CM' = 1$, $\beta = 1$, $\bar{n} = 100$ and $\Gamma'/\Gamma = 2001$. The optical input $\ain$ to be transferred is in the vacuum state. At the optimal gain $h=1$, where light-to-microwave transfer happens without squeezing, the contribution of the upper optical sideband is replaced by microwave noise. The remaining optical fluctuations are provided by the lower sideband, while feedback-cooled thermomechanical noise raises the total photocurrent spectrum above the quantum noise level of $1/2$. For higher values of gain $h$, noise squashing occurs where the total spectrum drops below the shot noise.}
    \label{fig:homodyne_spectrum}
\end{figure}

Next, to study the role of the feedback system, we plot the different contributions to $\bSYLout(\omega)$ in Fig.~\ref{fig:homodyne_spectrum}, for increasing values of the gain $h = G/8\CL$ (SI Section~\ref{sec:apx:homodyne_content_vacin_arbgain}). As an example, we consider the optical input $\ain$ in the ground state. At the optimal gain $h=1$ (left panel), where light-to-microwave transfer happens without squeezing, the noise contribution of the upper optical sideband is replaced by microwave noise. The remaining optical fluctuations are provided by the lower sideband, while feedback-cooled thermomechanical noise raises the total photocurrent spectrum above the quantum noise level of $1/2$. In the high cooperativity limit $\CL \to \infty$, this mechanical contribution is fully suppressed, so that the total spectrum $\bSYLout(\omega)$ is flat and, indeed, maximally uninformative.

Interestingly, increasing the gain $h > 1$ creates a dip in the photocurrent spectrum, as optical phase noise is imprinted strongly on the resonator and cancelled out in the photocurrent (centre and right panels). This phenomenon is known as `noise squashing'~\cite{lee_noise_squashing_2010} and our transducer operates optimally---without squeezing---precisely at the edge of this regime. The excess phase noise populates the resonator, so that the system emits microwave fluctuations above the vacuum level. In other words, the optical ground state is then transferred to the microwave output with squeezing. As a sidenote, feeding in a phase-squeezed optical input can counteract the transfer squeezing~\cite{schafermeier_squeezing_enhanced_cooling_2016}, so that vacuum fluctuations are emitted from the microwave port and the flatness of the photocurrent spectrum is restored.

\begin{figure}
    \centering
    \includegraphics{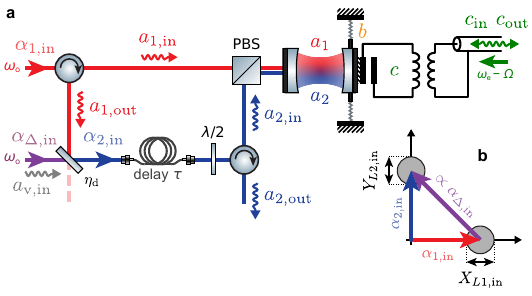}
    \caption{\textbf{Bidirectional state transfer between optical and microwave fields enabled by coherent feedback.} 
    \textbf{(a)}~System schematic. Two optical modes $a_1$, $a_2$ with orthogonal polarisation interact with mechanical resonator $b$. Light entering the system through port $\ainN1$ interacts with both cavity modes consecutively, while suitable tuning of the phase between coherent pump fields $\pumpinN1$ and $\pumpinN2$ allows both optical quadratures to interact with the resonator. Between passes, back-reflected light is combined with local oscillator field $\pumpinN\Delta$ on a beam-splitter with high reflectivity $\effd \sim 1$, delayed by time $\tau$, rotated in polarisation by a half-wave plate ($\lambda/2$) and coupled in to $a_2$ through a polarising beam-splitter (PBS). Light leaving the system through $\aoutN2$ carries the state of the microwave input $\cin$.
    \textbf{(b)}~Phase-space representation of the coherent pump amplitudes (arrows) and the input optical state (grey circles). Real $\ainN1$ transfers amplitude quadrature $\XLinN1$ on the first pass, while imaginary $\ainN2$ transfers phase quadrature $\YLinN2$ on the second pass. Conversely, the resonator state is imprinted on $\YLoutN1$ (first pass) and on $\XLoutN2$ (second pass). Local oscillator $\pumpinN\Delta$ provides the necessary phase space displacement between passes.%
    }
    \label{fig:coherent_feedback}
\end{figure}

\section{Bidirectional transfer by coherent feedback}
The observation that optical noise in the optical output is replaced by microwave fluctuations, as shown in Fig.~\ref{fig:homodyne_content}, offers a promising opportunity: While measurement-based feedback destroys the optical output in the detectors, what if we could keep those correlations in the feedback loop alive?
In this section, we show that this can be achieved using coherent optical feedback~\cite{ernzer_coherent-feedback_2023} and, moreover, that this allows bidirectional transfer of quantum states between the optical and microwave domains.
In this scheme, shown in Fig.~\ref{fig:coherent_feedback}, light entering the system interacts with the cavity twice, to facilitate transfer from and onto both optical quadratures. Conceptually similar protocols have been proposed using pulsed optomechanics and quantum non-demolition interactions~\cite{bennett_unresolved_sideband_pulsed_transfer_2016}.

The set-up features two cavity modes $a_1$ and $a_2$ with orthogonal polarisation, both coupled to the mechanical resonator $b$. Each is pumped by an individual coherent field $\pumpinN{1}$ to a coherent amplitude $\pumpN{j}$. For simplicity, we assume perfect couplings $\effL = \effM = 1$.
Light entering the transducer through port $\ainN1$ interacts first with $a_1$, couples out again, is delayed by a mechanical quarter-cycle $\tau=\pi/(2\Omega)$ and rotated in polarisation, and then couples into $a_2$. 
Importantly, as shown in Fig.~\ref{fig:coherent_feedback}b, the optical phase between $\pumpN1$ and $\pumpN2$ is adjusted so that, on the first pass, the amplitude quadrature $\XLinN1$ is imprinted on $b$ through radiation pressure and, on the second pass, $\YLinN2$ is. 
This mediates `forward' optical-to-microwave transfer in the same vein as measurement-based feedback did before. 

In fact, as detailed in SI Section~\ref{sec:apx:bidirectional_transfer}, the two can be treated as formally equivalent.
To phase-shift the pump field, light leaving $a_1$ is combined with an auxiliary pump on a beam-splitter with high reflectivity $\effd \sim 1$, at the cost of mixing an optical vacuum field $\ainN{\text{v}}$ into the feedback signal. Any losses in the optical feedback system can be collected in the `feedback efficiency' $\effd$, which plays the same role as detection efficiency did before.
Similarly, the gain ratio $h = G/8\CL$, which quantified the relative strength of the two quadrature imprints, is now replaced by the ratio $h=\sqrt{\effd \CLN2 / \CLN1}$ of the two cavity cooperativities $\CLN{j}$. 
To transfer without squeezing, i.e. for $h=1$, we note that $\CLN2 > \CLN1$ to compensate for feedback loop losses. 
With $\effd$ and $h$ thus redefined, the microwave output field $\cout$ is again characterised by Eq.~\eqref{eq:c_out_with_c_noise}, with corresponding transfer gains and added noise.

Next, we consider transfer in the `reverse' sense, starting with mechanical-to-optical conversion. Again, light enters the transducer from port $\ainN1$. On its first pass through the cavity, resonator position $Q(t)$ is imprinted on the output phase quadrature $\YLoutN1$, while, after a quarter cycle delay $\tau=\pi/(2\Omega)$, resonator momentum $P(t) \approx Q(t+\tau)$ is transferred, on the second pass, to the orthogonal quadrature $\XLoutN2$ as well. 
This combination enables mechanical-to-optical state transfer (SI Section~\ref{sec:apx:bidirectional_transfer}). 
Notably, the cooperatitivity ratio $s = \sqrt{\CLN2/(\effd CLN1)}$ that determines the level of squeezing applied during reverse transfer, is distinct from the forward ratio $h$. This can be understood from the optical circuit: In the forward direction, feedback loss $\effd$ affects the input to the \emph{second} transfer step mediated by $a_2$, whereas for reverse transfer, it is the output of $a_1$---mediating the \emph{first} transfer step---that is subject to loss. After turning on the electromechanical interaction, the optical output field is characterised, for no-squeezing $s=1$, by
\begin{eqnarray}
    &\aoutN2(\omega) = e^{\rmi\omega\tau} \sqrt{\effd} \times \nonumber \\
    & \quad \left(\frac{\effd-1+\taa(\omega)}{\effd} \XLinN1(\omega) + \rmi \taa(\omega) \YLinN1(\omega) \right) / \sqrt{2} \nonumber \\
    &\quad - e^{\rmi\omega\tau} \sqrt{1-\effd} \bigg(\XLv(\omega) + \rmi \taa(\omega) \YLv(\omega)\bigg) / \sqrt{2} \nonumber \\
    &\quad + \tba(\omega)\bin(\omega) - \rmi\tca(\omega) \cin(\omega-\Omega), \label{eq:aout_cfb}
\end{eqnarray}
featuring the input-to-light gains defined in Eq.~\eqref{eq:input_to_light_gains}. This central result describes the full microwave-to-optical transfer process.

At the lower (red) optical sideband $\omega\approx-\Omega$, the optical output
$\aoutN2(\omega\approx-\Omega) = e^{\rmi\omega\tau} \sqrt{\effd} \ainN{1}(\omega) - e^{\rmi\omega\tau} \sqrt{1-\effd} \ainN{\text{v}}(\omega)$
merely contains the reflected input field, after mixing with the added vacuum field and propagation through the feedback loop. 
In contrast, the upper (blue) optical sideband $\omega \approx +\Omega$ crucially carries the microwave input field $\cin$, with gain $\tca(\omega)$.
Each optical quadrature features in Eq.~\eqref{eq:input_to_light_gains} with a distinct prefactor, reflecting their different interactions in the feedback loop.
Nevertheless, the spectrum of the noise added to the output quadratures $\XLoutN{2}^{\SBpSymb}$ and $\YLoutN{2}^{\SBpSymb}$ of the upper sideband is the same, and given by
\begin{eqnarray}
    &\bSXoutCFB[|\text{noise}](\omega) = \bSYoutCFB[|\text{noise}](\omega) \nonumber \\
    &\hspace{2em} =
    \underbrace{\frac{1}{2} |\taa(\omega+\Omega)|^2}_{\text{optical}} 
    + \underbrace{\left(\nth + \frac{1}{2}\right)|\tba(\omega+\Omega)|^2}_{\text{mechanical}} \label{eq:main:noise_mw-to-opt} \\
    &\hspace{2em}+ \underbrace{\frac{1-\effd}{4\effd} \frac{\effbopt|\tba(\omega + \Omega)|^2}{\effbin}}_{\text{feedback}}. \nonumber
\end{eqnarray} 
In the limit of high cooperativity and matched transfer, when $|\taa(\omega+\Omega)|^2, |\tba(\omega+\Omega)|^2 \to 0$ and $\effbopt|\tba(\omega + \Omega)|^2 / \effbin \to 1$, the added noise reduces to $(\effd^{-1} - 1)/4$.
Interestingly, the added noise in the forward, optical-to-microwave direction can be expressed similar to Eq.~\eqref{eq:main:noise_mw-to-opt}  after exchanging $a \to c$ and $\omega + \Omega \to \omega$. This indicates that in both directions, the feedback loss $\effd$ imparts excess fluctuations onto the mechanical resonator, which then couple out into the relevant output mode. Together, these results show that coherent feedback enables bidirectional transfer between the optical and microwave ports of a bad-cavity transducer with similar noise performance.

\section{Discussion}
We have shown that continuous and bidirectional optical-microwave quantum state transfer can be performed using a ``bad" optical cavity. Traditionally, electro-optomechanical transducers operating in the ``good" cavity limit trade noise performance and scalability 
for high conversion efficiency. We have demonstrated that integrating feedback into an electro-optomechanical platform alleviates the need for a high-finesse optical cavity, while maintaining high conversion efficiency and quantum-compatible noise performance. 
To characterise performance in case of non-unity gain, we introduced the \textit{quantum transfer witness} \(\qtw\). This single-valued and input-independent figure of merit 
identifies both the transducer's capacity to preserve entanglement and benchmarks its performance against a classical detection and reconstruction scheme. Access to the bad cavity limit significantly broadens the classes of optomechanical systems which may serve as transducers, paving the way for compact on-chip converters with robust, quantum-compatible performance.

\section*{Data availability}
All data generated or analysed during this study are included in this published article and its supplementary information files.

\ack
The authors would like to thank James Bennett and Amy Navarathna for useful discussions. This research was primarily supported by the Australian Research Council Centres of Excellence for Engineered Quantum Systems (EQUS, CE170100009) and in Quantum Biotechnology (QUBIC, CE230100021). Support was also provided by the Air Force Office of Scientific Research under Grant No. FA9550-22-1-0047.

\section*{Author contributions}
The project was conceived by W.P.B. The manuscript was written by M.P.F. and J.J.S., with revisions from W.P.B. Theoretical results were derived by M.P.F. and J.J.S., with input from W.P.B. 

\section*{Competing interests}
All authors declare no financial or non-financial competing interests.

\section*{References}
\printbibliography[heading=none]

\clearpage
\jjsonecol
\markboth{Supplementary Information}{Supplementary Information}
\begin{refsection}
\appendix

\begin{center}
\large \textbf{Supplementary Information for ``\ourtitle''}
\end{center}

\startcontents[sections]
\printcontents[sections]{l}{1}{\setcounter{tocdepth}{2}}

I\section{System dynamics}
\label{sec:apx:system_dynamics}
In this section, we derive the dynamics of the transduction process in detail. We first model the optomechanical subsystem for arbitrary gain and coupling inefficiencies, and subsequently introduce the microwave cavity.

\subsection{Optomechanical subsystem with arbitrary gain}\label{sec:apx:optomechanical_subsystem_arbgain}
The (dimensioned) mechanical position $q$ shifts the resonance frequency $\omega_c(Q) = \omega_0 + \mathcal{G}q$ of the optical cavity, where $\mathcal{G} = \partial \omega_c / \partial q$ is the optomechanical coupling strength determined by the cavity geometry and $\omega_0$ is the nominal resonance frequency. This definition is used in Bowen \& Milburn~\cite{Warwick_Textbook}, while the review by Aspelmeyer et al.~\cite{Aspelmeyer2014cavity} defines $\mathcal{G}$ with opposite sign. The dimensioned mechanical position and momentum operators $q$, $p$ are related to the phonon creation ($b^\dagger$) and annihilation operators ($b$) through $q = \qzpf(b^\dagger + b)$ and $p = \rmi m \Omega \qzpf(b^\dagger - b)$, where $m$ and $\Omega$ are the effective mass and resonance frequency of the resonator, respectively, and $\qzpf = \sqrt{\hbar / (2m\Omega)}$ is the amplitude of the quantum zero-point fluctuations in the mechanical ground state. We use the latter to define the dimensionless position and momentum operators $Q = q/(\qzpf\sqrt{2}) = (b^\dagger + b) / \sqrt{2}$ and $P = p/(m\Omega \qzpf\sqrt{2}) = \rmi (b^\dagger - b)/\sqrt{2}$. These operators satisfy the commutation relation $[Q,P] = [q,p] / (2m\Omega\qzpf^2) = [q,p]/\hbar = \rmi$ and attain a ground-state variance of $\langle Q^2 \rangle = \langle P^2 \rangle = 1/2$.

The coherent dynamics of the optomechanical subsystem are governed, in the lab frame, by the Hamiltonian
\begin{eqnarray}
    H_{1,\text{lab}}/\hbar &= \omega_c(q) \bar{a}^\dagger \bar{a} + \Omega b^\dagger b - \Fclass q
    = \omega_0 \bar{a}^\dagger \bar{a} + \Omega b^\dagger b + \mathcal{G} q \bar{a}^\dagger \bar{a} - \Fclass q \\
    &=  \omega_0 \bar{a}^\dagger \bar{a} + \Omega b^\dagger b +  g_{\text{L},0} \bar{a}^\dagger \bar{a} (b^\dagger + b) - FQ
\end{eqnarray}
where $g_{\text{L},0} = \mathcal{G} \qzpf$ is the vacuum optomechanical coupling rate, $\Fclass$ is the (classical) force applied by the feedback system and $F = \sqrt{2}\Fclass \qzpf$. Next, we transform the lab frame optical field operator $\bar{a}$ into a frame that rotates along with the nominal optical frequency, $a = \bar{a} e^{\rmi \omega_0 t}$. The corresponding rotating frame Hamiltonian then reads
\begin{eqnarray}
    H_{1}/\hbar =  \Omega b^\dagger b +  g_{\text{L},0} a^\dagger a (b^\dagger + b) - FQ
\end{eqnarray}

The optical mode is subject to a drive laser that is resonant with $\omega_0$. This displaces the cavity field to a high amplitude $\alpha$, which we take to be real. Accordingly, we redefine the operator $a \to \alpha + a$, so that $a$ now describes the fluctuations on top of the strong carrier field $\alpha$. This results in the Hamiltonian
\begin{eqnarray}
    H_2/\hbar = \Omega b^\dagger b + g_{\text{L},0} (\alpha^2 + \alpha(a^\dagger + a) + a^\dagger a) (b^\dagger + b) - FQ.
\end{eqnarray}
Assuming $\alpha \gg \langle a \rangle$, we can linearise the optomechanical interaction by neglecting the quadratic fluctuation term $a^\dagger a$. In addition, as the quadratic carrier term $\alpha^2$ applies only a static radiation pressure force on the mechanical resonator, we redefine the origin of the mechanical position $q$ accordingly to remove it. We then obtain the Hamiltonian
\begin{eqnarray}
    H_3/\hbar = \Omega b^\dagger b + \Hom / \hbar - FQ
\end{eqnarray}
with interaction term
\begin{eqnarray}
    \Hom/\hbar = \gL (a^\dagger + a) (b^\dagger + b) = 2\gL \XL Q.
\end{eqnarray}
Here, $\gL = \alpha g_{\text{L},0}$ is the coherent-amplitude boosted optomechanical coupling rate.
In addition, we have defined the operators $\XL = (a^\dagger + a) / \sqrt{2}$ and $\YL = \rmi(a^\dagger - a) / \sqrt{2}$ for the amplitude and phase quadratures of the optical fluctuation field $a$, respectively. These operators satisfy the commutation relations $[\XL, \YL] = \rmi$ and attain a ground-state variance of $\langle \XL^2 \rangle = \langle \YL^2 \rangle = 1/2$.

Next, we model the open dynamics of the optomechanical subsystem using quantum Heisenberg-Langevin equations. For a mechanical or optical operator $\mathcal{O}$, these are given by~\cite{Warwick_Textbook}
\begin{eqnarray}
    \frac{\rmd \mathcal{O}}{\rmd t} = \frac{[\mathcal{O}, H_3]}{\rmi \hbar} - [\mathcal{O}, c^\dagger] \left( \frac{\gamma}{2}c - \sqrt{\gamma} c_\text{in}(t) \right) + \left( \frac{\gamma}{2} c^\dagger - \sqrt{\gamma}c_\text{in}^\dagger(t) \right) [\mathcal{O},c], \label{eq:sup:qle}
\end{eqnarray}
where $c = a$ in case $\mathcal{O}$ is an optical operator and $c = b$ in case $\mathcal{O}$ is a mechanical operator. In addition, $\gamma$ is the relevant decay rate and $c_\text{in}(t)$ the corresponding input field operator. By using Eq.~\eqref{eq:sup:qle} to model the open mechanical dynamics, we have effectively performed a rotating-wave approximation on the mechanical bath~\cite{Warwick_Textbook,Doherty2012quantum}. 

The resulting time-domain equations of motion (EOMs) read
\begin{eqnarray}
    \dot{\XL} &= -\frac{\kL}{2} \XL + \sqrt{\kL} \XLin, \\
    \dot{\YL} &= -\frac{\kL}{2} \YL + \sqrt{\kL} \YLin - 2 \gL Q, \\
    \dot{Q} &= \Omega P - \frac{\Gamma}{2} Q + \sqrt{\Gamma} \Qin, \\
    \dot{P} &= -\Omega Q - \frac{\Gamma}{2} P + \sqrt{\Gamma} \Pin - 2 \gL \XL + F,
\end{eqnarray}
where $\XLin$, $\YLin$, $\Qin$, $\Pin$ are optical and mechanical input operators that satisfy $[\XLin(t), \YLin(t')] = [\Qin(t), \Pin(t')] = \rmi \delta(t-t')$.
As the timescale of the optical dynamics $ \kL^{-1} \ll \Omega^{-1}, \gL^{-1}, \Gamma^{-1}$ is much shorter than the system's other dynamical timescales, we adiabatically eliminate the optical evolution~\cite{Warwick_Textbook}. In doing so, we assume that the optical quadrature amplitudes are always in their steady state ($\dot{\XL} = \dot{\YL} = 0$) on mechanical timescales, so that we have
\begin{eqnarray}
    \XL(t) &= 2\XLin(t) / \sqrt{\kL}, \\
    \YL(t) &= 2\YLin(t) / \sqrt{\kL} - 4 (\gL/\kL) Q(t).
\end{eqnarray}
Plugging this into the momentum EOM results in
\begin{eqnarray}
    \dot{P} &= -\Omega Q - \frac{\Gamma}{2} P + \sqrt{\Gamma} \Pin - 2 \sqrt{\Gamma \CL} \XLin + F,
\end{eqnarray}
where we have defined the optomechanical cooperativity $\CL = 4 \gL^2 / \Gamma \kL$. Importantly, we see that the mechanical resonator is driven directly by the input amplitude quadrature $\XLin$.

Next, we analyse the optical output using the general input-output relation
\begin{eqnarray}
    \mathcal{O}_\text{out}(t) = \mathcal{O}_\text{in}(t) - \sqrt{\gamma}\mathcal{O}(t).
\end{eqnarray}
From this, we obtain the optical output phase quadrature 
\begin{equation}
    \YLout(t) = -\sqrt{\effd} \YLin(t) + 2\sqrt{\effd \Gamma \CL} Q(t) + \sqrt{1-\effd} \YLv(t), \label{eq:sup:apx-yout}
\end{equation}
which we detect using a homodyne interferometer. Here, imperfect detection is expressed by the efficiency $\effd$ and corresponding vacuum noise $\YLv$ arising from detection loss~\cite{Warwick_Textbook, Warwick_Paper}, while optical coupling inefficiency will be introduced in the next section. Crucially, the homodyne photocurrent encodes information about the input phase quadrature $\YLin(t)$, which we will transfer onto the mechanical resonator via the feedback force $F$. 

To do so, we filter the (classical) photocurrent $\YLout$ using a real, causal filter function $f(t)$, amplify this signal by a tuneable gain factor $G$ and feed it into the feedback actuator to generate a force
\begin{eqnarray}
    F(t) = G \cdot \frac{\Gamma}{2} \left( f(t) \circledast \frac{\YLout(t)}{2\sqrt{\effd \Gamma \CL}} \right).
\end{eqnarray}
Here, we have chosen to rescale the photocurrent so that $\YLout(t)/2\sqrt{\effd \Gamma \CL} = Q(t) + \cdots$ constitutes an estimate of oscillator position $Q(t)$ in the presence of optical input terms, and scaled the filtered signal by $\Gamma/2$, a system parameter, to ensure that $G$ is dimensionless. 

As mentioned in the main text, we restrict our analysis to the particular choice of filter function $f(t) = \delta(t-\tau)$ that delays the measured photocurrent by one-quarter of the mechanical resonance period $\tau = \pi/2\Omega$. An analysis for arbitrary $f(t)$ can be found in~\cite{Warwick_Paper}. Substituting the feedback force for this filter function into the momentum EOM results in
\begin{eqnarray}
    \dot{P} = - \Omega Q - \frac{\Gamma}{2}P + \sqrt{\Gamma} \Pin - 2\sqrt{\Gamma\CL} \XLin + \frac{G \Gamma}{4} \frac{\YLout(t-\tau)}{\sqrt{\effd \Gamma \CL}}. \label{eq:eom-P-SI}
\end{eqnarray}
Next, we combine the EOMs for $Q$ and $P$ into the EOM for the mechanical annihilation operator $b = (Q + \rmi P)/\sqrt{2}$,
\begin{eqnarray}
    \dot{b} = -\rmi \Omega b - \frac{\Gamma}{2} b + \sqrt{\Gamma} \bin - \rmi \sqrt{2\Gamma\CL} \XLin + \frac{\rmi G \Gamma}{4} \frac{\YLout(t-\tau)}{\sqrt{2\effd \Gamma \CL}}, \label{eq:sup:apx-eom-b-1}
\end{eqnarray}
given in the main text as Eq.~\eqref{eq:eom-b-1}. We substitute the photocurrent $\YLout$ in Eq.~\eqref{eq:sup:apx-yout} into Eq.~\eqref{eq:sup:apx-eom-b-1} to obtain, after some algebra,
\begin{eqnarray}
     \dot{b} = &-\rmi \Omega b - \frac{\Gamma}{2} b + \sqrt{\Gamma} \bin - \rmi \sqrt{2\Gamma\CL} \XLin(t) + \frac{\rmi G \Gamma}{4} [b(t-\tau) + b^\dagger(t-\tau)] \nonumber \\
     &-\rmi \sqrt{2\Gamma \CL} \left[ \frac{G}{8\CL} \YLin(t-\tau) - \frac{G}{8\CL} \sqrt{\frac{1-\effd}{\effd}} \YLv(t-\tau) \right]. \label{eq:dotb-1}
\end{eqnarray}

At this point, we deviate from the analysis in ref.~\cite{Warwick_Paper} in two ways. Firstly, we apply an additional rotating wave approximation on the mechanical degree of freedom to neglect the counter-rotating term $b^\dagger(t-\tau)$. Secondly, we approximate the effect of the delay filter by the free evolution of the resonator $b(t-\tau) \approx e^{\rmi \Omega t} b(t) = \rmi b(t)$, neglecting back-action and damping within a single cycle. This is valid when the resonator is weakly coupled to the cavity ($\gL \ll \Omega$) and retains a high quality factor $\mathcal{Q}' \gg 1$ even with feedback cooling on. Effectively, this replaces the frequency response $f(\omega) = e^{\rmi \omega \tau}$ of the filter, given by the Fourier transform of $f(t)$, by $f(\omega) = \rmi \sign(\omega)$. We can then write
\begin{eqnarray}
    \dot{b}(t) = &-\rmi\Omega b(t) - \frac{\Gamma'}{2} b(t) + \sqrt{\Gamma}\bin(t) \\
    &- \rmi \sqrt{2\Gamma \CL} \Bigg( \XLin(t) + \frac{G}{8\CL} \YLin(t-\tau) - \frac{G}{8\CL} \varepsilon \YLv(t-\tau)  \Bigg) \nonumber,
\end{eqnarray}
given in the main text as Eq.~\eqref{eq:eom-b-2}. Here, we have absorbed the mechanical feedback term $\rmi\Gamma G b(t-\tau) / 4 \approx -\Gamma G b(t) / 4$ into the feedback-broadened linewidth $\Gamma' = (1+G/2)\Gamma$ and defined $\varepsilon = \sqrt{(1-\effd)/\effd}$. 

In the frequency domain, the corresponding mechanical response is given in the main text in Eq.~\eqref{eq:eom-b-freq} and repeated here for completeness,
\begin{eqnarray}
    b(\omega) &= \sqrt{\Gamma} \chib(\omega) \bigg[ \bin(\omega) \nonumber \\
    &- \rmi \sqrt{2\CL} \left( \XLin(\omega) + h  e^{\rmi \omega \tau} \YLin(\omega) - h \varepsilon e^{\rmi \omega \tau} \YLv(\omega) \right) \bigg].
\end{eqnarray}
Here, we define the ratio $h=G/8\CL$ and the mechanical susceptibility 
\begin{equation}
    \chib(\omega) = \frac{1}{\Gamma'/2 - \rmi(\omega - \Omega)}.
\end{equation}
The susceptibility is sharply peaked around $+\Omega$ in the high-quality regime $\mathcal{Q}' = \Omega / \Gamma' \gg 1$, so that we can make the approximation $e^{\rmi \omega \tau} \approx e^{\rmi \Omega \tau} = \rmi$.
Formally, the same condition underlies the approximation $b(t-\tau) \approx \rmi b(t)$ that we made before, as well as both RWAs.

In contrast to the main text, here we will keep an arbitrary feedback gain ratio $h$. After some algebra, the mechanical response can then be written as
\begin{eqnarray}
    b(\omega) = \chib(\omega) \bigg[ 
        & \sqrt{\Gamma} \bin(\omega) \nonumber \\
        &- \rmi \sqrt{4 h \CL \Gamma} \left( \frac{1}{\sqrt{h}} \XLin(\omega) + \rmi \sqrt{h} \YLin(\omega) \right) / \sqrt{2} \label{eq:sup:eom-b-freq-h-1} \\
        &- \sqrt{4 h \CL \Gamma} \sqrt{h} \varepsilon \YLv(\omega) / \sqrt{2}
    \bigg] \nonumber
\end{eqnarray}

In this expression, we recognise the optical mode
\begin{eqnarray}
    \ainT(\omega) = \left( \frac{1}{\sqrt{h}} \XLin(\omega) + \rmi \sqrt{h} \YLin(\omega) \right) / \sqrt{2}. \label{eq:sup:ainT-h}
\end{eqnarray}
that is transferred onto the mechanical resonator. This mode satisfies the commutator $[\ainT(\omega), \ainT(\omega')^\dagger] = \delta(\omega - \omega')$
and is thus a canonically normalised annihilation operator. 
Noting that for arbitrary gain, the broadened linewidth is given by $\Gamma'/\Gamma = 1 + 4h\CL$, we now define the effective optical and mechanical coupling efficiencies 
\begin{eqnarray}
    \effbopt &= 4h\CL \Gamma / \Gamma' = 4h\CL / (1+4h\CL), \\
    \effbin &= \Gamma / \Gamma' = 1/(1+4h\CL),
\end{eqnarray}
respectively. These efficiencies $\effbopt + \effbin = 1$ add up to one and, in the case $h=1$, reduce to the effective efficiencies derived in the main text. Importantly, after multiplying Eq.~\eqref{eq:sup:eom-b-freq-h-1} by $\sqrt{\Gamma'/\Gamma'}$, these allow us to write the mechanical response as
\begin{eqnarray}
    b(\omega) &= \sqrt{\Gamma'} \chib(\omega) \bigg[ 
        &\sqrt{\effbin} \bin(\omega) + \sqrt{\effbopt} \left(- \rmi \ainT(\omega) - \sqrt{h/2} \varepsilon \YLv(\omega) \right)
    \bigg] \label{eq:sup:b-freq-2} \\
    &= \sqrt{\Gamma'} \chib(\omega) \bigg[ 
        &\sqrt{\effbin} \bin(\omega) + \sqrt{\effbopt} \bopt(\omega) \bigg]
\end{eqnarray}
This expression has the same form as Eq.~\eqref{eq:mech-eff-channels} in the main text. However, as the optical input channel is now given by
\begin{eqnarray}
    \bopt(\omega) = - \rmi \ainT(\omega) - \sqrt{h} \varepsilon \YLv(\omega) / \sqrt{2} 
\end{eqnarray}
with transferred optical mode $\ainT(\omega)$ given by Eq.~\eqref{eq:sup:ainT-h}, we see that if $h>1$, the optical quadrature $\YLin$ is squeezed as it transferred, while $\XLin$ is anti-squeezed, and vice-versa if $h<1$.

\subsection{Lower sideband cancellation of radiation pressure and feedback forces}
\label{sec:lower_sideband_cancel}
When the feedback gain is set to $G=8\CL$ ($h=1$), so that the optical quadratures are transferred without squeezing, the feedback force exactly cancels the radiation pressure generated by the lower optomechanical sideband. 
To see this, we take the Fourier transform of the equation of motion~\eqref{eq:eom-P-SI} for $P$ to obtain
\begin{eqnarray}
    -\rmi \omega P(\omega) = &-\Omega Q(\omega) - \frac{\Gamma}{2} P(\omega) + \sqrt{\Gamma} \Pin(\omega) \\
    &- 2\sqrt{\Gamma\CL}\XLin(\omega) + \frac{G\sqrt{\Gamma}}{4\sqrt{\CL}} f(\omega) \YLout(\omega),
\end{eqnarray}
where we have assumed perfect detection ($\effd = 1$) for simplicity. Here, we still assume a general feedback filter function $f(t)$ with Fourier transform $f(\omega)$. After plugging in the frequency response of the output phase quadrature, given by the Fourier transform of Eq.~\eqref{eq:sup:apx-yout}, we find
\begin{eqnarray}
     -\rmi \omega P(\omega) = &-\Omega Q(\omega) - \frac{\Gamma}{2} \bigg[ P(\omega) + G f(\omega)Q(\omega) \bigg] + \sqrt{\Gamma} \Pin(\omega) \\
    &- 2\sqrt{\Gamma\CL} \bigg[ \XLin(\omega) + \frac{G}{8\CL} f(\omega) \YLin(\omega) \bigg].
\end{eqnarray}
We rewrite this in terms of the optical input field operator $\ain$ to get
\begin{eqnarray}
    -\rmi \omega &P(\omega) = \-\Omega Q(\omega) - \frac{\Gamma}{2} \bigg[ P(\omega) + G f(\omega)Q(\omega) \bigg] + \sqrt{\Gamma} \Pin(\omega) \\
    &- \sqrt{2\Gamma\CL} \underbrace{\bigg[ \ain(-\omega)^\dagger \big( 1 + \rmi h f(\omega) \big) + \ain(\omega) \big(1 -  \rmi h f(\omega) \big) \bigg]}_{P_\text{opt}(\omega)}.
\end{eqnarray}
At this point, we recall that $f(\pm\Omega) = \pm \rmi$, so that at the mechanical frequency, the optical driving term $P_\text{opt}(\omega)$ in the expression above is given by
\begin{eqnarray}
    P_\text{opt}(\omega = \pm \Omega) &= \ain(\mp\Omega)^\dagger \big( 1 \mp h \big) + \ain(\pm\Omega) \big(1 \pm h\big) \\
    &= \cases{\ain(+\Omega) & $\omega = +\Omega$ \\ \ain(+\Omega)^\dagger & $\omega = -\Omega$}.
\end{eqnarray}
For $h=1$, we find that the radiation pressure and the feedback force contributions to $P_\text{opt}(\omega)$, proportional to $1$ and $h$, respectively, precisely cancel for the lower optical sideband $\ain(-\Omega)$. As a result, the optical driving term then only depends on the upper optical sideband $\ain(+\Omega)$.

In the Supplemental Material to ref.~\cite{Warwick_Paper}, this cancellation can be observed in the insets in Fig. S1, where the spectral mode of the signal at $\omega < 0$ exhibits a sharp dip at the frequency $\omega = -\Omega$ of the lower optomechanical sideband.

Alternatively, it is instructive to analyse the optomechanical system's response to classical driving of the upper or lower sideband. To do so, we substitute the input field $\ain(t) = \asb e^{-\rmi \nu \Omega t}$ by a coherent drive with complex amplitude $\asb$ and detuning $\nu \Omega$ from the cavity resonance, where $\nu = 1$ ($\nu=-1$) corresponds to the upper (lower) optomechanical sideband. For simplicity, we neglect quantum fluctuations.

For this drive, the (classical) optical quadratures evolve as
\begin{eqnarray}
    \XLin(t) &= \Re \left[ \asb e^{-\rmi \nu \Omega t} \right] = \Re \left[ \asb \right] \cos(\Omega t) + \nu \Im \left[ \asb \right] \sin(\Omega t), \\
    \YLin(t) &= \Im \left[ \asb e^{-\rmi \nu \Omega t} \right] = \Im\left[\asb\right] \cos(\Omega t) - \nu \Re[\asb] \sin(\Omega t).
\end{eqnarray}
After the quarter-cycle feedback filter delay of $\tau = \pi / 2\Omega $, the optical contribution to the feedback signal is given by
\begin{eqnarray}
    \YLin(t-\tau) &= \Im\left[\asb\right] \cos(\Omega t - \pi/2) - \nu \Re[\asb] \sin(\Omega t - \pi/2) \\
    &= \Im\left[\asb\right] \sin(\Omega t) + \nu \Re\left[\asb\right] \cos(\Omega t) = \nu \XLin(t).
\end{eqnarray}
We can plug these into the equation of motion for $P$,
\begin{eqnarray}
    \dot{P}(t) = &-\Omega Q(t) - \frac{\Gamma}{2} \bigg[ P(t) + G Q(t-\tau) \bigg] + \sqrt{\Gamma}\Pin(t) \\
    &- 2\sqrt{\Gamma \CL} \bigg[ \XLin(t) + \frac{G}{8\CL} \YLin(t - \tau) \bigg],
\end{eqnarray}
to obtain
\begin{eqnarray}
    \dot{P}(t) = &-\Omega Q(t) - \frac{\Gamma}{2} \bigg[ P(t) + G Q(t-\tau) \bigg] + \sqrt{\Gamma}\Pin(t) \\
    &- 2\sqrt{\Gamma \CL} \bigg[ \underbrace{\XLin(t)}_{\text{rad. press.}} + \underbrace{\nu h \XLin(t)}_{\text{feedback}} \bigg].
\end{eqnarray}
Again, for $h=1$ and $\nu = -1$, at the lower sideband, we see that the radiation pressure force is exactly cancelled by the feedback force coming from the optical drive.

\subsection{Optical coupling inefficiency}
We will now introduce an optical loss port $\al$ to model coupling inefficiency. This port couples to the cavity with efficiency $1-\effL$, so that the coupling efficiency of the signal input port is reduced to $\effL$. The input optical quadratures are then replaced by a linear combination of signal and noise
\begin{eqnarray}
    \XLin(t) &\to \sqrt{\effL} \XLin(t) + \sqrt{1-\effL} \XLloss(t), \label{eq:sup:XLin-ineff} \\
    \YLin(t) &\to \sqrt{\effL} \YLin(t) + \sqrt{1-\effL} \YLloss(t), \label{eq:sup:YLin-ineff}
\end{eqnarray}
with $\XLloss, \YLloss$ corresponding to the amplitude and phase quadratures of the loss port.

The reduced coupling efficiency $\effL$ also applies to the output port, so that the measured homodyne signal is now given by
\begin{eqnarray}
    \YLout(t) &= 2\sqrt{\effd \effL \Gamma \CL} Q(t) - \sqrt{\effd}(2\effL - 1)\YLin(t) \nonumber \\ 
    &- 2\sqrt{\effd\effL(1-\effL)}\YLloss(t) + \sqrt{1-\effd}\YLv(t). \label{eq:sup:YLout-ineff}
\end{eqnarray}
We plug Eqs.~\eqref{eq:sup:XLin-ineff} and~\eqref{eq:sup:YLout-ineff} into the mechanical evolution Eq.~\eqref{eq:sup:apx-eom-b-1} to obtain
\begin{eqnarray}
    \dot{b} = &-\rmi \Omega b - \frac{\Gamma}{2} b + \sqrt{\Gamma} \bin - \rmi \sqrt{2\Gamma\CL\effL} \XLin(t) 
    - \rmi \sqrt{2\Gamma\CL(1-\effL)} \XLloss(t) \nonumber \\
    &+ \frac{\rmi G \Gamma}{4} \sqrt{2\effL} Q(t - \tau) - \rmi \frac{G}{8\CL} \sqrt{2 \Gamma \CL} \bigg[ (2\effL - 1)\YLin(t-\tau) \nonumber \\
    &+ \sqrt{4\effL(1-\effL)}\YLloss(t-\tau) - \varepsilon \YLv(t-\tau) \bigg].
\end{eqnarray}
As before, after applying the additional RWA to neglect $b^\dagger(t-\tau)$ in $Q(t-\tau)$ and the high-quality approximation $b(t-\tau) \approx \rmi b(t)$, we are left with 
\begin{eqnarray}
    \dot{b} = &-\rmi \Omega b - \frac{\Gamma'}{2} b + \sqrt{\Gamma} \bin - \rmi \sqrt{2\Gamma\CL\effL} \XLin(t) 
    - \rmi \sqrt{2\Gamma\CL(1-\effL)} \XLloss(t) \nonumber \\
    &- \rmi \frac{G}{8\CL} \sqrt{2 \Gamma \CL} \bigg[ (2\effL - 1)\YLin(t-\tau) \nonumber \\
    &+ \sqrt{4\effL(1-\effL)}\YLloss(t-\tau) - \varepsilon \YLv(t-\tau) \bigg],
\end{eqnarray}
where the feedback-broadened linewidth is now given by $\Gamma' / \Gamma = 1 + \sqrt{\effL} G / 2$. This reflects the fact that for a given position $Q$, the mechanical contribution to $\YLout$ -- and therefore the feedback signal -- is weaker by a factor $\sqrt{\effL}$.

Moreover, the phase quadrature $\YLin$ of the optical input field now contributes with prefactor $\lambda = 2\effL - 1$, from which we identify two regimes. We will soon see that, when the cavity is overcoupled ($\effL >  1/2$) and the sign of $\lambda > 0$ is positive, the upper mechanical sideband is transferred at the optimal operating point. Conversely, for an undercoupled cavity ($\effL < 1/2$), the sign of $\lambda < 0$ is reversed, coupling the lower sideband to the resonator instead. At critical coupling ($\effL = 1/2$), the input field is fully lost into the loss port. In that case, the reflected output field holds no information on $\YLin$ and transfer of the phase quadrature is ruled out.

We reorder the mechanical evolution as 
\begin{eqnarray}
    \dot{b} = &-\rmi \Omega b - \frac{\Gamma'}{2} b + \sqrt{\Gamma} \bin - \rmi \sqrt{4\Gamma\CL} \times \nonumber \\
    &\Bigg[ \sqrt{\effL} \XLin(t) / \sqrt{2} + \frac{G(2\effL - 1)}{8\CL} \YLin(t-\tau) / \sqrt{2} \nonumber \\
    &+ \sqrt{1-\effL}\XLloss(t) / \sqrt{2}  + \frac{G\sqrt{4\effL(1-\effL)}}{8\CL} \YLloss(t-\tau) / \sqrt{2} \\
    &- \frac{G}{8\CL}  \varepsilon \YLv(t-\tau)/ \sqrt{2} \Bigg]
\end{eqnarray}
From this expression, we see that both optical input quadratures are transferred with equal strength if 
\begin{eqnarray}
    G_\text{sym} = 8 \CL \frac{\signY \sqrt{\effL}}{2\effL - 1} = 8 \CL \frac{\sqrt{\effL}}{|2\effL - 1|}.
\end{eqnarray}
where the sign factor $\signY = \{ +1 \,\, \text{ for } \,\, \effL>1/2,\, -1 \,\, \text{ for } \,\, \effL<1/2 \}$ ensures that we select a positive, noise-suppressing feedback gain. We note that for $1/4 < \effL < 1$, the optimal gain is larger than it is in the absence of coupling inefficiency. Moreover, at critical coupling $\effL = 1/2$ the optimal gain diverges.

Again, we keep the gain arbitrary and define the ratio $\hineff = G/G_\text{sym}$. In terms of this ratio, the feedback-broadened linewidth is given by
\begin{eqnarray}
    \Gamma' = \Gamma \left( 1 + \sqrt{\effL} G / 2 \right) = \Gamma \left( 1 + 4 \hineff \CL \frac{\signY \effL}{2\effL - 1} \right) = \Gamma \left( 1 + 4 \hineff \lwfact \CL \right), 
\end{eqnarray}
where we have defined the ratio $\lwfact = \effL / |2\effL - 1|$.
Substituting these into the mechanical evolution yields
\begin{eqnarray}
    \dot{b} = &-\rmi \Omega b - \frac{\Gamma'}{2} b + \sqrt{\Gamma} \bin - \rmi \sqrt{4\Gamma\CL} \times \nonumber \\
    &\Bigg[ \sqrt{\hineff \effL} \bigg( \frac{1}{\sqrt{\hineff}} \XLin(t) + \signY \sqrt{\hineff} \YLin(t-\tau) \bigg) / \sqrt{2} \nonumber \\
    &+ \sqrt{2 \hineff \lwfact(1-\effL)} \left( \frac{1}{\sqrt{2 \hineff \lwfact}} \XLloss(t) + \sqrt{2 \hineff \lwfact} \YLloss(t-\tau) \right) / \sqrt{2} \\
    &- \sqrt{\hineff \effL} \frac{\signY \varepsilon\sqrt{\hineff}}{2\effL - 1} \YLv(t-\tau)/ \sqrt{2} \Bigg].
\end{eqnarray}

At this point, we move to the frequency domain. As in the main text, we apply the high-quality approximation to the frequency response of the delay filter $f(\omega) = -e^{\rmi \omega \tau} \approx -e^{\rmi \Omega \tau} = -\rmi$ around $\omega \approx \Omega$ to obtain the mechanical response
\begin{eqnarray}
    b(\omega) = &\sqrt{\Gamma} \chib(\omega) \Bigg[
    \bin(\omega) - \rmi \sqrt{4 \hineff\effL\CL} \left( \frac{1}{\sqrt{\hineff}} \XLin(\omega) + \rmi \signY \sqrt{\hineff} \YLin(\omega) \right) / \sqrt{2} \nonumber \\
    &-\rmi \sqrt{8\hineff\CL \lwfact (1-\effL)} \left( \frac{1}{\sqrt{2\hineff\lwfact}} \XLloss(\omega) + \rmi \sqrt{2\hineff \lwfact} \YLloss(\omega) \right) / \sqrt{2} \nonumber \\
    &- \sqrt{4\hineff\effL\CL} \frac{\signY \varepsilon \sqrt{\hineff}}{2\effL - 1} \YLv(\omega) / \sqrt{2} \Bigg].
\end{eqnarray}
From this expression, we recognise the mode that is transferred onto the mechanical resonator from the optical signal port,
\begin{eqnarray}
    \ainT(\omega) = \left( \frac{1}{\sqrt{\hineff}} \XLin(\omega) + \rmi \signY \sqrt{\hineff} \YLin(\omega) \right) / \sqrt{2},
\end{eqnarray}
and from the optical loss port,
\begin{eqnarray}
    \alT(\omega) = \left( \frac{1}{\sqrt{2\hineff\lwfact}} \XLloss(\omega) + \rmi \sqrt{2\hineff \lwfact} \YLloss(\omega) \right) / \sqrt{2}. \label{eq:sup:loss_sqz_mode}
\end{eqnarray}
The latter satisfies the commutator $[\alT(\omega), \alT(\omega')^\dagger] = \delta(\omega - \omega')$ and is a canonically normalised annihilation operator. Meanwhile, the former satisfies the commutator 
\begin{eqnarray}
    [\ainT(\omega), \ainT(\omega')^\dagger] = \signY \delta(\omega - \omega') \label{eq:sup:ainT_commutator}
\end{eqnarray}
and thus represents either an annihilation operator ($\effL > 1/2$) or creation operator ($\effL < 1/2$), depending on the whether the cavity is over- or undercoupled. We will discuss the consequences of this later.

In addition, we define the corresponding effective coupling efficiencies for the  mechanical, optical signal and optical loss input channels
\begin{eqnarray}
    \effbin &= \frac{\Gamma}{\Gamma'} = \frac{1}{1+4\hineff\lwfact\CL}, \\
    \effbopt &= \frac{4\hineff\effL \CL \Gamma}{\Gamma'} = \frac{4\hineff\effL\CL}{1+4\hineff\lwfact\CL}, \\
    \effbol &= \frac{8\hineff\CL\lwfact(1-\effL)\Gamma}{\Gamma'} = \frac{4\hineff\CL(\lwfact - \signY \effL)}{1+4\hineff \lwfact \CL}, \label{eq:sup:channel_efficiencies_general}
\end{eqnarray}
respectively. If the cavity is overcoupled ($\signY = 1$), the efficiencies $\effbin + \effbopt + \effbol = 1$ sum to one. However, if $\effL < 1/2$ ($\signY = -1$), the efficiencies sum to the larger-than-unity value $1 + 8 \hineff \effL \CL / (1+4\hineff\lwfact\CL)$.

With these operators and efficiencies, we can conveniently write the mechanical response as
\begin{eqnarray}
    b(\omega) = &\sqrt{\Gamma'} \chib(\omega) \Bigg[ \sqrt{\effbin}
    \bin(\omega) - \rmi \sqrt{\effbopt} \ainT(\omega) -\rmi \sqrt{\effbol} \alT(\omega) \nonumber \\
    &- \sqrt{\effbopt} \frac{\signY \varepsilon \sqrt{\hineff}}{2\effL - 1 } \YLv(\omega) / \sqrt{2} \Bigg] = \sqrt{\Gamma'}\chib(\omega) \bdrv(\omega),
\end{eqnarray}
where we collect all the mechanical driving terms in
\begin{eqnarray}
    \bdrv(\omega) &= \sqrt{\effbin}
    \bin(\omega) - \rmi \sqrt{\effbopt} \ainT(\omega) -\rmi \sqrt{\effbol} \alT(\omega) \nonumber \\ 
    &- \sqrt{\effbopt} \frac{\signY \varepsilon \sqrt{\hineff}}{2\effL-1} \YLv(\omega) / \sqrt{2}. \label{eq:sup:bdrv_gain_ineff} 
\end{eqnarray}

This concludes our analysis of the response of the optomechanical subsystem for arbitrary gain and incoupling efficiency. Importantly, the mechanical response still has the same form as Eq.~\eqref{eq:mech-eff-channels} in the main text, with updated definitions for the input channels and their effective coupling efficiencies.

\subsection{Microwave coupling inefficiency}
\label{sec:apx:microwave_coupling_ineff}
Next, we reintroduce the electromechanical subsystem, by replacing the mechanical drive term $\bdrv$ in Eq.~\eqref{eq:em-time-evo} with Eq.~\eqref{eq:sup:bdrv_gain_ineff}. We also include a microwave loss port $\cl$, so that the microwave input in Eq.~\eqref{eq:c_freq_response_perfect} is replaced by
\begin{eqnarray}
    \cin(\omega) \to \sqrt{\effM} \cin(\omega) + \sqrt{1-\effM} \cl(\omega)
\end{eqnarray}
with coupling efficiency $\effM$. The microwave output port is similarly affected. The frequency response of the mechanical and microwave resonator is then given by
\begin{eqnarray}
     b(\omega) &= \chibld(\omega) \sqrt{\Gamma'}\bdrv(\omega)  \\
     &+ \rmi  \chix(\omega - \Omega) \left[ \sqrt{\effM\kM} \cin(\omega-\Omega) + \sqrt{(1-\effM)\kM} \cl(\omega - \Omega) \right] \nonumber  \\
    c(\omega) &= \chicld(\omega) \left[ \sqrt{\effM\kM} \cin(\omega) + \sqrt{(1-\effM)\kM} \cl(\omega) \right]  \\ 
    &+ \rmi \chix(\omega) \sqrt{\Gamma'} \bdrv(\omega) \nonumber
\end{eqnarray}
From the microwave input-output relation, we then obtain the frequency response
\begin{eqnarray}
    \cout(\omega) = &\tac(\omega) \ainT(\omega+\Omega) + \talc(\omega) \alT(\omega + \Omega) + \tbc(\omega) \bin(\omega + \Omega) \nonumber \\
    &+ \tcc(\omega) \cin(\omega) + \tclc(\omega) \cl(\omega) + \tvc(\omega) \YLv(\omega+\Omega). \label{eq:sup:cout_inefficient}
\end{eqnarray}
for the microwave output port.
The transfer gains for each channel are given by
\begin{eqnarray}
    &\tac(\omega) = \sqrt{\effM \effbopt} \tinf(\omega), 
    \quad&\tcc(\omega) = 1 - \effM \kM \chicld(\omega) \nonumber \\
    &\talc(\omega) = \sqrt{\effM \effbol} \tinf(\omega), 
    &\tclc(\omega) = -\sqrt{\effM(1-\effM)}\kM \chicld(\omega), \label{eq:sup:tgains_inefficient} \\
    &\tbc(\omega) = \rmi \sqrt{\effM \effbin} \tinf(\omega),
    &\tvc(\omega) = -\rmi \sqrt{\effM\effbopt} \frac{\signY\varepsilon\sqrt{\hineff/2}}{2\effL -1}\tinf(\omega), \nonumber
\end{eqnarray}
in terms of the ideal gain $\tinf(\omega)$ and susceptibility $\chicld(\omega)$, as defined before in the main text. 

As a sanity check, we will verify that $\cout(\omega)$ is canonically normalised. We compute the commutator
\begin{eqnarray}
    [\cout(\omega), \cout(\omega')^\dagger] &= \delta(\omega - \omega') \bigg( \signY \Tac(\omega) + \Talc(\omega) \nonumber \\
    &+ \Tbc(\omega) + \Tcc(\omega) + \Tclc(\omega) \bigg)
\end{eqnarray}
Here, we have defined the transduction coefficients $T_j(\omega) = |t_j(\omega)|^2$ and used the fact that operators of different inputs commute. Furthermore, we have used commutator~\eqref{eq:sup:ainT_commutator} and $[\YLv(\omega), \YLv(\omega')] = 0$, noting that $\YLv$ does not contribute to the quantum correlations in $\cout$ and acts as a purely classical noise source.

The transduction coefficients 
\begin{eqnarray}
    &\Tac(\omega) = \effM\effbopt \Tinf(\omega), \quad \Talc(\omega) = \effM \effbol \Tinf(\omega) \\
    &\Tbc(\omega) = \effM \effbin \Tinf(\omega), \\
    &\Tcc(\omega) = 1 - \big( \effM (1-\effM)(1+(2\omega/\Gamma')^2) + \effM \big) \Tinf(\omega), \\
    &\Tclc(\omega) = \effM(1-\effM)(1+(2\omega/\Gamma')^2)\Tinf(\omega),
\end{eqnarray}
all depend on the ideal transfer transduction $\Tinf(\omega) = |\tinf(\omega)|^2$. We also note that $\Tcc(\omega) + \Tclc(\omega) = 1-\effM \Tinf(\omega)$ and, from the definitions in Eq.~\eqref{eq:sup:channel_efficiencies_general}, $\signY\effbopt + \effbol +\effbin = 1$. We can then evaluate
\begin{eqnarray}
    &\left[ \cout(\omega), \cout(\omega')^\dagger \right] = 
    \\
    &\quad \delta(\omega-\omega') \Big( 1 + \effM \Tinf(\omega) \Big[-1  +  \signY\effbopt + \effbol + \effbin \Big] \Big) = \delta(\omega - \omega')
\end{eqnarray}
to confirm that $\cout(\omega)$ describes a canonically normalised annihilation operator. This is true for every value of $\effL$ except $\effL = 1/2$, when the output phase quadrature carries no information on the input state and the factor $\lwfact$ diverges. In addition, 
\begin{eqnarray}
    \left[ \cout(\omega), \cout(\omega') \right] = \left[ \cout(\omega)^\dagger, \cout(\omega')^\dagger \right] = 0
\end{eqnarray}

\section{Noise calculations}
In this section, we calculate in detail the variance of the various noise contributions to the microwave output. We start by expanding the microwave amplitude quadrature
\begin{eqnarray}
    \XMout(\omega) = &[\cout(-\omega)^\dagger + \cout(\omega)] / \sqrt{2} = \\
    &\tac(\omega)[\ainT(-\omega + \Omega)^\dagger + \ainT(\omega + \Omega)] / \sqrt{2} \\
    &+ \talc(\omega)[\alT(-\omega + \Omega)^\dagger + \alT(\omega + \Omega)] / \sqrt{2} \\
    &+ \tbc(\omega)[\bin(-\omega + \Omega)^\dagger + \bin(\omega + \Omega)] / \sqrt{2} \\
    &+ \tcc(\omega)[\cin(-\omega)^\dagger + \cin(\omega)] / \sqrt{2} \\
    &+ \tclc(\omega)[\cl(-\omega)^\dagger + \cl(\omega)] / \sqrt{2} \\
    &+ \tvc(\omega)[\YLv(-\omega + \Omega)^\dagger + \YLv(\omega + \Omega)] / \sqrt{2},
\end{eqnarray}
using the fact that the transfer gains $t_{j\ell}(\omega) = t_{j\ell}^*(-\omega)$ are all Hermitian. As our protocol transfers the frequency component at $+\Omega$ of $\ainT$, $\alT$, and $\bin$ onto the microwave resonance, we define the frequency-shifted (sideband) operators $\SBp{\mathcal{O}}(\omega) = \mathcal{O}(\omega + \Omega)$ for each of these inputs, along with their quadratures 
$\pXLjT(\omega) = [a_{j}^{T,\SBpSymb}(-\omega)^\dagger + a_{j}^{T,\SBpSymb}(\omega)] / \sqrt{2}$, $\pYLjT(\omega) = \rmi[a_{j}^{T,\SBpSymb}(-\omega)^\dagger - a_{j}^{T,\SBpSymb}(\omega)] / \sqrt{2}$ and similarly $\SBp{\Qin}, \SBp{\Pin}$ for $\SBp{\bin}$. These allow us to write
\begin{eqnarray}
    \XMout(\omega) &= \tac(\omega) \pXLinT(\omega) + \talc(\omega) \pXLlossT(\omega) + \tbc(\omega) \SBp{\Qin}(\omega) \nonumber \\
    &+ \tcc(\omega) \XMin(\omega) + \tclc(\omega) \XMloss(\omega) \nonumber \\
    &+ \tvc(\omega)[\YLv(\omega - \Omega) + \YLv(\omega + \Omega)] / \sqrt{2} \label{eq:sup:mw_quadrature}
\end{eqnarray}

To quantify the different noise contributions, we calculate the power spectral density~\cite{Warwick_Textbook} of the microwave amplitude quadrature, given by
\begin{eqnarray}
    \SXMout(\omega) = \int_{-\infty}^\infty \rmd \omega' \langle \XMout(\omega) \XMout(\omega') \rangle. \label{eq:sup:mw_PSD_definition}
\end{eqnarray}
Assuming that all input operators in Eq.~\eqref{eq:sup:mw_quadrature} are uncorrelated, we can split the contributions $\SopXMout(\omega)$ to the microwave output PSD by source operator $\mathcal{O}$. 

\subsection{Optical loss noise}
\label{sec:apx:optical_loss_noise}
We start by calculating the optical loss contribution. We recall the definition of the transferred loss mode $\alT(\omega) = [\XLloss^T(\omega) + \rmi \YLloss^T(\omega)] / \sqrt{2}$, writing it in terms of the squeezed and anti-squeezed loss port quadratures
\begin{eqnarray}
    \XLloss^T(\omega) = \frac{1}{\sqrt{2\hineff\lwfact}} \XLloss(\omega), \quad \YLloss^T(\omega) = \sqrt{2\hineff\lwfact} \YLloss(\omega),
\end{eqnarray}
respectively. We use these to rewrite the amplitude quadrature of $a_\ell^{T,\SBpSymb}(\omega)$, the sideband of $\alT(\omega)$ that is imprinted on the microwave output, as
\begin{eqnarray}
    \pXLlossT(\omega) &= [\alT(-\omega + \Omega)^\dagger + \alT(\omega + \Omega)] / \sqrt{2}  \\
    &= \frac{1}{2} \left[ \XLloss^T(\omega - \Omega) + \XLloss^T(\omega + \Omega) - \rmi \YLloss^T(\omega - \Omega) + \rmi \YLloss^T(\omega + \Omega) \right]. \nonumber
\end{eqnarray}
From this expression, we see that both the negative and positive optomechanical sidebands of the loss port field contribute to $\pXLlossT(\omega)$, and thus $\XMout(\omega)$. 

Next, we calculate the contribution of the optical loss to the microwave output noise spectrum 
\begin{eqnarray}
    &\SalTXMout(\omega) = \int_{-\infty}^\infty \rmd \omega' \langle \talc(\omega) \pXLlossT(\omega)\, \talc(\omega') \pXLlossT(\omega') \rangle \nonumber \\
    &= \frac{1}{4} \int_{-\infty}^\infty \rmd \omega' \talc(\omega) \talc(\omega') \Big\langle \XLloss^T(\omega-\Omega) \XLloss^T(\omega' - \Omega) + \XLloss^T(\omega+\Omega) \XLloss^T(\omega' + \Omega) \Big\rangle \nonumber \\
    &- \frac{1}{4} \int_{-\infty}^\infty \rmd \omega' \talc(\omega) \talc(\omega') \Big\langle \YLloss^T(\omega-\Omega) \YLloss^T(\omega' - \Omega) + \YLloss^T(\omega+\Omega) \YLloss^T(\omega' + \Omega) \Big\rangle \nonumber \\
    &+ \frac{1}{4} \int_{-\infty}^\infty \rmd \omega' \talc(\omega) \talc(\omega') \Big\langle \XLloss^T(\omega-\Omega) \XLloss^T(\omega' + \Omega)  + \XLloss^T(\omega+\Omega) \XLloss^T(\omega' - \Omega) \Big \rangle \nonumber \\
    &+ \frac{1}{4} \int_{-\infty}^\infty \rmd \omega' \talc(\omega) \talc(\omega') \Big\langle \YLloss^T(\omega-\Omega) \YLloss^T(\omega' + \Omega)  + \YLloss^T(\omega+\Omega) \YLloss^T(\omega' - \Omega) \Big \rangle, \nonumber
\end{eqnarray}
noting that $\langle \XLloss^T(\omega) \YLloss^T(\omega') \rangle = -\langle \YLloss^T(\omega) \XLloss^T(\omega) \rangle$, so that all cross-quadrature terms cancel out. We further note that, in the optical ground state, the \emph{inter}-sideband $(\pm\Omega,\pm\Omega)$-correlators given by
\begin{eqnarray}
    \left\langle \XLloss^T(\omega\pm\Omega) \XLloss^T(\omega' \pm \Omega) \right\rangle
    &= \left(\frac{1}{2\hineff\lwfact}\right) \frac{1}{2} \delta(\omega + \omega' \pm 2\Omega), \\
    \left\langle \YLloss^T(\omega\pm\Omega) \YLloss^T(\omega' \pm \Omega) \right\rangle
    &= \left(2\hineff\lwfact\right) \frac{1}{2} \delta(\omega + \omega' \pm 2\Omega),
\end{eqnarray}
appearing in the first two terms peak at frequencies $\omega' = -\omega \pm 2\Omega$, while the \emph{intra}-sideband $(\pm\Omega,\mp\Omega)$-correlators
\begin{eqnarray}
    \left\langle \XLloss^T(\omega\pm\Omega) \XLloss^T(\omega' \mp \Omega) \right\rangle &= \left(\frac{1}{2\hineff\lwfact}\right) \frac{1}{2} \delta(\omega + \omega'), \\
    \left\langle \YLloss^T(\omega\pm\Omega) \YLloss^T(\omega' \mp \Omega) \right\rangle 
    &= \left(2\hineff\lwfact\right) \frac{1}{2} \delta(\omega + \omega'),
\end{eqnarray}
appearing in the last two terms peak at $\omega' = -\omega$. We can thus write
\begin{eqnarray}
    &\SalTXMout(\omega) = \nonumber \\
    &\frac{1}{4} \talc(\omega)\left[\talc(\omega - 2\Omega) + \talc(\omega + 2\Omega)\right] \int_{-\infty}^\infty \rmd \omega' \left\langle \XLloss^T(\omega) \XLloss^T(\omega') \right\rangle - \left\langle \YLloss^T(\omega) \YLloss^T(\omega') \right\rangle \nonumber \\
    &+ \frac{1}{2}\talc(\omega)\talc(-\omega) \int_{-\infty}^\infty \rmd \omega' \left\langle \XLloss^T(\omega) \XLloss^T(\omega') \right\rangle + \left\langle \YLloss^T(\omega) \YLloss^T(\omega') \right\rangle \nonumber.
\end{eqnarray}
The first term represents correlations between the upper and lower sidebands, filtered by the resonator response. In the high-quality limit $\mathcal{Q}' \gg 1$, the transfer gain $|\talc(\omega)|$ has negligible magnitude around $\omega \approx \pm 2\Omega$, so that we can neglect this cross-sideband contribution to obtain
\begin{eqnarray}
    \SalTXMout(\omega) &= \Talc(\omega) \left( \frac{1}{2}\spectSymb[\XLloss^T](\omega) + \frac{1}{2} \spectSymb[\YLloss^T](\omega) \right) \\
    &= \Talc(\omega) \frac{1}{4} \left( \frac{1}{2\hineff\lwfact} + 2\hineff\lwfact \right) \\
    &= \effM \effbol \Tinf(\omega) \frac{1}{4} \left( \frac{1}{2\hineff\lwfact} + 2\hineff\lwfact \right) \label{eq:sup:optical_loss_noise}
\end{eqnarray}
We recognise the optical loss noise in Eq.~\eqref{eq:sup:optical_loss_noise} as the average variance of the squeezed and anti-squeezed quadratures $\XLloss^T$ and $\YLloss^T$, which attains a minimum of $1/2$ for $\hineff = 1/(2\lwfact)$, multiplied by the transfer coefficient $\Talc(\omega)$. 

\subsection{Detection inefficiency}
\label{sec:apx:detection_noise}
Next, we calculate the spectrum of the noise contributed by detection inefficiency. This is given by
\begin{eqnarray}
    &\SYLvXMout(\omega) = \frac{1}{2} \int_{-\infty}^\infty \rmd \omega' \tvc(\omega) \tvc(\omega') \times \nonumber \\
    & \hspace{5em} \big\langle [\YLv(\omega - \Omega) + \YLv(\omega + \Omega)] 
    [\YLv(\omega' - \Omega) + \YLv(\omega' + \Omega)] 
    \big\rangle \nonumber \\
    &= \frac{1}{2} \int_{-\infty}^\infty \rmd \omega' \tvc(\omega) \tvc(\omega') \big\langle \YLv(\omega-\Omega) \YLv(\omega' - \Omega) + \YLv(\omega' + \Omega) \YLv(\omega + \Omega)  \big\rangle \nonumber \\
    &+ \frac{1}{2} \int_{-\infty}^\infty \rmd \omega' \tvc(\omega) \tvc(\omega') \big\langle \YLv(\omega-\Omega) \YLv(\omega' + \Omega) + \YLv(\omega' + \Omega) \YLv(\omega - \Omega)  \big\rangle. \nonumber 
\end{eqnarray}
As before for the optical loss contribution, the first term represents filtered correlations between the sidebands. We can invoke the same high-quality approximation to neglect it, and obtain
\begin{eqnarray}
    \SYLvXMout(\omega) &= \Tvc(\omega) \spectSymb[\YLv](\omega) = \Tvc(\omega) / 2 \\
    &= \frac{\effM \effbopt}{(2\effL - 1)^2} \frac{1-\effd}{\effd}  \frac{\hineff}{4} \Tinf(\omega). \nonumber
\end{eqnarray}
Here, we have assumed shot-noise-limited detection so that $\spectSymb[\YLv](\omega) = 1/2$.

\subsection{Mechanical noise}
The spectrum of the mechanical noise contribution is given by
\begin{eqnarray}
    &\SbinXMout(\omega) = \int_{-\infty}^\infty \rmd \omega' \tbc(\omega)\tbc(\omega') \big\langle \SBp{\Qin}(\omega) \SBp{\Qin}(\omega') \big\rangle \nonumber \\
    &= \frac{1}{2}\int_{-\infty}^\infty \rmd \omega' \tbc(\omega)\tbc(\omega') \big\langle 
        \bin(-\omega+\Omega)^\dagger \bin(\omega' + \Omega) + \bin(\omega+\Omega) \bin(-\omega' + \Omega)^\dagger
    \big\rangle \nonumber \\
    &= \left( \bar{n} + \frac{1}{2} \right) \int_{-\infty}^\infty \rmd \omega' \tbc(\omega)\tbc(\omega') \delta(\omega + \omega') =  \tbc(\omega)\tbc(-\omega) \left( \bar{n} + \frac{1}{2} \right) \\
    &= \Tbc(\omega) \left( \bar{n} + \frac{1}{2} \right) = \effM \effbin \Tinf(\omega) \left( \bar{n} + \frac{1}{2} \right).
\end{eqnarray}

\subsection{Microwave noise}
At temperatures relevant for quantum information processing, the microwave environment comprising $\cin$ and $\cl$ is in its ground state, so that $\SXMin(\omega) = \SXMl(\omega) = 1/2$. The microwave contributions to the output noise are thus given by
\begin{eqnarray}
    &\ScinXMout(\omega) + \SclXMout(\omega) = (\Tcc(\omega) + \Tclc(\omega))/2 \nonumber \\
    &= \frac{1-\effM\Tinf(\omega)}{2}.
\end{eqnarray}

\subsection{Total noise}
The spectrum of all noise contributions to the microwave output is then given by
\begin{eqnarray}
    &\spectSymb[\XMout|\text{noise}] = \frac{1}{2}\Big[
    \overbrace{\effM \effbol \frac{1}{2} (2\hineff\lwfact + 1/2\hineff\lwfact) \Tinf(\omega)}^\text{optical} 
    + \overbrace{\effM \effbin (2\bar{n} + 1) \Tinf(\omega)}^\text{mechanical} 
    \nonumber \\
    &\hspace{5em} + \underbrace{(1-\effM \Tinf(\omega))}_\text{microwave}  
    + \underbrace{\frac{\effM \effbopt}{(2\effL - 1)^2} \frac{1 - \effd}{\effd} \frac{\hineff}{2} \Tinf(\omega) }_\text{detection}
    \Big], \label{eq:sup:output_noise_hineff}
\end{eqnarray}
which reduces to Eq.~\eqref{eq:output-noise-specific} in the main text for $\hineff = 1$.

\subsection{Noise in the output phase quadrature}
The phase quadrature of the microwave output can be decomposed as
\begin{eqnarray}
    \YMout(\omega) &= \tac(\omega) \pYLinT(\omega) + \talc(\omega) \pYLlossT(\omega) + \tbc(\omega) \SBp{\Pin}(\omega) \nonumber \\
    &+ \tcc(\omega) \YMin(\omega) + \tclc(\omega) \YMloss(\omega) \nonumber \\
    &+ \rmi\tvc(\omega)[\YLv(\omega - \Omega) - \YLv(\omega + \Omega)] / \sqrt{2}.
\end{eqnarray}
Under the (rotating-wave and high-quality) approximations that we have made in our analysis, our transduction protocol transfers amplitude and phase quadratures symmetrically. Therefore, similar calculations show that the contribution of each noise source to $\SYMout(\omega)$ is exactly the same as for $\SXMout(\omega)$. The total noise spectrum of the phase quadrature $\spectSymb[\YMout|\text{noise}] = \spectSymb[\XMout|\text{noise}]$ is thus given by Eq.~\eqref{eq:sup:output_noise_hineff} as well.

\section{Content of the homodyne photocurrent}
\label{sec:apx:homodyne_content}
In this section, we analyse the content of the homodyne photocurrent obtained by measuring the output phase quadrature. For simplicity, we assume perfect detection and coupling efficiencies $\effd = \effL = \effM = 1$, so that its frequency response reads
\begin{eqnarray}
    \YLout(\omega) = \sqrt{4\Gamma\CL}Q(\omega) - \YLin(\omega) = \sqrt{\effbopt \Gamma' / h} Q(\omega) - \YLin(\omega). \label{eq:sup:ylout-cnt-1}
\end{eqnarray}
Here, we have used the fact that $4\Gamma\CL = \Gamma' 4\CL / (1 + 4 h \CL) = \effbopt \Gamma' / h$. Next, we express $Q(\omega) = [b(-\omega)^\dagger + b(\omega)]/ \sqrt{2}$ in terms of the mechanical response
\begin{eqnarray}
    b(\omega) &= \chibld(\omega) \sqrt{\Gamma'} \bdrv(\omega) + \rmi \chix(\omega-\Omega) \sqrt{\kM} \cin(\omega-\Omega) \label{eq:sup:b-cnt-1} \\
    &= \chibld(\omega) \sqrt{\Gamma'} \left[ \sqrt{\effbin} \bin(\omega) - \rmi \sqrt{\effbopt} \ainT(\omega) \right] + \rmi \chix(\omega-\Omega) \sqrt{\kM} \cin(\omega-\Omega) \nonumber,
\end{eqnarray}
where the transferred mode $\ainT(\omega)$ is given in Eq.~\eqref{eq:sup:ainT-h}.

We combine Eqs.~\eqref{eq:sup:ylout-cnt-1} and~\eqref{eq:sup:b-cnt-1} to obtain
\begin{eqnarray}
    \YLout(\omega) = &-\rmi \left[ \ain(-\omega)^\dagger - \ain(\omega) \right] / \sqrt{2} \nonumber \\
    &+ \rmi \frac{\effbopt}{\sqrt{h}} \left[ \Gamma' \chibld^*(-\omega) \ainT(-\omega)^\dagger - \Gamma' \chibld(\omega) \ainT(\omega) \right] / \sqrt{2} \nonumber \\
    &+ \frac{\sqrt{\effbopt \effbin}}{\sqrt{h}} \left[ \Gamma' \chibld^*(-\omega) \bin(-\omega)^\dagger + \Gamma' \chibld(\omega) \bin(\omega) \right] / \sqrt{2} \nonumber \\
    &-\rmi \frac{\sqrt{\effbopt}}{\sqrt{h}} \bigg[ \sqrt{\Gamma' \kM} \chix^*(-\omega-\Omega) \cin(-\omega-\Omega)^\dagger \nonumber \\
    &\hspace{5em} - \sqrt{\Gamma'\kM} \chix(\omega-\Omega) \cin(\omega-\Omega) \bigg] / \sqrt{2} \label{eq:sup:ylout-cnt-2}
\end{eqnarray}
For convenience, we rewrite the bare optical input mode operator in terms of $\ainT(\omega)$,
\begin{eqnarray}
    \ain(\omega) = \frac{1+h}{2\sqrt{h}} \ainT(\omega) - \frac{1-h}{2\sqrt{h}} \ainT(-\omega)^\dagger,
\end{eqnarray}
so that we can write the bare input phase quadrature as
\begin{eqnarray}
    \YLin(\omega) = \rmi \left[ \ainT(-\omega)^\dagger - \ainT(\omega) \right] / \sqrt{2h}.
\end{eqnarray}
Plugging this into Eq.~\eqref{eq:sup:ylout-cnt-2} yields
\begin{eqnarray}
     \YLout(\omega) = &-\frac{\rmi}{\sqrt{h}} \left[ \taa^*(-\omega) \ainT(-\omega)^\dagger - \taa(\omega) \ainT(\omega) \right] / \sqrt{2} \nonumber \\
    &+ \frac{1}{\sqrt{h}} \left[\tba^*(-\omega) \bin(-\omega)^\dagger + \tba(\omega) \bin(\omega) \right] / \sqrt{2} \nonumber \\
    &+ \frac{\rmi}{\sqrt{h}} \bigg[\tca^*(-\omega) \cin(-\omega-\Omega)^\dagger - \tca(\omega) \cin(\omega-\Omega) \bigg] / \sqrt{2}, \label{eq:sup:ylout-cnt-3}
\end{eqnarray}
where we have defined the input-to-light transfer gains
\begin{eqnarray}
    \taa(\omega) &= 1 - \effbopt \Gamma' \chibld(\omega), \label{eq:taa_def} \\
    \tba(\omega) &= \sqrt{\effbopt \effbin} \Gamma' \chibld(\omega), \\
    \tca(\omega) &= -\sqrt{\effbopt} \sqrt{\Gamma' \kM} \chix(\omega-\Omega) =\sqrt{\effbopt} \tinf(\omega-\Omega).
\end{eqnarray}
These transfer gains exhibit either a dip (for $\taa(\omega)$) or a peak (for $\tba(\omega), \tca(\omega)$) around the positive mechanical frequency $\omega \approx +\Omega$. Consequently, none of these functions is Hermitian.
For $h=1$, the expression~\eqref{eq:sup:ylout-cnt-3} for the output phase quadrature reduces to Eq.~\eqref{eq:homodyne_from_inputs} in the main text.

The corresponding transfer coefficients $T_j(\omega) = |t_j(\omega)|^2$ can be written as 
\begin{eqnarray}
    \Taa(\omega) &= 1 - \effbopt \effbin (1+u(\omega-\Omega)^2) \frac{\Tinf(\omega-\Omega)}{\CM'} - \effbopt \Tinf(\omega-\Omega), \label{eq:Taa_def} \\
    \Tba(\omega) &= \effbopt \effbin (1+u(\omega-\Omega)^2) \frac{\Tinf(\omega-\Omega)}{\CM'}, \\
    \Tca(\omega) &= \effbopt \Tinf(\omega-\Omega),
\end{eqnarray}
where $u(\omega) = 2\omega/\kM$ is the relative detuning. It follows immediately that the coefficients $\Taa(\omega) + \Tba(\omega) + \Tca(\omega) = 1$ sum to one. We recall that the ideal transfer coefficient can be written as
\begin{eqnarray}
    \Tinf(\omega) = \frac{4\CM' \beta^2}{u(\omega)^4 + u(\omega)^2 (1-2 \CM' \beta + \beta^2) + \beta^2 (1+\CM')^2},
\end{eqnarray}
where $\beta = \Gamma' / \kM$ and $\CM' = 4 \gM^2 / \Gamma' \kM$. In addition, we list the useful identities
\begin{eqnarray}
    \Re[\Gamma' \chibld(\Delta+\Omega)] &= [1+u(\Delta)^2+\CM']\frac{\Tinf(\Delta)}{2\CM'}, \label{eq:Re_GammaChiM} \\
    \Im[\Gamma' \chibld(\Delta+\Omega)] &= u(\Delta)[1+u(\Delta)^2 - \CM'\beta]\frac{\Tinf(\Delta)}{2\CM'\beta}, \\
    |\Gamma' \chibld(\Delta+\Omega)|^2 &= [1+u(\Delta)^2]\frac{\Tinf(\Delta)}{\CM'}.
\end{eqnarray}

\subsection{Mechanical and microwave contributions}
To quantify the statistics of the classical photocurrent, readily observed in experiment, we calculate the power spectral density of $\YLout(\omega)$ in Eq.~\eqref{eq:sup:ylout-cnt-3}. This is given by
\begin{eqnarray}
    \spectSymb[\YLout](\omega) = \int_{-\infty}^{\infty} \rmd \omega' \left\langle \YLout(\omega) \YLout(\omega') \right \rangle.
\end{eqnarray}
Assuming the optical, mechanical and microwave input fields are uncorrelated, we can split $\spectSymb[\YLout](\omega)$ into three corresponding contributions. 

As we expect the mechanical and microwave environments to be a thermal state or ground state, respectively, their contributions are readily calculated, and given by
\begin{eqnarray}
    \spectSymb[\YLout|\bin](\omega) &= \frac{1}{2h} \int_{-\infty}^\infty \rmd \omega' \bigg\langle \left[\tba^*(-\omega) \bin(-\omega)^\dagger + \tba(\omega) \bin(\omega) \right]  \nonumber \\
    &\hspace{4em} \times \left[\tba^*(-\omega') \bin(-\omega')^\dagger + \tba(\omega') \bin(\omega') \right] \bigg \rangle \nonumber \\
    &= \frac{1}{2h} \int_{-\infty}^\infty \rmd \omega'  \tba^*(-\omega) \tba(\omega') \langle \bin(-\omega)^\dagger  \bin(\omega') \rangle \nonumber \\
    &+ \frac{1}{2h} \int_{-\infty}^\infty \rmd \omega' \tba(\omega) \tba^*(-\omega') \langle \bin(\omega)  \bin(-\omega')^\dagger \rangle \nonumber \\
    &= \frac{\bar{n} \Tba(-\omega) + (\bar{n} + 1) \Tba(\omega) }{2h}, 
\end{eqnarray}
and
\begin{eqnarray}
     \spectSymb[\YLout|\cin](\omega) &= \frac{-1}{2h}  \int_{-\infty}^\infty \rmd \omega' \bigg\langle \left[\tca^*(-\omega) \cin(-\omega-\Omega)^\dagger - \tca(\omega) \cin(\omega-\Omega) \right] \nonumber \\
    &\hspace{4em} \times \left[\tca^*(-\omega') \cin(-\omega'-\Omega)^\dagger - \tca(\omega') \cin(\omega'-\Omega) \right] \bigg \rangle \nonumber \\
    &= \frac{1}{2h} \int_{-\infty}^\infty \rmd \omega' \tca(\omega) \tca^*(-\omega') \langle \cin(\omega - \Omega)  \cin(-\omega' - \Omega)^\dagger \rangle \nonumber \\
    &= \frac{\Tca(\omega)}{2h}.
\end{eqnarray}

\subsection{Optical contribution from correlators}
The optical contribution to $\spectSymb[\YLout]$ is given by
\begin{eqnarray}
    \spectSymb[\YLout|\ainT](\omega) &= \frac{-1}{2h} \int_{-\infty}^\infty \rmd \omega' \bigg\langle \left[\taa^*(-\omega) \ainT(-\omega)^\dagger - \taa(\omega) \ainT(\omega) \right]  \nonumber \\
    &\hspace{4em} \times \left[\taa^*(-\omega') \ainT(-\omega')^\dagger - \taa(\omega') \ainT(\omega') \right] \bigg \rangle \nonumber \\
    &= \frac{1}{2h} \int_{-\infty}^\infty \rmd \omega' \taa^*(-\omega) \taa(\omega') 
    \underbrace{\langle \ainT(-\omega)^\dagger \ainT(\omega') \rangle}_{\CA} \nonumber \\
    &+ \frac{1}{2h} \int_{-\infty}^\infty \rmd \omega' \taa(\omega) \taa^*(-\omega') 
    \underbrace{\langle \ainT(\omega) \ainT(-\omega')^\dagger \rangle}_{\CB} \nonumber \\
    &- \frac{1}{2h} \int_{-\infty}^\infty \rmd \omega' \taa^*(-\omega) \taa^*(-\omega') 
    \underbrace{\langle \ainT(-\omega)^\dagger \ainT(-\omega')^\dagger \rangle}_{\CC} \nonumber \\
    &- \frac{1}{2h} \int_{-\infty}^\infty \rmd \omega' \taa(\omega) \taa(\omega') 
    \underbrace{\langle \ainT(\omega) \ainT(\omega') \rangle}_{\CD} \label{eq:opt_spectrum_from_Cs}
\end{eqnarray}
The four indicated correlators $\mathcal{C}_j$ that appear in this contribution are more cumbersome to calculate, for two reasons. Firstly, for $h\neq 1$, $\ainT(\omega)$ does not represent the input field directly but rather the (squeezed) Bogoliubov mode given by Eq.~\eqref{eq:sup:ainT-h}, and secondly, we cannot assume the optical environment to be in the ground state, as it embeds the arbitrary quantum state to be transferred. 

\subsection{Constructing an optical input}
\label{sec:apx:construct_optical_input}
To specify the input state, we define a single-photon input pulse as a coherent superposition of environmental excitations, weighted by a spectral mode shape $\xi(\omega)$. This definition follows section 6.2 in ref.~\cite{Warwick_Textbook} and reads 
\begin{eqnarray}
    \ket{1_\xi} = \int_{-\infty}^\infty \rmd\omega\, \xi(\omega) \ain(\omega)^\dagger \ket{0}.
\end{eqnarray}
As $\ain(\omega)^\dagger$ has units of $1/\sqrt{\text{Hz}}$, the spectral mode shape must have units of $1/\sqrt{\text{Hz}}$ to ensure that $\ket{1_\xi}$ is dimensionless.
The state is normalised when
\begin{eqnarray}
    \braket{1_\xi}{1_\xi} &= \intinf \rmd\omega'\, \intinf \rmd\omega\, \xi^*(\omega')\xi(\omega) \sbraket{0\middle|\ain(\omega')\ain(\omega)^\dagger\middle|0} \\
    &= \intinf \rmd\omega'\, \intinf \rmd\omega\, \xi^*(\omega')\xi(\omega) \delta(\omega - \omega') \\
    &= \intinf \rmd\omega\, |\xi(\omega)|^2 = 1.
\end{eqnarray}
Accordingly, we define the creation operator $Z^\dagger$ that corresponds to the input pulse as
\begin{eqnarray}
    Z^\dagger =  \int_{-\infty}^\infty \rmd\omega\, \xi(\omega) \ain(\omega)^\dagger
\end{eqnarray}
This operator is also dimensionless. It satisfies the commutator
\begin{eqnarray}
    [Z, Z^\dagger] &= ZZ^\dagger - Z^\dagger Z = \intinf \rmd \omega'\, \intinf \rmd\omega\, \xi^*(\omega')\xi(\omega) \ain(\omega')\ain(\omega)^\dagger \\
    &- \intinf \rmd \omega'\, \intinf \rmd\omega\, \xi^*(\omega')\xi(\omega) \ain(\omega)^\dagger\ain(\omega') \\
    &= \intinf \rmd \omega'\, \intinf \rmd\omega\, \xi^*(\omega')\xi(\omega) [\ain(\omega'), \ain(\omega)^\dagger] \\
    &= \intinf \rmd \omega'\, \intinf \rmd\omega\, \xi^*(\omega')\xi(\omega) \delta(\omega - \omega') \\
    &= \intinf \rmd\omega\, |\xi(\omega)|^2 = 1,
\end{eqnarray}
as expected for a bosonic mode. In addition, it inherits from $\ain(\omega')$ the usual ladder operator properties
\begin{eqnarray}
    Z\ket{n_\xi} = \sqrt{n} \ket{n-1_\xi}, \quad\quad Z^\dagger\ket{n_\xi} = \sqrt{n+1} \ket{n+1_\xi},
\end{eqnarray}
where $\ket{n_\xi} = (Z^\dagger)^n / \sqrt{n!} \ket0$ is the Fock state with $n$ photons in the pulse. Using $Z^\dagger$, we define the operator
\begin{eqnarray}
    \Psi^\dagger = \sum_n \frac{\psi_n^*}{\sqrt{n!}} (Z^\dagger)^n
\end{eqnarray}
that populates our pulse with an arbitrary quantum state, specified by the Fock state amplitudes $\psi_n$. The populated input state is then given by
\begin{eqnarray}
    \ket{\psi} = \Psi^\dagger\ket{0} = \sum_n \psi_n^* \ket{n_\xi}.
\end{eqnarray}

Next, we determine how the environmental annihilation operator $\ain(\omega')$ acts on the input state $\ket\psi$. We will prove by induction that
\begin{eqnarray}
    \ain(\omega') \ket{n_\xi} = \xi(\omega') Z \ket{n_\xi}. \label{eq:ain_on_psi}
\end{eqnarray}
We first prove this expression for the lowest two Fock states $n=0, 1$ explicitly,
\begin{eqnarray*}
    \ain(\omega') \ket{0_\xi} &= \ain(\omega') \ket{0} = 0 \\
    &= \xi(\omega') Z \ket{0_\xi}, \\
    \ain(\omega') \ket{1_\xi} &= \ain(\omega') \intinf \rmd\omega\, \xi(\omega) \ain(\omega)^\dagger \ket{0} = \intinf \rmd\omega\, \xi(\omega)  \ain(\omega') \ain(\omega)^\dagger \ket{0} \\
    &= \intinf \rmd\omega\, \xi(\omega) \bigg[ \delta(\omega-\omega') + \ain(\omega)^\dagger \underbrace{\ain(\omega') \bigg] \ket{0}}_{=0} \\
    &= \intinf \rmd\omega\, \xi(\omega) \delta(\omega-\omega') \ket{0} \\
    &= \xi(\omega') \ket{0} = \xi(\omega') \ket{0_\xi} \\
    &= \xi(\omega') Z \ket{1_\xi}.
\end{eqnarray*}
We then prove the induction step
\begin{eqnarray*}
    \ain(\omega') \ket{n_\xi} &= \ain(\omega') \frac{Z^\dagger}{\sqrt{n}} \ket{n-1_\xi} \\
    &= \frac{1}{\sqrt{n}} \ain(\omega') \intinf \rmd\omega\, \xi(\omega) \ain(\omega)^\dagger\ket{n-1_\xi} \\
    &= \frac{1}{\sqrt{n}}\intinf \rmd\omega\, \xi(\omega) \ain(\omega') \ain(\omega)^\dagger\ket{n-1_\xi} \\
    &= \frac{1}{\sqrt{n}} \intinf \rmd\omega\, \xi(\omega) \bigg[ \delta(\omega - \omega') + \ain(\omega)^\dagger \underbrace{\ain(\omega') \bigg] \ket{n-1_\xi}}_{=\sqrt{n-1}\xi(\omega')\ket{n-2_\xi}} \\
    &= \frac{\xi(\omega')}{\sqrt{n}} \ket{n-1_\xi} + \frac{\xi(\omega')\sqrt{n-1}}{\sqrt{n}} \underbrace{ \intinf \rmd\omega\, \xi(\omega) \ain(\omega)^\dagger}_{=Z^\dagger} \ket{n-2_\xi} \\
    &= \frac{\xi(\omega')}{\sqrt{n}} \ket{n-1_\xi} + \frac{\xi(\omega')\sqrt{n-1}}{\sqrt{n}} \underbrace{Z^\dagger \ket{n-2_\xi}}_{=\sqrt{n-1}\ket{n-1_\xi}} \\
    &= \frac{\xi(\omega')}{\sqrt{n}} \ket{n-1_\xi} + \frac{\xi(\omega')(n-1)}{\sqrt{n}} \ket{n-1_\xi} 
    \\
    &= \sqrt{n} \xi(\omega') \ket{n-1_\xi} \\ 
    &= \xi(\omega') Z \ket{n_\xi},
\end{eqnarray*}
using the induction hypothesis Eq.~\eqref{eq:ain_on_psi} for $n-1$ on the fourth line. 
Finally, the action of $\ain(\omega')$ on $\ket\psi$ follows straightforwardly as
\begin{eqnarray}
    \ain(\omega') \ket\psi = \sum_n \psi_n^* \ain(\omega') \ket{n_\xi} = \xi(\omega') Z \sum_n \psi_n^* \ket{n_\xi} = \xi(\omega') Z\ket\psi.
\end{eqnarray}

\subsection{Correlators of the optical input}
Next, we turn to calculating the four correlators $\mathcal{C}_j$ that appear in Eq.~\eqref{eq:opt_spectrum_from_Cs}. We start with the special case $h=1$, where $\ainT(\omega) = \ain(\omega)$. For the first correlator, we then find
\begin{eqnarray}
    \CA' &= \bra\psi \ain(-\omega)^\dagger \ain(\omega') \ket\psi = \Big(\ain(-\omega) \ket\psi\Big)^\dagger \Big(\ain(\omega') \ket\psi\Big) \nonumber \\
    &= \xi^*(-\omega) \xi(\omega') \bra\psi Z^\dagger Z \ket\psi = \xi^*(-\omega) \xi(\omega') \nxi,
\end{eqnarray}
where $\nxi = \bra\psi Z^\dagger Z \ket\psi = \sum_n n |\psi_n|^2$ is the mean photon number in the pulse.
The second correlator follows similarly as
\begin{eqnarray}
    \CB' &= \bra\psi \ain(\omega) \ain(-\omega')^\dagger \ket\psi = \bra\psi \delta(\omega + \omega') + \ain(-\omega')^\dagger \ain(\omega) \ket\psi \nonumber \\
    &= \delta(\omega + \omega') + \xi(\omega) \xi^*(-\omega') \bra\psi Z^\dagger Z \ket\psi \nonumber \\
    &= \delta(\omega + \omega') + \xi(\omega) \xi^*(-\omega') \nxi.
\end{eqnarray}
The third correlator evaluates to
\begin{eqnarray}
    \CC' &= \bra\psi \ain(-\omega)^\dagger \ain(-\omega')^\dagger \ket\psi = \Big(\ain(-\omega') \ain(-\omega) \ket\psi \Big)^\dagger \ket\psi \\
    &= \xi^*(-\omega) \xi^*(-\omega') \bra\psi Z^\dagger Z^\dagger \ket\psi = \xi^*(-\omega) \xi^*(-\omega') \zsqxicj,
\end{eqnarray}
where $\zsqxi = \bra\psi ZZ \ket\psi = \sum_n \sqrt{(n+1)(n+2)} \psi_n \psi_{n+2}^*$.
Finally, the fourth correlator is given by
\begin{eqnarray}
    \CD' &= \bra\psi \ain(\omega) \ain(\omega') \ket\psi = \xi(\omega) \xi(\omega') \bra\psi ZZ \ket\psi \\
    &= \xi(\omega) \xi(\omega') \zsqxi.
\end{eqnarray}

To calculate the correlators for arbitrary gain, we recall that the transferred optical field is given by Eq.~\eqref{eq:sup:ainT-h},
\begin{eqnarray}
    \ainT(\omega) &= \left( \frac{1}{\sqrt{h}} \XLin(\omega) + \rmi \sqrt{h} \YLin(\omega) \right) / \sqrt{2} \\
    &= \frac{1+h}{2\sqrt{h}} \ain(\omega) + \frac{1-h}{2\sqrt{h}} \ain(-\omega)^\dagger
\end{eqnarray}
After tedious but straightforward calculations, we find that the optical correlators for $h\neq 1$ are combinations of the correlators calculated above for $h=1$, and given by
\begin{eqnarray}
    \CA &= \frac{(1+h)^2}{4h} \CA' + \frac{1-h^2}{4h} (\CC' + \CD') + \frac{(1-h)^2}{4h} \CB', \\
    \CB &= \frac{(1+h)^2}{4h} \CB' + \frac{1-h^2}{4h} (\CC' + \CD') + \frac{(1-h)^2}{4h} \CA', \\
    \CC &= \frac{(1+h)^2}{4h} \CC' + \frac{1-h^2}{4h} (\CA' + \CB') + \frac{(1-h)^2}{4h} \CD', \\
    \CD &= \frac{(1+h)^2}{4h} \CD' + \frac{1-h^2}{4h} (\CA' + \CB') + \frac{(1-h)^2}{4h} \CC'.
\end{eqnarray}

\subsection{Homodyne spectrum for optimal gain}\label{sec:apx:homodyne_spectrum_pulse_optgain}
We can now use these correlators to calculate the optical contribution to the homodyne spectrum. 
Here, we do that for the optimal (no-squeezing) gain setting $h=1$ and find
\begin{eqnarray*}
    \spectSymb[\YLout|\ainT](\omega) 
    &= \frac{\Taa(\omega)}{2} + \frac{1}{2} \int_{-\infty}^\infty \rmd \omega' \taa^*(-\omega) \taa(\omega') 
    \xi^*(-\omega)\xi(\omega') \langle Z^\dagger Z \rangle  \\
    &+ \frac{1}{2} \int_{-\infty}^\infty \rmd \omega' \taa(\omega) \taa^*(-\omega') 
    \xi(\omega)\xi^*(-\omega') \langle Z^\dagger Z \rangle  \\
    &- \frac{1}{2} \int_{-\infty}^\infty \rmd \omega' \taa^*(-\omega) \taa^*(-\omega') 
    \xi^*(-\omega)\xi^*(-\omega') \langle Z^\dagger Z^\dagger \rangle  \\
    &- \frac{1}{2} \int_{-\infty}^\infty \rmd \omega' \taa(\omega) \taa(\omega') 
    \xi(\omega)\xi(\omega') \langle Z Z \rangle, \\
\end{eqnarray*}
where the ground state fluctuations of the continuum of optical modes entering the input, including the mode to be transferred, are contained in the first term $\Taa(\omega)/2$. This expression is valid for any spectral mode shape $\xi(\omega)$. 

However, for high-fidelity state transfer, our input pulse (with centre frequency $\omega_p$) should have a narrow frequency distribution, well within the bandwidth of the transfer process. We will therefore assume that $\xi(\omega) = e^{\rmi \theta_0} \xi_0(\omega)$, where $\theta_0$ is the optical phase of the pulse and the real-valued envelope $\xi_0(\omega)$ is much narrower than the variation in the transfer gain $\taa(\omega)$. 
We can then approximate $\xi(\omega)\taa(\omega) \approx e^{\rmi \theta_0} \xi_0(\omega)\taa(\omega_p)$ to obtain
\begin{eqnarray*}
    \spectSymb[\YLout|\ainT](\omega) 
    &= \frac{\Taa(\omega)}{2} + \frac{\Taa(\omega_p)}{2} \int_{-\infty}^\infty \rmd \omega'  
    \xi_0(-\omega)\xi_0(\omega') \langle Z^\dagger Z \rangle  \\
    &+ \frac{\Taa(\omega_p)}{2} \int_{-\infty}^\infty \rmd \omega' 
    \xi_0(\omega)\xi_0(\omega') \langle Z^\dagger Z \rangle  \\
    &- \frac{\Taa(\omega_p)}{2} e^{-2\rmi\faa(\omega_p)} \int_{-\infty}^\infty \rmd \omega' 
    \xi_0(-\omega)\xi_0(\omega') e^{-2\rmi \theta_0} \langle Z^\dagger Z^\dagger \rangle  \\
    &- \frac{\Taa(\omega_p)}{2} e^{2\rmi\faa(\omega_p)} \int_{-\infty}^\infty \rmd \omega'
    \xi_0(\omega)\xi_0(\omega') e^{2\rmi \theta_0} \langle Z Z \rangle, \\
\end{eqnarray*}
where we also use the identity $\intinf f(-x)\,\rmd x = \intinf f(x)\,\rmd x$.
Here, $\faa(\omega_p) = \arg \taa(\omega_p)$. Informed by this expression, we define the rotated pulse operator $Z_\theta = e^{\rmi[\theta_0 + \faa(\omega_p)]} Z$. We also note that
\begin{eqnarray*}
    \int_{-\infty}^\infty \rmd \omega' \xi_0(\omega') = 2\pi \xi_t(0),
\end{eqnarray*}
where $\xi_t(\omega)$ is the inverse Fourier transform of $\xi_0(\omega)$. Moreover, $\xi_t(0)$ is real.

Our classical homodyne detector measures the symmetrised spectrum 
\begin{eqnarray}
    \bar\spectSymb[\YLout](\omega) = \frac{\spectSymb[\YLout](\omega) + \spectSymb[\YLout](-\omega)}{2}.
\end{eqnarray}
The optical contribution to the symmetrised spectrum reads
\begin{eqnarray*}
    \bar\spectSymb[\YLout|\ainT](\omega) 
    &= \frac{\Taa(\omega) + \Taa(-\omega)}{4} + \frac{\Taa(\omega_p)}{4} [\xi_0(-\omega) + \xi_0(\omega)] 2\pi \xi_t(0) \langle Z_\theta^\dagger Z_\theta \rangle \\
    &+ \frac{\Taa(\omega_p)}{4} [\xi_0(-\omega) + \xi_0(\omega)] 2\pi \xi_t(0) \langle Z_\theta^\dagger Z_\theta \rangle  \\
    &- \frac{\Taa(\omega_p)}{4}  [\xi_0(-\omega) + \xi_0(\omega)] 2\pi \xi_t(0) \langle Z_\theta^\dagger Z_\theta^\dagger \rangle  \\
    &- \frac{\Taa(\omega_p)}{4}  [\xi_0(-\omega) + \xi_0(\omega)] 2\pi \xi_t(0) \langle Z_\theta Z_\theta \rangle. 
\end{eqnarray*}
Next, we return to the positive frequency part of $\bar\spectSymb[\YLout|\ainT](\omega)$, from which the negative frequency part can be obtained trivially by negating the frequency axis. We approximate $\Taa(-\omega) \approx 1$ and $\xi_0(-\omega) \approx 0$ to get the single-sided spectrum
\begin{eqnarray*}
    \spectSymb[\YLout|\ainT](\omega) 
    &= \frac{1}{4} + \frac{\Taa(\omega)}{4} + \Taa(\omega_p) x(\omega) \left\langle Z_\theta^\dagger Z_\theta - \frac{1}{2} Z_\theta^\dagger Z_\theta^\dagger - \frac{1}{2} Z_\theta Z_\theta \right\rangle, \\
\end{eqnarray*}
with modeshape function $x(\omega) = \pi \xi_0(\omega) \xi_t(0)$. This can be conveniently expressed as
\begin{eqnarray*}
    \spectSymb[\YLout|\ainT](\omega) 
    &= \frac{1}{4} + \frac{\Taa(\omega)}{4} + \Taa(\omega_p) x(\omega) \bra\psi B_\theta^2 - 1/2 \ket\psi, \\
\end{eqnarray*}
where
\begin{eqnarray*}
    B_\theta = \rmi (Z_\theta^\dagger - Z_\theta) / \sqrt{2}
\end{eqnarray*}
is a quadrature of the pulse operator.

Finally, the total single-sided homodyne spectrum for optimal gain is given by
\begin{eqnarray}
    \bar\spectSymb[\YLout](\omega) = &\frac{1}{4} + \frac{\Taa(\omega)}{4} + \frac{\Tca(\omega)}{4} 
    + \frac{(2\bar n + 1) \Tba(\omega)}{4} \\
    &+ \Taa(\omega_p) x(\omega) \bra\psi B_\theta^2 - 1/2 \ket\psi
\end{eqnarray}
After applying the identity $\Taa(\omega) + \Tba(\omega) + \Tca(\omega) = 1$, this simplifies to
\begin{eqnarray}
    \spectSymb[\YLout](\omega) = &\frac{1}{2} + 
     \frac{\bar n}{2} \Tba(\omega) 
    + \Taa(\omega_p) x(\omega) \bra\psi B_\theta^2 - 1/2 \ket\psi \label{eq:optimal_homodyne_spectrum}
\end{eqnarray}
given in the main text as Eq.~\eqref{eq:homodyne_spectrum_main}. Note that for an input pulse squeezed in $B_\theta$, the final term is negative, as the input mode's ground state fluctuations are already incorporated in the first term $1/2$.

From Eq.~\eqref{eq:optimal_homodyne_spectrum}, we recognise that information about the input state, as encoded in the quadrature variance $\langle B_\theta^2 \rangle$, is imprinted on the spectrum of the classical photocurrent with intensity $\Taa(\omega_p)$. Importantly, as we increase the optomechanical cooperatitivity, the efficiency $\effbopt$ tends to one so that $\Taa(\omega_p)$ is minimised. In the limit $\effbopt \to 1$, we see from Eq.~\eqref{eq:Taa_def} that the light-to-light transfer coefficient reduces to $\Taa(\omega_p) = 1 - \Tinf(\omega_p-\Omega)$. Then, by matching the pulse centre frequency $\omega_p = \Omega + \sigdet$ to value close to the positive sideband where the light-to-microwave transfer efficiency $\Tinf(\sigdet) \approx 1$ is high, we find that $\Taa(\omega_p) \approx 0$ -- indicating that, indeed, the photocurrent does not reveal the input state.

\subsection{Vacuum input homodyne spectrum for arbitrary gain}
\label{sec:apx:homodyne_content_vacin_arbgain}
In the case of arbitrary gain $h \neq 1$, the transfer dynamics include squeezing and are harder to model for an arbitrary state in the input pulse -- which should now be spread over both the positive and negative sidebands. We therefore restrict our calculations to transfer of the optical vacuum state.

We then have $\CA' = \CC' = \CD' = 0$ and $\CB' = \delta(\omega+\omega')$, so that the correlators read
\begin{eqnarray}
    \CA &= \bra 0 \ainT(-\omega)^\dagger \ainT(\omega') \ket 0 &=  \frac{(1-h)^2}{4h} \delta(\omega+\omega'), \\
    \CB &= \bra 0 \ainT(\omega) \ainT(-\omega')^\dagger \ket 0 &= \frac{(1+h)^2}{4h} \delta(\omega+\omega'), \\
    \CC &= \bra 0 \ainT(-\omega)^\dagger \ainT(-\omega')^\dagger \ket 0 &= \frac{1-h^2}{4h} \delta(\omega+\omega'), \\
    \CD &= \bra 0 \ainT(\omega) \ainT(\omega') \ket 0 &= \frac{1-h^2}{4h} \delta(\omega+\omega').
\end{eqnarray}
We plug these into the optical contribution to get 
\begin{eqnarray*}
    \spectSymb[\YLout|\ainT](\omega)
    &= \frac{1}{2h} \int_{-\infty}^\infty \rmd \omega' \taa^*(-\omega) \taa(\omega') 
    \frac{(1-h)^2}{4h} \delta(\omega+\omega')  \\
    &+ \frac{1}{2h} \int_{-\infty}^\infty \rmd \omega' \taa(\omega) \taa^*(-\omega') 
    \frac{(1+h)^2}{4h} \delta(\omega+\omega')  \\
    &- \frac{1}{2h} \int_{-\infty}^\infty \rmd \omega' \taa^*(-\omega) \taa^*(-\omega') 
    \frac{1-h^2}{4h} \delta(\omega+\omega')  \\
    &- \frac{1}{2h} \int_{-\infty}^\infty \rmd \omega' \taa(\omega) \taa(\omega') 
    \frac{1-h^2}{4h} \delta(\omega+\omega') \\
    &= \frac{1}{8h^2} \bigg[(1-h)^2 \Taa(-\omega) + (1+h)^2 \Taa(\omega) \\
    &- (1-h^2) \taa^*(-\omega) \taa^*(\omega) - (1-h^2) \taa(\omega) \taa(-\omega) \bigg] \\
    &= \frac{1}{8h^2} \bigg[(1-h)^2 \Taa(-\omega) + (1+h)^2 \Taa(\omega) \\
    &- 2 (1-h^2) \Re[ \taa(-\omega) \taa(\omega)] \bigg]
\end{eqnarray*}
As $\taa(\omega)$ has a narrow dip around $\omega \approx +\Omega$ from its base value of $\taa(\omega \neq \Omega) = 1$, we can approximate 
\begin{eqnarray*}
    \taa(-\omega)\taa(\omega) &= ([\taa(-\omega) - 1] + 1)([\taa(\omega) - 1] + 1) \\
    &= \underbrace{[\taa(-\omega) - 1][\taa(\omega) - 1]}_{\text{at\ least\ one\ factor\  zero}} + \taa(-\omega) + \taa(\omega) - 1 \\
    &\approx \taa(-\omega) + \taa(\omega) - 1.
\end{eqnarray*}
The corresponding symmetrised spectrum simplifies to
\begin{eqnarray*}
    \bar\spectSymb[\YLout|\ainT](\omega)
    &= \frac{1+h^2}{8h^2} \bigg[ \Taa(\omega) + \Taa(-\omega) \bigg] \\
    &- \frac{2(1-h^2)}{8h^2} \Re[\taa(-\omega) + \taa(\omega) - 1] \\
    &= \frac{1+h^2}{8h^2} \bigg[ \Taa(\omega) + \Taa(-\omega) \bigg] \\
    &- \frac{2(1-h^2)}{8h^2} \left[1 - \frac{\Tba(\omega) + \Tba(-\omega)}{2\effbin}  - \frac{\Tca(\omega) + \Tca(-\omega)}{2}\right],
\end{eqnarray*}
after substituting Eq.~\eqref{eq:Re_GammaChiM}. Adding the mechanical and microwave contributions, we arrive at the full symmetrised homodyne spectrum
\begin{eqnarray*}
    \bar\spectSymb[\YLout](\omega) &= \frac{1+h^2}{8h^2} \bigg[ \Taa(\omega) + \Taa(-\omega) \bigg] \\
    &- \frac{(1-h^2)}{8h^2} \left[2 - \frac{\Tba(\omega) + \Tba(-\omega)}{\effbin}  - (\Tca(\omega) + \Tca(-\omega))\right] \\
    &+ \frac{(2\bar{n} + 1) \Tba(-\omega) + (2\bar{n} + 1) \Tba(\omega) + \Tca(\omega) + \Tca(-\omega)}{4h} \\
    &= \frac{\Taa(\omega) + \Taa(-\omega)}{4} + \frac{\Tca(\omega) + \Tca(-\omega)}{4h} \\
    &+ \frac{1-\effbin - h^2 + \effbin h (h + 4\bar n + 2)}{8h^2\effbin} [\Tba(\omega) + \Tba(-\omega)].
\end{eqnarray*}
Note that for $\omega \not\approx \pm \Omega$, the spectrum $\bar\spectSymb[\YLout](\omega) = 1/2$ indicates optical shot noise, as expected.

From now on, we restrict ourselves to positive frequencies $\omega>0$ for simplicity. The full spectrum can then be approximated as
\begin{eqnarray*}
    \bar\spectSymb[\YLout](\omega)
    &= \frac{1}{4} + \frac{\Taa(\omega)}{4} + \frac{\Tca(\omega)}{4h} \\
    &+ \frac{1-\effbin - h^2 + \effbin h (h + 4\bar n + 2)}{8h^2\effbin} \Tba(\omega).
\end{eqnarray*}
The equivalent form
\begin{eqnarray}
    \bar\spectSymb[\YLout](\omega) &= \frac{1}{2} + \frac{(4\CL + 1)\effbin - 1}{4} \Tinf(\omega- \Omega) \\
    &+ [4\CL(1/2 + \bar n + \CL) - 2 h \CL (1+2 h \CL)] \effbin^2 \\
    &\hspace{2em} \times (1+u(\omega-\Omega)^2) \frac{\Tinf(\omega-\Omega)}{2\CM'} \label{eq:spectrum_arbgain_symm_h0}
\end{eqnarray}
is amenable to taking the limit $h\to 0$, while the expression
\begin{eqnarray}
    \bar\spectSymb[\YLout](\omega) &= \frac{1}{2} - \left( 1-\frac{1}{h} \right) \frac{\Tinf(\omega-\Omega)}{4} \\
    &- \effbopt^2 (1+u(\omega-\Omega)^2) \frac{\Tinf(\omega-\Omega)}{8\CM'} \\
    &+ \frac{(1/2+\bar n + \CL)/h-1/2}{h\CL} \effbopt^2 \\
    &\hspace{2em} \times (1+u(\omega-\Omega)^2) \frac{\Tinf(\omega-\Omega)}{8\CM'}
    \label{eq:spectrum_arbgain_symm_hinf}
\end{eqnarray}
is amenable to taking the limit $h \to \infty$.

\subsubsection{Sanity checks}
As a sanity check, we calculate the spectrum in some limits. We start with $h\to 0$, when $\effbin \to 1$. In that case, Eq.~\eqref{eq:spectrum_arbgain_symm_h0} reduces to
\begin{eqnarray}
    \bar\spectSymb[\YLout](\omega) &= \frac{1}{2} + 4\CL \Tinf(\omega- \Omega) \\
    &+ [4\CL(1/2 + \bar n + \CL)] (1+u(\omega-\Omega)^2) \frac{\Tinf(\omega-\Omega)}{2\CM'}.
\end{eqnarray}
Next, we check the spectrum in the absence of the microwave cavity. In that case, we just optically detect the mechanical motion. We start by recalling
\begin{eqnarray*}
    \Tinf(\omega-\Omega) &= \Gamma' \kM |\chix(\omega - \Omega)|^2 = \Gamma' \kM \gM^2 |\chic(\omega-\Omega)|^2 |\chibld(\omega)|^2 \\
    &= \CM' \frac{1}{1+u(\omega-\Omega)^2} |\Gamma'\chibld(\omega)|^2, \\
\end{eqnarray*}
so that in the limit $\CM' \to 0$, we can write $\Tinf(\omega - \Omega) =0$ and
\begin{eqnarray*}
    (1+u(\omega-\Omega)^2) \frac{\Tinf(\omega-\Omega)}{\CM'} &= |\Gamma' \chib(\omega)|^2 = \frac{4}{1+(2(\omega-\Omega)/\Gamma')^2} \\
    &= 4 L_{\Gamma'}(\omega - \Omega),
\end{eqnarray*}
where $L_{\Gamma'}(\omega)$ is a Lorentzian function with FWHM of $\Gamma'$ and a peak value $L_{\Gamma'}(0) = 1$. The spectrum thus reads
\begin{eqnarray*}
     \bar\spectSymb[\YLout](\omega) &= \frac{1}{2} 
     + 2\CL(1/2 + \bar n + \CL) |\Gamma' \chib(\omega)|^2\\
     &= \frac{1}{2} + 8\CL(1/2 + \bar n + \CL) L_{\Gamma'}(\omega - \Omega),
\end{eqnarray*}
as expected for an optomechanical oscillator with thermal phonon occupancy $\bar n$, heated by $\CL$ additional back-action phonons.

Conversely, we check the limit $h\to\infty$, when $\effbopt \to 1$. In that case, Eq.~\eqref{eq:spectrum_arbgain_symm_hinf} reduces to
\begin{eqnarray*}
    \bar\spectSymb[\YLout](\omega) &= \frac{1}{2} - \frac{\Tinf(\omega-\Omega)}{4} 
    - (1+u(\omega-\Omega)^2) \frac{\Tinf(\omega-\Omega)}{8\CM'}.
\end{eqnarray*}
In the absence of the microwave cavity, this simplifies to 
\begin{eqnarray*}
    \bar\spectSymb[\YLout](\omega) &= \frac{1}{2} 
    - (1+u(\omega-\Omega)^2) \frac{\Tinf(\omega-\Omega)}{8\CM'} = \frac{1}{2}-\frac{1}{2}L_{\Gamma'}(\omega-\Omega),
\end{eqnarray*}
reflecting the complete cancellation of optical noise at the mechanical sideband by noise squashing.

\section{Bidirectional transfer by coherent feedback}
\label{sec:apx:bidirectional_transfer}
In this section, we show that bidirectional transfer of quantum states between the optical and microwave ports can be achieved if we replace measurement-based feedback by coherent optical feedback~\cite{ernzer_coherent-feedback_2023}. In this scheme, shown in Fig.~\ref{fig:coherent_feedback}, light entering the system interacts with the cavity twice to facilitate transfer from and onto both optical quadratures. We model this process in the following, again starting with the optomechanical subsystem while the electromechanical interaction is off ($\gM = 0$). For simplicity, we assume perfect coupling efficiencies $\effL = \effM = 1$.

Specifically, we use two cavity modes $a_1$ and $a_2$ with orthogonal polarisation, both coupled to the mechanical resonator $b$. Each optical mode  is pumped, on resonance, by an individual coherent field $\pumpinN{1}$ to a coherent amplitude $\pumpN{j}$. Importantly, the optical phase between $\pumpN1 = |\pumpinN1|$ and $\pumpN2 = -\rmi |\pumpinN2|$ is adjusted so that, for $a_1$, the amplitude quadrature $\XLN1$ is imprinted on $b$ through radiation pressure and, for $a_2$, phase quadrature $-\YLN2$ is. The linearised Hamiltonian that describes this two-part interaction reads
\begin{eqnarray}
    \Hom &= \hbar \gLN1 (a_1^\dagger + a_1)(b^\dagger + b) - \rmi \hbar \gLN2  (a_2^\dagger - a_2) (b^\dagger + b) \nonumber \\
    &= 2\hbar\gLN1 \XLN1 Q - 2\hbar \gLN2 \YLN2 Q, \label{eq:cfb-interaction-Ham}
\end{eqnarray}
where $\XLN{j}, \YLN{j}$ are the amplitude and phase quadratures, respectively, of the optical mode $a_j$, and $\gLN{j} = |\pumpinN{j}| g_{\text{L}j,0}$ are  pump-induced optomechanical coupling rates, which, in turn, depend on the vacuum optomechanical coupling rate $g_{\text{L}j,0}$ of each cavity mode.

After combining Hamiltonian~\eqref{eq:cfb-interaction-Ham} with the free mechanical evolution $H_b = \hbar \Omega b^\dagger b$, we derive the Heisenberg-Langevin equations of motions (EOMs) for the mechanical operators
\begin{eqnarray}
    \dot{Q}(t) &= \Omega P(t) - \frac{\Gamma}{2}Q(t) + \sqrt{\Gamma}\Qin(t), \\
    \dot{P}(t) &= -\Omega P(t) - \frac{\Gamma}{2}P(t) + \sqrt{\Gamma}\Pin(t) 
    - 2 \gLN1 \XLN1(t) + 2 \gLN2 \YLN2(t). \label{eq:cfb-dotP}
\end{eqnarray}
As before, the mechanical resonator is assumed to interact with its thermal environment under a rotating wave approximation.
After adiabatic elimination of the photon dynamics, the optical quadrature amplitudes are given by
\begin{eqnarray}
    \XLN1(t) &= 2 \XLinN1(t) / \sqrt{\kLN1}, \label{eq:cfb-xl1} \\
    \YLN1(t) &= 2 \YLinN1(t) / \sqrt{\kLN1} - 4(\gLN1 / \kLN1) Q(t), \\
    \XLN2(t) &= 2 \XLinN2(t) / \sqrt{\kLN2} - 4(\gLN2 / \kLN2) Q(t), \\
    \YLN2(t) &= 2 \YLinN2(t) / \sqrt{\kLN2}. \label{eq:cfb-yl2}
\end{eqnarray}
Likewise, from input-output theory, we retrieve the corresponding quadrature amplitudes for the output fields
\begin{eqnarray}
    \YLoutN1(t) &= -\YLinN1(t) + \sqrt{4\Gamma\CLN1} Q(t), \label{eq:cfb-YLout1} \\
    \XLoutN2(t) &= -\XLinN2(t) + \sqrt{4\Gamma\CLN2} Q(t).
\end{eqnarray}
We plug the quadratures Eq.~\eqref{eq:cfb-xl1} and~\eqref{eq:cfb-yl2} into Eq.~\eqref{eq:cfb-dotP} to obtain the optically driven momentum EOM
\begin{eqnarray}
    \dot{P}(t) &= -\Omega P(t) - \frac{\Gamma}{2}P(t) + \sqrt{\Gamma}\Pin(t) - \sqrt{4\Gamma\CLN1} \XLinN1(t) + \sqrt{4\Gamma\CLN2} \YLinN2(t) \label{eq:cfb-dotP-2}
\end{eqnarray}
with cooperativities $\CLN{j} = 4 \gLN{j}^2 /(\Gamma \kLN{j})$.

Until now, the optical modes $a_1, a_2$ and their inputs were considered independently. In our set-up, however, the output of $a_1$ is related to the input of $a_2$ by the optical feedback loop. As shown in Fig.~\ref{fig:coherent_feedback}, light leaving $a_1$ is combined with an auxiliary pump on a beam-splitter with high reflectivity $\effd \sim 1$, delayed by time $\tau$, rotated in polarisation, and finally coupled into $a_2$. The coherent amplitude $\pumpinN{\Delta}$ of the auxiliary pump beam serves to rotate the carrier to
\begin{eqnarray}
    \pumpinN{2} &= \sqrt{\effd} \pumpoutN{1} + \sqrt{1-\effd} \pumpinN{\Delta} = -\sqrt{\effd} \pumpinN{1} + \sqrt{1-\effd} \pumpinN{\Delta}
\end{eqnarray}
on the second pass. Accordingly, we set $\pumpinN{\Delta} = (\sqrt{\effd}|\pumpinN{1}| - \rmi |\pumpinN{2}|) / \sqrt{1-\effd}$ to ensure that the coherent driving fields $\pumpinN{1}$ and $\pumpinN{2}$ are $-\pi/2$ out of phase. However, combining those fields on a beam-splitter comes at the cost of mixing in an optical vacuum field $\ainN{\text{v}}$ into the optical signal 
\begin{eqnarray}
    \aoutN{1} \to \sqrt{\effd} \aoutN{1} + \sqrt{1-\effd} \ainN{\text{v}}
\end{eqnarray}
propagating in the feedback loop. In fact, any losses in the optical feedback system can be collected in the `feedback efficiency' $\effd$, which plays the same role as detection efficiency in the measurement-based feedback loop. Nevertheless, in principle, zero feedback loss $\effd \to 0$ can be attained by exchanging the beam-splitter with a dichroic mirror that separates pump and sidebands. Finally, after accounting for the delay, we obtain the relation 
\begin{eqnarray}
    \ainN{2}(t) = \sqrt{\effd} \aoutN{1}(t-\tau) + \sqrt{1-\effd} \ainN{\text{v}}(t-\tau) \label{eq:cfb-feedback-relation}
\end{eqnarray}
between the cavity output on the first pass, and its input on the second. In this way, the coherent feedback loop cascades the optomechanical system with itself.

Next, we plug Eqs.~\eqref{eq:cfb-feedback-relation} and~\eqref{eq:cfb-YLout1} into the momentum EOM~\eqref{eq:cfb-dotP-2} to obtain
\begin{eqnarray}
    \dot{P}(t) &= -\Omega P(t) - \frac{\Gamma}{2}P(t) + \sqrt{\Gamma}\Pin(t) - \sqrt{4\Gamma\CLN1} \XLinN1(t) 
    - \sqrt{4 \effd \Gamma\CLN2} \YLinN1(t-\tau) \nonumber \\
    &+ 4\Gamma\sqrt{\effd\CLN1 \CLN2} Q(t-\tau) + \sqrt{(1-\effd)4\Gamma\CLN2} \YLv(t-\tau), \nonumber
\end{eqnarray}
where $\YLv$ is the phase quadrature of the vacuum field $\ainN{\text{v}}$.
For ease of notation, we define the cooperativity ratio $h = \sqrt{\effd \CLN2 / \CLN1}$, so that we can write
\begin{eqnarray}
    \dot{P}(t) &= -\Omega P(t) - \frac{\Gamma}{2}P(t) + \sqrt{\Gamma}\Pin(t) + 4h\Gamma \CLN1 Q(t-\tau) \\
    &- \sqrt{4\Gamma\CLN1} \big[ \XLinN1(t) + h \YLinN1(t-\tau) - h \varepsilon \YLv(t-\tau) \big],
\end{eqnarray}
which, as before, features the ratio $\varepsilon = \sqrt{(1-\effd)/\effd}$. We continue by combining $\dot{b} = (\dot{Q} + \rmi \dot{P})/\sqrt{2}$ into an EOM for the mechanical annihilation operator
\begin{eqnarray}
    \dot{b} &= -\rmi \Omega b - \frac{\Gamma}{2}b + \sqrt{\Gamma} \bin + 4\rmi h \Gamma \CLN1 Q(t-\tau)/\sqrt{2} \nonumber \\
    &- \rmi \sqrt{2\Gamma\CLN1} \big[ \XLinN1(t) + h \YLinN1(t-\tau) - h \varepsilon \YLv(t-\tau) \big]. \label{eq:cfb-dotb-1}
\end{eqnarray}

Notably, this equation is formally the same as Eq.~\eqref{eq:dotb-1} upon substituting $G \to 8h\CL$ and $\CL \to \CLN1$, conveniently permitting us to follow the same analysis as in~\ref{sec:apx:optomechanical_subsystem_arbgain}. In particular, we arrive at the same mechanical frequency response 
\begin{eqnarray}
    b(\omega) &= \sqrt{\Gamma'} \chib(\omega) \bigg[ 
        &\sqrt{\effbin} \bin(\omega) + \sqrt{\effbopt} \left(- \rmi \ainTN1(\omega) - \sqrt{h/2} \varepsilon \YLv(\omega) \right)
    \bigg] \label{eq:cfb_mech_freq_1}
\end{eqnarray}
as given in Eq.~\eqref{eq:sup:b-freq-2}, again for broadened linewidth $\Gamma'/\Gamma = 1 + 4h\CLN1$, effective optical and mechanical coupling efficiencies $\effbopt = 4h\CLN1/(1+4h\CLN1)$, $\effbin = 1/(1+4h\CLN1)$ and transferred optical mode 
\begin{eqnarray}
    \ainTN1(\omega) = \left( \frac{1}{\sqrt{h}} \XLinN1(\omega) + \rmi \sqrt{h} \YLinN1(\omega) \right) / \sqrt{2}
\end{eqnarray}
similar to Eq.~\eqref{eq:sup:ainT-h}. 
Evidently, the optomechanical subsystem with all-optical coherent feedback responds exactly as before with measurement-based feedback. By extension, after turning on the electromechanical interaction, the optical-to-microwave transfer process is again described by the expression
\begin{eqnarray}
    \cout(\omega) = &\tac(\omega)\ainT(\omega+\Omega) + \tbc(\omega)\bin(\omega+\Omega) + \tcc(\omega) \cin(\omega) \label{eq:cfb_cout_v1} \\
    &+ \rmi \tac(\omega) \varepsilon \sqrt{h/2} \YLv(\omega) \nonumber
\end{eqnarray}
for the microwave output field.
Notably, this features the same transfer gain $\tac(\omega)$ and is characterised by the same noise performance, concluding our analysis of state transfer in the forward direction.

\subsection{Microwave-to-optical transfer}
As coherent feedback keeps quantum correlations alive in the optical signal traversing the loop, coherent transfer is now also possible in the reverse direction, from microwave input $\cin$ to optical output $\aoutN2$. We model this process in the following section.

First, we recall expressions for the two cavity amplitudes
\begin{eqnarray}
    a_1(t) &= \frac{2\ainN1(t)}{\sqrt{\kLN1}} - 4\rmi \left(\frac{\gLN1}{\kLN1}\right) \frac{Q(t)}{\sqrt{2}}, \\
    a_2(t) &= \frac{2\ainN2(t)}{\sqrt{\kLN2}} - 4 \left(\frac{\gLN2}{\kLN2}\right) \frac{Q(t)}{\sqrt{2}}.
\end{eqnarray}
We trace the propagation of light entering the transducer through $\ainN1$, interacting with $a_1$, traversing the feedback loop, interacting with $a_2$, and finally leaving the circuit through $\aoutN2$. Anchouring the time variable $t$ to light leaving the system and accounting for the feedback delay $\tau$, we obtain from input-output theory the expressions
\begin{eqnarray}
    \aoutN1(t-\tau) &= -\ainN1(t-\tau) + \rmi \sqrt{4\Gamma\CLN1} Q(t-\tau)/\sqrt{2}, \\
    \ainN2(t) &= \sqrt{\effd} \aoutN1(t-\tau) + \sqrt{1-\effd} \ainN{\text{v}}(t-\tau), \\
    \aoutN2(t) &= -\ainN2(t) + \sqrt{4\Gamma\CLN2} Q(t)/\sqrt{2} \\
    &=  \sqrt{\effd}\ainN1(t-\tau) - \sqrt{1-\effd} \ainN{\text{v}}(t-\tau) \\
    &- \rmi \sqrt{\effd} \sqrt{4\Gamma\CLN1} Q(t-\tau)/\sqrt{2} + \sqrt{4\Gamma\CLN2} Q(t)/\sqrt{2}. \nonumber
\end{eqnarray}
Assuming that the mechanical resonator still has a high quality factor $\mathcal{Q}' = \Omega/\Gamma' \gg 1$ after feedback, we can approximate $Q(t-\tau) \approx -P(t)$. This shows that the double-pass optical circuit, indeed, facilitates transfer of both mechanical quadratures onto the optical output. After applying this approximation---effectively a rotating wave approximation---we obtain
\begin{eqnarray}
    \aoutN2(t) 
    &=  \sqrt{\effd}\ainN1(t-\tau) - \sqrt{1-\effd} \ainN{\text{v}}(t-\tau) \\
    &+ \sqrt{4h\Gamma\CLN1} \left[ \sqrt{h/\effd}  Q(t) + \frac{\rmi}{\sqrt{h/\effd}} P(t)\right]/\sqrt{2} \nonumber \\
    &= \sqrt{\effd}\ainN1(t-\tau) - \sqrt{1-\effd} \ainN{\text{v}}(t-\tau) + \sqrt{4h\Gamma\CLN1} b^S(t),
\end{eqnarray}
where we have reintroduced the cooperatitivity ratio $h = \sqrt{\effd \CLN2 / \CLN1}$.
Moreover, we recognise that an imbalance in the strength of the two optomechanical transfer steps, again, leads to squeezing---in this case as the mechanical state is imprinted on the optical output. We thus define the mechanical Bogoliubov mode
\begin{eqnarray}
    b^S(t) = \left[ \sqrt{h/\effd}  Q(t) + \frac{\rmi}{\sqrt{h/\effd}} P(t)\right]/\sqrt{2} = \frac{1+s}{2\sqrt{s}} b(t) - \frac{1-s}{2\sqrt{s}} b^\dagger(t),
\end{eqnarray}
where $s = h/\effd = \sqrt{\CLN2/(\effd\CLN1)}$. 
The inverse of this transformation is given by
\begin{eqnarray}
    b(t) = \frac{1+s}{2\sqrt{s}} b^S(t) + \frac{1-s}{2\sqrt{s}} b^S(t)^\dagger.
\end{eqnarray}
This informs our choice to study the optical output Bogoliubov mode
\begin{eqnarray}
    \aoutRN2(t) &= \frac{1+s}{2\sqrt{s}} \aoutN2(t) + \frac{1-s}{2\sqrt{s}} \aoutN2(t)^\dagger \nonumber \\
    &= \sqrt{\effd} \ainRN{1}(t-\tau) - \sqrt{1-\effd} \ainRN{\text{v}}(t-\tau) \label{eq:cfb_aoutR_1} + \sqrt{4h\Gamma\CLN1} b(t), \nonumber
\end{eqnarray}
to which the mechanical state is transferred without squeezing. 
For convenience, we have also defined the similarly transformed Bogoliubov modes
\begin{eqnarray}
    \ainRN{j}(t) &= \frac{1+s}{2\sqrt{s}} \ainN{j}(t) + \frac{1-s}{2\sqrt{s}} \ainN{j}(t)^\dagger = \left(\frac{1}{\sqrt{s}}\XLinN{j}(\omega) + \rmi \sqrt{s} \YLinN{j}(\omega)\right)/\sqrt{2}.
\end{eqnarray}
for the optical input and added vacuum fields.

Notably, for $\effd < 1$, the two cooperatitivity ratios $h \neq s$ that determine the level of squeezing applied during forward and reverse transfer, respectively, are distinct. This can be understood intuitively by inspecting the optical circuit: in the forward direction (optical to microwave), the feedback loss $\effd$ affects the input to the \emph{second} transfer step mediated by $a_2$, whereas in the reverse direction (microwave to optical), it is the output of $a_1$, mediating the \emph{first} transfer step, that is subject to loss.

Next, we take the Fourier transform of Eq.~\eqref{eq:cfb_aoutR_1} to obtain the frequency response
\begin{eqnarray}
    \aoutRN2(\omega) &= e^{\rmi\omega\tau} \sqrt{\effd} \ainRN{1}(\omega) - e^{\rmi\omega\tau} \sqrt{1-\effd} \ainRN{\text{v}}(\omega) + \sqrt{4h\Gamma\CLN1} b(\omega), \label{eq:cfb_aoutR_freq_1}
\end{eqnarray}
of the output mode. Plugging in the mechanical response with electromechanical interaction on Eq.~\eqref{eq:sup:b-cnt-1}, reproduced here as
\begin{eqnarray*}
b(\omega) &= \sqrt{\Gamma'} \chibld(\omega) \bigg[ 
        \sqrt{\effbin} \bin(\omega) + \sqrt{\effbopt} \left(- \rmi \ainTN1(\omega) - \sqrt{h/2} \varepsilon \YLv(\omega) \right)
    \bigg] \\
    &+ \rmi \chix(\omega-\Omega) \sqrt{\kM} \cin(\omega-\Omega)
\end{eqnarray*}
then results in
\begin{eqnarray*}
    \aoutRN2(\omega) &= e^{\rmi\omega\tau} \sqrt{\effd} \ainRN{1}(\omega) - e^{\rmi\omega\tau} \sqrt{1-\effd} \ainRN{\text{v}}(\omega) \nonumber \\
    &+ \sqrt{\effbopt} \Gamma' \chibld(\omega) \bigg[ 
        \sqrt{\effbin} \bin(\omega) + \sqrt{\effbopt} \left(- \rmi \ainTN1(\omega) - \sqrt{h/2} \varepsilon \YLv(\omega) \right)
    \bigg] \\
    &+ \rmi \sqrt{\effbopt} \sqrt{\Gamma'\kM} \chix(\omega-\Omega) \cin(\omega-\Omega), \nonumber
\end{eqnarray*}
after substituting $4h\Gamma\CLN1 = \effbopt\Gamma'$. We simplify further using the input-to-light transfer gains defined in Eq.~\eqref{eq:taa_def},
\begin{eqnarray*}
    \aoutRN2(\omega) &= e^{\rmi\omega\tau} \sqrt{\effd} \ainRN{1}(\omega) - e^{\rmi\omega\tau} \sqrt{1-\effd} \ainRN{\text{v}}(\omega) \nonumber \\
    &- \rmi \effbopt \Gamma' \chibld(\omega) \ainTN1(\omega) - \effbopt \Gamma' \chibld(\omega) \sqrt{h/2} \varepsilon \YLv(\omega) \\
    &+ \tba(\omega)\bin(\omega) 
    -\rmi \tca(\omega) \cin(\omega-\Omega). \nonumber
\end{eqnarray*}

In the high-quality limit $\mathcal{Q}' \gg 1$ that we work in, the susceptibilities $\chibld(\omega)$, $\tba(\omega)$, $\tca(\omega)$ are sharply peaked around the upper mechanical sideband $\omega \approx -\Omega$. 
For frequencies outside that sideband, including at the lower sideband $\omega \approx -\Omega$, we thus find that the Bogoliubov-transformed output field
\begin{eqnarray*}
    \aoutRN2(\omega \not\approx \Omega) &= e^{\rmi\omega\tau} \sqrt{\effd} \ainRN{1}(\omega) - e^{\rmi\omega\tau} \sqrt{1-\effd} \ainRN{\text{v}}(\omega)
\end{eqnarray*}
merely contains the reflected input field, after mixing with the added vacuum field and propagation through the feedback loop. 

Next, we apply the approximate identity $1 = -\rmi e^{\rmi \Omega \tau} \approx -\rmi e^{\rmi \omega \tau}$ to those sharply peaked susceptibility terms to obtain
\begin{eqnarray*}
    \aoutRN2(\omega) 
%
%
    &= e^{\rmi\omega\tau} \Bigg(\sqrt{\effd} \ainRN{1}(\omega) -  \sqrt{1-\effd} \ainRN{\text{v}}(\omega) \nonumber \\
    &\hspace{4em}- \effbopt \Gamma' \chibld(\omega) \ainTN1(\omega) + \rmi \effbopt \Gamma' \chibld(\omega) \sqrt{h/2} \varepsilon \YLv(\omega) \Bigg) \\
    &+ \tba(\omega)\bin(\omega) 
    - \rmi \tca(\omega) \cin(\omega-\Omega). \nonumber \\
\end{eqnarray*}
It is now useful to express $\ainTN1(\omega)$ in terms of $\ainRN1(\omega)$,
\begin{eqnarray}
    \ainTN1(\omega) &= \frac{s+h}{2\sqrt{hs}} \ainRN1(\omega) + \frac{s-h}{2\sqrt{hs}} \ainRN1(-\omega)^\dagger \\
    &= \frac{1+\effd}{2\sqrt{\effd}} \ainRN1(\omega) + \frac{1-\effd}{2\sqrt{\effd}} \ainRN1(-\omega)^\dagger, \nonumber
\end{eqnarray}
and the vacuum field phase quadrature as
\begin{eqnarray*}
    \YLv(\omega) = \frac{\rmi}{\sqrt{2s}} \left( \ainRN{\text{v}}(-\omega)^\dagger - \ainRN{\text{v}}(\omega) \right).
\end{eqnarray*}
We apply these along with the substitutions $h = s\effd$ and $\varepsilon = \sqrt{(1-\effd)/\effd}$ to get
\begin{eqnarray*}
    \aoutRN2(\omega) &= e^{\rmi\omega\tau} \Bigg( \sqrt{\effd} \ainRN{1}(\omega) - \effbopt \Gamma' \chibld(\omega)\left[ \frac{1+\effd}{2\sqrt{\effd}} \ainRN1(\omega) + \frac{1-\effd}{2\sqrt{\effd}} \ainRN1(-\omega)^\dagger \right] \\
    &- \sqrt{1-\effd} \ainRN{\text{v}}(\omega)
    + \sqrt{1-\effd} \effbopt \Gamma' \chibld(\omega) \frac{\ainRN{\text{v}}(\omega) - \ainRN{\text{v}}(-\omega)^\dagger}{2} \Bigg) \\
    &+ \tba(\omega)\bin(\omega) 
    - \rmi \tca(\omega) \cin(\omega-\Omega) .
\end{eqnarray*}

Using the light-to-light transfer gain $\taa(\omega) = 1 - \effbopt \Gamma' \chibld(\omega)$ defined in Eq.~\eqref{eq:taa_def}, we obtain the final expression
\begin{eqnarray}
    \aoutRN2(\omega) &= e^{\rmi\omega\tau} \Bigg( 
    \sqrt{\effd} \bigg[ \frac{ \effd-1+(1+\effd)\taa(\omega)}{2\effd} \ainRN1(\omega) \\
    &\hspace{4em} + \frac{(1-\effd)(\taa(\omega)-1)}{2\effd} \ainRN1(-\omega)^\dagger \bigg] \label{eq:cfb_aoutR_v1} \nonumber \\
    &\hspace{4em} - \sqrt{1-\effd} \left[ \frac{\taa(\omega)+1}{2} \ainRN{\text{v}}(\omega)
    - \frac{\taa(\omega)-1}{2}\ainRN{\text{v}}(-\omega)^\dagger \right] \Bigg) \nonumber \\
    &+ \tba(\omega)\bin(\omega)
    - \rmi \tca(\omega) \cin(\omega-\Omega) \nonumber
\end{eqnarray}
that describes the full microwave-to-optical transfer process.
As a sanity check, we evaluate the commutator of the mode $\aoutRN2(\omega)$ as given in Eq.~\eqref{eq:cfb_aoutR_v1}. A tedious but straightforward calculation shows that
\begin{eqnarray*}
    \left[\aoutRN2(\omega), \aoutRN2(\omega')^\dagger\right] = \Big(|\taa(\omega)|^2 + |\tba(\omega)|^2 + |\tca(\omega)|^2\Big)\,\delta(\omega-\omega') = \delta(\omega-\omega'),
\end{eqnarray*}
indicating proper normalisation of the output mode for all $\effd$.
Finally, in terms of optical input quadratures, the output field can be expressed as
\begin{eqnarray}
    \aoutRN2(\omega) = 
    &e^{\rmi\omega\tau} \Bigg( \sqrt{\effd} \left[\frac{\effd-1+\taa(\omega)}{\effd \sqrt{s}} \XLinN1(\omega) + \rmi \sqrt{s} \taa(\omega) \YLinN1(\omega) \right] / \sqrt{2} \\
    &\hspace{4em}- \sqrt{1-\effd} \left[\frac{1}{\sqrt{s}} \XLv(\omega) + \rmi \sqrt{s}\taa(\omega) \YLv(\omega) \right] / \sqrt{2} \Bigg) \nonumber \\
    &+ \tba(\omega)\bin(\omega) - \rmi\tca(\omega) \cin(\omega-\Omega) \nonumber
\end{eqnarray}

\subsection{Performance of microwave-to-optical transfer}
Compared to transfer in the forward direction, i.e. from optical to microwave fields as described by Eq.~\eqref{eq:cfb_cout_v1}, the equivalent expression Eq.~\eqref{eq:cfb_aoutR_v1} that applies to transfer in the reverse, microwave-to-optical direction is considerably more intricate. In particular, every optical quadrature features with a different prefactor, reflecting their different interactions as light traces the feedback loop. In this section, we evaluate the performance of the microwave-to-optical transfer process in various limits.

First, we assume no losses in the feedback circuit, so that $\effd=1$. In that case, the cooperatitivity ratios $h = \sqrt{\effd \CLN2 / \CLN1}$ and $s=\sqrt{\CLN2 / \effd \CLN1}$ have the same value, indicating that the transferred state experiences the same level of squeezing in both directions. Indeed, we find that the Bogoliubov-transformed modes $\ainTN{j}$ and $\ainRN{j}$ are identical. In this limit, we find that the microwave input $\cin(\omega)$ is imprinted on the upper sideband of the Bogoliubov-transformed optical output field
\begin{eqnarray}
    \aoutRN2(\omega \approx \Omega) = \rmi \taa(\omega) \ainRN1(\omega) + \tba(\omega) \bin(\omega) - \rmi \tca(\omega) \cin(\omega-\Omega)
\end{eqnarray}
with gain $\tca(\omega)$. Moreover, for $h = s = 1$, we find $\aoutRN{2} = \aoutN{2}$, so that the microwave input is transferred without squeezing.

Next, we allow for arbitrary feedback efficiency $\effd$ but assume cooperativity ratio $s=\sqrt{\CLN2 / \effd \CLN1}=1$, so that the microwave input is mapped onto the optical output without squeezing. Note that for $\effd < 1$, this implies that simultaneous optical-to-microwave transfer necessarily involves squeezing. For $s=1$, we find $\aoutRN{2} = \aoutN{2}$ and $\ainRN{1} = \ainN{1}$, so that we can write
\begin{eqnarray}
    \aoutN2(\omega) = 
    &e^{\rmi\omega\tau} \sqrt{\effd} \left[\frac{\effd-1+\taa(\omega)}{\effd} \XLinN1(\omega) + \rmi \taa(\omega) \YLinN1(\omega) \right] / \sqrt{2}\\
    &- e^{\rmi\omega\tau} \sqrt{1-\effd} \frac{\XLv(\omega) + \rmi \taa(\omega) \YLv(\omega)}{\sqrt{2}} \\
    &+ \tba(\omega)\bin(\omega) - \rmi\tca(\omega) \cin(\omega-\Omega)
\end{eqnarray}
In the high-cooperativity limit $\CLN1, \CLN2 \to \infty$ and for sufficient electromechanical cooperativity, we find $\taa(\sigdet) \to 0, \tba(\sigdet) \to 0$ for signals at the optimal detuning $\sigdet$ (see Fig.~\ref{fig:transfer-coeff-freq}), while $\tca(\sigdet) \to -1$. In that case, the optical output field 
\begin{eqnarray}
    \aoutN2(\omega \approx \Omega) = \rmi \cin(\omega-\Omega) - \rmi \frac{1-\effd}{\sqrt{2\effd}} \XLinN{1}(\omega) - \rmi \frac{\sqrt{1-\effd}}{\sqrt{2}} \XLv(\omega).
\end{eqnarray}
features noise contributions from the optical input and vacuum fields, arising from the feedback inefficiency $\effd < 1$.

Next, we calculate the noise power spectral density (PSD) of the optical output for $s=1$. To do so, we first recall that input fields resonant with the microwave cavity are imprinted onto the upper optomechanical sideband of the output field. Accordingly, we define the frequency-shifted (sideband) operator of the optical output $\aoutN{2}^{\SBpSymb}(\omega) = \aoutN2(\omega + \Omega)$, along with its quadratures $\XLoutN{2}^{\SBpSymb}(\omega) = [\aoutN{2}^{\SBpSymb}(-\omega)^\dagger + \aoutN{2}^{\SBpSymb}(\omega)]/\sqrt{2}$ and $\YLoutN{2}^{\SBpSymb}(\omega) = \rmi[\aoutN{2}^{\SBpSymb}(-\omega)^\dagger - \aoutN{2}^{\SBpSymb}(\omega)]/\sqrt{2}$. We employ the computer algebra system \texttt{Mathematica} with package \texttt{NCAlgebra} to evaluate the individual contributions of the input fields $\ainN1$, $\ainN{\text{v}}$ and $\bin$ to the symmetrised output quadrature PSDs $\bSXoutCFB(\omega)$ and $\bSYoutCFB(\omega)$, and find
\begin{eqnarray}
    \hspace{-4em}\bSXoutCFB[|\ainN1](\omega) &= 
        \frac{1 + \effd^2}{4 \effd} |\taa(\omega+\Omega)|^2 
        + \frac{1-\effd}{4\effd} \Big(1 - 2\effd + (\effd - 2)\Re[\taa(\omega+\Omega)]\Big), \\
    \hspace{-4em}\bSYoutCFB[|\ainN1](\omega) &= 
        \frac{1 + \effd^2}{4 \effd} |\taa(\omega+\Omega)|^2 
        + \frac{1-\effd}{4\effd} \Big(1 - (\effd + 2)\Re[\taa(\omega+\Omega)]\Big), \\
    \hspace{-4em}\bSXoutCFB[|\ainN{\text{v}}](\omega) &= 
        \frac{1-\effd}{4}\Big( 2 + |\taa(\omega+\Omega)|^2 + \Re[\taa(\omega+\Omega)] \Big), \\ 
    \hspace{-4em}\bSYoutCFB[|\ainN{\text{v}}](\omega) &= 
        \frac{1-\effd}{4}\Big( |\taa(\omega+\Omega)|^2 - \Re[\taa(\omega+\Omega)] \Big), \\
    \hspace{-4em}\bSXoutCFB[|\bin](\omega) &= \bSYoutCFB[|\bin](\omega) = \left( \nth + \frac{1}{2} \right) |\tba(\omega+\Omega)|^2.
\end{eqnarray}
In these calculations, we have applied the high-quality approximation $\taa(\omega-\Omega) \approx 1$. Interestingly, while the individual noise contributions of the optical modes $\ainN1$, $\ainN{\text{v}}$ to both output quadratures are different, the total optical noise, given by their sum,
\begin{eqnarray}
    \bSXoutCFB[|\text{opt.}](\omega) &= \bSYoutCFB[|\text{opt.}](\omega) \\
    &= \frac{1-\effd}{4\effd} + \frac{1+\effd}{4\effd} |\taa(\omega+\Omega)|^2 - \frac{1-\effd}{2\effd} \Re[\taa(\omega+\Omega)] \nonumber
\end{eqnarray}
is the same for both quadratures, as is the mechanical contribution. The total noise added by the transfer protocol is thus quadrature-symmetric. In the limit of high optomechanical cooperatitivity, sufficient electromechanical cooperatitivity and optimal detuning, the total added noise reduces to
\begin{eqnarray}
    \bSXoutCFB[|\text{noise}](\omega) = \bSYoutCFB[|\text{noise}](\omega) = \frac{1-\effd}{4\effd},
\end{eqnarray}
exactly the same number one finds for ideal, no-squeezing transfer in the opposite direction. 

Finally, we point out a general symmetry in noise performance for both directions. The total added noise in the microwave-to-optical direction can be expressed as
\begin{eqnarray}
    &\bSXoutCFB[|\text{noise}](\omega) = \bSYoutCFB[|\text{noise}](\omega) \nonumber \\
    &\hspace{2em} =
    \underbrace{\frac{1}{2} |\taa(\omega+\Omega)|^2}_{\text{optical}} 
    + \underbrace{\left(\nth + \frac{1}{2}\right)|\tba(\omega+\Omega)|^2}_{\text{mechanical}} 
    + \underbrace{\frac{1-\effd}{4\effd} \effbopt\frac{|\tba(\omega + \Omega)|^2}{\effbin}}_{\text{feedback}}. \label{eq:noise_mw-to-opt}
\end{eqnarray}  
The added noise in the opposite, optical-to-microwave direction is given by exactly the same expression after exchanging the indices $a \leftrightarrow c$,
\begin{eqnarray}
    &\bSXMout[|\text{noise}](\omega) = \bSYMout[|\text{noise}](\omega) \nonumber \\
    &\hspace{2em} =
    \underbrace{\frac{1}{2} |\tcc(\omega)|^2}_{\text{optical}} 
    + \underbrace{\left(\nth + \frac{1}{2}\right)|\tbc(\omega)|^2}_{\text{mechanical}} 
    + \underbrace{\frac{1-\effd}{4\effd} \frac{\effbopt}{\effbin} |\tbc(\omega)|^2}_{\text{feedback}}, \label{eq:noise_opt-to-mw}
\end{eqnarray}  
and $\omega + \Omega \leftrightarrow \omega$ to account for different frequency offsets. From Eqs.~\eqref{eq:noise_mw-to-opt} and~\eqref{eq:noise_opt-to-mw} we recognize that, in both cases, the feedback loss $\effd$ imparts excess fluctuations onto the mechanical resonator, which then couple out into the output mode.

\section{Correlators for thermal states}
For convenience, here we list the correlation relations for thermal inputs. For a general input field operator $\ain$ in a thermal state with bath occupancy $\bar{n}$, these are given by~\cite{Warwick_Textbook}
\begin{eqnarray}
    \langle \ain(\omega)^\dagger \ain(\omega') \rangle &= \bar{n}\delta(\omega - \omega'), \\
    \langle \ain(\omega) \ain(\omega')^\dagger \rangle &= (\bar{n} + 1) \delta(\omega - \omega'), \\
    \langle \ain(\omega) \ain(\omega') \rangle &= 0, \\
    \langle \ain(\omega)^\dagger \ain(\omega')^\dagger \rangle &= 0.
\end{eqnarray}
We define the corresponding input quadratures $\XinG(\omega) = [\ain(-\omega)^\dagger + \ain(\omega)]/\sqrt{2}$ and $\YinG(\omega) = \rmi [\ain(-\omega)^\dagger - \ain(\omega)] / \sqrt{2}$. These are Hermitian operators, so that $\mathcal{O}(\omega) = \mathcal{O}(-\omega)^\dagger$. Their correlators read
\begin{eqnarray}
    \langle \XinG(\omega)^\dagger \XinG(\omega') \rangle = \langle \YinG(\omega)^\dagger \YinG(\omega') \rangle &= (\bar{n} + 1/2) \delta(\omega - \omega'), \\
    \langle \XinG(\omega)^\dagger \YinG(\omega') \rangle = - \langle \YinG(\omega)^\dagger \XinG(\omega') \rangle &= \rmi \delta(\omega - \omega') / 2,
\end{eqnarray}
and
\begin{eqnarray}
    \langle \XinG(\omega) \XinG(\omega') \rangle = \langle \YinG(\omega) \YinG(\omega') \rangle &= (\bar{n} + 1/2) \delta(\omega + \omega') \\
    \langle \XinG(\omega) \YinG(\omega') \rangle = -\langle \YinG(\omega) \XinG(\omega') \rangle &= \rmi \delta(\omega + \omega')/2.
\end{eqnarray}

\section{Computing variances for narrowband input signals}
\label{sec:apx:variance_narrowband_sigs}
Here, we justify Eq.~\eqref{eq:V_PSD_relation} in the main text. We show that the variance of an observable in a narrowband mode is given by sampling the corresponding power spectral density at the mode's centre frequency.

In the main text, we quantify the transducer's performance using the noise variances added to the output microwave field quadratures. 
In practice, matched filtering is performed on the transducer output to extract the signal. Furthermore, to enable high-fidelity state transfer, the input pulse should have a narrow frequency distribution well within the bandwidth of the transfer process. Therefore, in this section we are interested in computing the variance of an observable \(A\) within a modeshape \(\xi\) that has vanishingly narrow spectral width. 

The value of the observable within the modeshape is given by the convolution
\begin{eqnarray}
    \tilde{A}(t) &= (\xi \circledast A)(t),
\end{eqnarray}
where we consider a rectangular input pulse of duration \(T\) in the time domain
\begin{eqnarray}
    \xi(t) = \frac{\rect(t/T)}{\sqrt{T}} e^{\mathrm{i}\Omega t}
\end{eqnarray}
to represent a spectral mode with centre-frequency \(\Omega\) and finite frequency resolution defined by \(1/T\). Here, the exponential selects the centre frequency of interest and the \(\rect\) function is unity for \(-T/2 \leq t \leq T/2\) and zero elsewhere. The modeshape is appropriately normalised to ensure \(\int_{-\infty}^\infty \rm{d}t \, |\xi(t)|^2 = 1\). Ultimately, we wish to take the limit of an infinitely long input pulse \(T\rightarrow \infty\) to model a single-frequency mode.

In the frequency domain, we have from the convolution theorem \(\tilde A(\omega) = \xi(\omega) A(\omega)\), where
\begin{eqnarray}
    \xi(\omega) = \sqrt{\frac{T}{2\pi}} \sinc\bigg[ \frac{(\omega-\Omega) T }{2\pi } \bigg] \label{eq:u(w)_defn}
\end{eqnarray}
and \(\sinc(x) = \sin(\pi x)/(\pi x)\). The additional factor of \(1/\sqrt{2\pi}\) preserves the normalisation \(\int_{-\infty}^\infty \rm{d}\omega \, |\xi(\omega)|^2 = 1\). We now consider the limit \(T\rightarrow \infty\) of \eqref{eq:u(w)_defn}. Noting 
\begin{eqnarray}
    \delta(\omega) = \lim_{\varepsilon\rightarrow0} \frac{\sinc(\omega/\varepsilon)}{\varepsilon},
\end{eqnarray}
we may re-rewrite the long time limit of Eq.~\eqref{eq:u(w)_defn} in terms of the delta function
\begin{eqnarray}
    \xi(\omega) = \lim_{T\rightarrow\infty} \sqrt{\frac{2\pi}{T}} \,\delta(\omega-\Omega) \, . \label{eq:u(w)_limit}
\end{eqnarray}
For a stationary system, the variance of a  hermitian variable \(\tilde A\) is given by \cite{Warwick_Paper} 
\begin{eqnarray}
    V[\tilde A] &= \frac{1}{2\pi} \int\int_{-\infty}^\infty \rm{d} \omega \, \rm{d} \omega' \, \langle \tilde{A}(\omega) \tilde{A}(\omega')\rangle. \label{eq:def_V[X]_PSD}
\end{eqnarray}
Substituting our filtered observable \(\tilde A\) in terms of the limiting spectral modeshape \(\xi\) in Eq.~\eqref{eq:u(w)_limit} and \(A\) yields
\begin{eqnarray}
    V[\tilde{A}] &= \lim_{T \rightarrow \infty} \frac{1}{T} \int\int_{-\infty}^\infty \rm{d} \omega \, \rm{d} \omega' \, \delta(\omega-\Omega) \delta(\omega'-\Omega) \langle A(\omega)A(\omega')\rangle \\
    &= \lim_{T \rightarrow \infty} \frac{1}{T} \langle A(\Omega) A(\Omega)\rangle \\
    &= S[A](\Omega),
\end{eqnarray}
where \(S[A](\omega)\) is the power spectral density of the observable \(A\). We have shown that, in the limit of a single frequency input pulse, the variance of an observable \(A\) in the matched output is found by evaluating the corresponding power spectral density at the pulse frequency. Note that we do not distinguish the variable in the matched output mode \(\tilde A \leftrightarrow A\) in the main text.  Indeed, \(V[\tilde{A}] \approx S[A](\Omega)\) remains a valid approximation in the finite-width case provided that the input pulse has a spectral width well within the transducer's bandwidth.


\section{State-of-the-art electro-optomechanical components}
\label{sec:apx:state_ot_art_params}
Here, we elaborate in more detail on our choice of system parameters in Sections \ref{sec:state_of_the_art_params} and \ref{sec:quantum_transfer_witness}. 

We consider a \(\Omega/2\pi = 1\) MHz mechanical oscillator intermediary with thermal occupancy \(\nth = 10^3\) corresponding to a \(T\sim 50\) mK bath temperature. The MHz mechanical frequency is characteristic of suspended-membrane~\cite{andrews_bidirectional_membrane_2014,higginbotham_harnessing-electro-optic-correlations_2018,delaney_qubit-readout-via-electro-optic-transducer_2022,brubaker_optomechanical-ground-state-cooling-electro-optic-transducer_2022} and zipper photonic crystal \cite{arnold_electro-opto-mechanical_2020} transducers, and jointly coincides with the mechanical frequencies used in feedback cooling demonstrations~\cite{Groblacher_feedback-cooling-ground-state_2019, rossi_measurement-based-mech-q-control_2018}. The bath temperature choice is conservative for dilution refrigerators to account for deleterious heating from photon absorption. We assume a bare mechanical quality factor \(\mathcal{Q} = 10^7\), which has been exceeded in both silicon nitride membrane transducers~\cite{delaney_qubit-readout-via-electro-optic-transducer_2022, rossi_measurement-based-mech-q-control_2018} and on-chip silicon nitride defect modes used in feedback cooling~\cite{Groblacher_feedback-cooling-ground-state_2019}. This choice permits optomechanical cooperativities up to \(\CL = 10\, \nth\) (\(\mathcal{Q}' \gtrsim 250\)) before mode mismatch noise (see Ref.~\cite{Warwick_Paper}) becomes appreciable (more than \(\sim 2\%\) of the total). Optomechanical cooperativities satisfying \(\CL \gg \nth\) are indeed readily attainable with only moderate pump powers, given demonstrations of single-photon cooperativities~\(4(g_{L,0})^2/\kappa_L\Gamma \sim 10^2\)--\(10^3\) in relevant photonic crystal designs~\cite{Groblacher_feedback-cooling-ground-state_2019, leijssen_nonlinear-optomech-high-single-photon-C_2017}. With increasing \(\CL\), the simultaneous requirement of matched cooperativities \(\CM'=1\) (\(\CM \sim \CL\)) to enable \(\Tinf(\sigdet)=1\) becomes more challenging. However, encouraging electro-mechanical cooperativities with a silicon nitride membrane \(\CM \sim \mathcal{O}(10^5)\) have been demonstrated~\cite{yuan_large-Cem-bulk-superconducting-cav_2015}, though compromising in scalability through use of a power resilient bulk superconducting cavity. Operating at a lower dilution fridge temperature can relieve the stringency of the requirements \(\CM \sim \CL\) and \(\{\CM,\,\CL\} \gg \nth\).

Given routine demonstrations of highly over-coupled superconducting microwave resonators, we take \(\eta_M = 0.98\) \cite{xu_radiative-cooling_2020, Han_microwave-optical-review_2021, wu_mech-supermode-piezo-transducer-params_2020}. 

Comparatively, it is more difficult to achieve a highly over-coupled optical cavity \(\eta_L \rightarrow 1\) in traditional sideband resolved MHz electro-optomechanical transducers  without degrading the sideband resolution. Critically, our protocol alleviates the resolved sideband constraint, in principle allowing the external coupling rate to be made arbitrarily large. For this reason, we take \(\eta_L = 0.95\) which has been realised in both membrane and photonic crystal cavity feedback-cooling experiments \cite{rossi_measurement-based-mech-q-control_2018, Groblacher_feedback-cooling-ground-state_2019}. We also consider a homodyne detection efficiency \(\eta_d = 0.85\) as reported in the former (after separating from the cavity outcoupling efficiency) \cite{rossi_measurement-based-mech-q-control_2018}.

\section{Computing the transferred Wigner function}
\label{sec:apx:transferred_wigner}
Here, we show how a single mode Wigner function transforms under a linear transformation of its quadrature operators, following closely the discussion in Ref.~\cite{bennett_PhD_2017}. 

In terms of setup, a state with density matrix \(\hat \rho\) has a corresponding characteristic function \(\chi(\beta) = \text{Tr}[\hat \rho \, \hat D(\beta)]\), where \(\hat D(\beta) = \exp(\beta \hat a^\dag - \beta^* \hat a)\) denotes the displacement operator, \(\hat a\) is the mode annihilation operator, and \(\beta\) is a complex number. The real-valued Wigner function can be obtained via the symplectic Fourier transform of the characteristic function
\begin{eqnarray}
    W(\bi{r}) = \frac{1}{(2\pi)^2} \int \rm{d}^2 \mathcal{B} \,\, \chi(\mathcal{B}) \, \exp(- \rm{i} \bi{r} \cdot \varpi \mathcal{B}), \label{eq:sympletic_FT}
\end{eqnarray}
where \(\bi{r}=(X,Y)^T\), \(\mathcal{B}=(\Re(\beta), \Im(\beta))^T\) and
\begin{eqnarray}
    \varpi = \bigg(\begin{array}{cc}
       0  & 1 \\
       -1  & 0
    \end{array}\bigg)
\end{eqnarray}
is a symplectic matrix satisfying \(\varpi \varpi^T = \varpi^T \varpi = \mathbb{I}\). Using the quadrature convention \(\hat a = (\hat X+\rm{i}\hat Y)/2\) and defining \(\bi X = (X, Y)^T\), we may re-write the displacement operator \(\hat D(\mathcal{B}) = \exp(\rm{i} \bi X \cdot \varpi \mathcal{B})\) so that the characteristic function may instead be written in terms of the quadrature operators as
\begin{eqnarray}
    \chi(\mathcal{B}) = \text{Tr}\bigg[\hat \rho \, \exp\big(\rm{i} \bi{X} \cdot \varpi \mathcal{B}\big)\bigg]. \label{eq:chi_quadrature_form}
\end{eqnarray}

We now begin by considering a linear transformation of an input mode \(\bi{X}_\text{in} = (X_\text{in}, Y_\text{in})^T\) into an output mode \(\bi{X}_\text{out} = (X_\text{out}, Y_\text{out})^T\) via
\begin{eqnarray}
    \bi{X}_\text{out} = \mathbf{G} \bi{X}_\text{in} + \bi{N}, \label{eq:linear_transformation}
\end{eqnarray}
where \(\bi{N} = (X_\text{noise}, Y_\text{noise})^T\) denotes the added noise on each quadrature and \(\mathbf{G}\) is a matrix encoding the transfer gains. Using Eq.~\eqref{eq:chi_quadrature_form}, we may compute the characteristic function of the output state as
\begin{eqnarray}
    \chi_\text{out}(\mathcal{B}) &= \text{Tr}\bigg[\hat \rho \, \exp\big(\rm{i}\bi{X}_\text{out} \cdot \varpi \mathcal{B}\big)\bigg] \\
    &= \text{Tr}\bigg[\hat \rho \, \exp\big(\rm{i} \mathbf{G} \bi{X}_\text{in} \cdot \varpi \mathcal{B}\big) \exp\big(\rm{i} \bi{N} \cdot \varpi \mathcal{B}\big) \bigg],
\end{eqnarray}
where we have separated the exponent using the Baker–Campbell–Hausdorff formula under the assumption that the noise is uncorrelated with the input. Given the same assumption, the noise and system expectation values also factorise, permitting
\begin{eqnarray}
    \chi_\text{out}(\mathcal{B}) &= \text{Tr}\bigg[\hat \rho \, \exp\big(\rm{i} \mathbf{G} \bi{X}_\text{in} \cdot \varpi \mathcal{B}\big)\bigg] \text{Tr}\bigg[ \hat \rho  \, \exp\big(\rm{i} \bi{N} \cdot \varpi \mathcal{B}\big) \bigg]. \label{eq:chi_out}
\end{eqnarray}
Using properties of the symplectic \(\varpi\), we may equivalently write the argument of the first exponential as \(\rm{i}\bi{X}_\text{in} \cdot \varpi (\varpi^T \mathbf{G}^T \varpi \mathcal{B})\). Recognising that the characteristic function of the input state is \(\chi_\text{in}(\mathcal{B}) = \text{Tr}[\hat \rho \, \exp(\rm{i}\bi{X}_\text{in} \cdot \varpi \mathcal{B})]\) and defining \(\chi_\text{noise}(\mathcal{B}) = \text{Tr}[\hat \rho \, \exp(\rm{i}\bi{N} \cdot \varpi \mathcal{B})]\), we find
\begin{eqnarray}
    \chi_\text{out}(\mathcal{B}) = \chi_\text{in}\bigg( (\varpi^T \textbf{G}^T \varpi) \mathcal{B}\bigg) \, \chi_\text{noise}(\mathcal{B}). \label{eq:chi_out_simple}
\end{eqnarray}
Therefore, the output characteristic function is simply the product of the gain-transformed input  characteristic function and that corresponding to the noise contribution, \(\bi N\). We may then compute the output Wigner function from Eq.~\eqref{eq:chi_out_simple} via the symplectic Fourier transform \eqref{eq:sympletic_FT} and the convolution theorem, yielding Eq.~\eqref{eq:def_W_out} in the main text 
\begin{eqnarray}
    W_\text{out}(\mathbf{r}) = (W_\text{in}'\circledast \mathcal{G})(\mathbf{r}), \label{eq:def_W_out_SI}
\end{eqnarray}
where
\begin{eqnarray}
    W_\text{in}'(\bi r) = \frac{1}{(2\pi)^2} \int \rm{d}^2 \mathcal{B} \,\, \chi_\text{in}\bigg( (\varpi^T \mathbf{G}^T \varpi) \mathcal{B} \bigg) \, \exp(-\rm{i} \bi r \cdot \varpi \mathcal{B}).
\end{eqnarray}
For Gaussian added noise \(\bi N\), \(\mathcal{G}(\bi r)\) is a Gaussian kernel given by
\begin{eqnarray}
    \mathcal{G}(\bi r) &= \bigg(2\pi \det(\mathbf{V})^{1/2}\bigg)^{-1} \exp\bigg( - \frac{1}{2} (\delta \bi{r})^T \, \mathbf{V}^{-1} \, \delta \bi{r} \bigg),
\end{eqnarray}
where \(\delta \bi{r} = \bi r - \langle \bi N \rangle\) and \(\mathbf{V}_{ij} = \langle \bi N_i \bi N_j + \bi N_j \bi N_i \rangle/2 - \langle \bi N_i \rangle \langle \bi N_j \rangle\) is the noise covariance matrix.

In the case of a channel with real quadrature transfer gains \(g_X, g_Y\) and no cross-quadrature transfer
\begin{eqnarray}
    \mathbf{G} = \bigg(\begin{array}{cc}
        g_X & 0 \\
        0 & g_Y
    \end{array}\bigg),
\end{eqnarray}
one can readily verify that \(W_\text{in}'(\bi r)\) is simply rescaled in each direction by the respective gain, i.e. \(W_\text{in}'(X,Y) = (g_X g_Y)^{-1} \cdot W_\text{in}\big(X/g_X, \, Y/g_Y\big)\). Therefore, for real \(g_X = g_Y = \tac\) in the main text, we retrieve the simple expression \(W_\text{in}'(\mathbf{r})=W_\text{in}(\mathbf{r}/\tac)/(\tac)^2\). For complex quadrature-symmetric gain \(\tac\), note that one can simply rotate the input quadratures by the same phase and retrieve an equivalent result.  We learn from Eq.~\eqref{eq:def_W_out_SI} that a channel of the form \(\bi{X}_\text{out} = \mathbf{G} \bi{X}_\text{in} + \bi{N}\) re-scales the input Wigner function according to \(\mathbf{G}\) and performs a Gaussian blur with magnitude determined by the statistics of \(\bi N\). \\

In Fig.~\ref{fig:CL_noise_analysis} and Fig.~\ref{fig:detection_loss_analysis_SI}, we model the performance of the transducer by considering specific fock (\(\ket{n}\)) and cat (\(\ket{\text{cat}}_\alpha\)) state inputs. The Wigner function \(W_\text{in}(\bi r)\) used for each is listed below for the mode convention \(\hat a = (\hat X + \rm{i} \hat Y)/2\) and \(\bi r = (X,Y)^T\)
\begin{eqnarray}
    \ket{n}: \quad &W_\text{in}(\bi r) = (2\pi)^{-1}(-1)^n\exp(-\bi{r}\cdot \bi{r}/2) \mathcal{L}_n(\bi{r}\cdot \bi{r}), \label{eq:wigner_fock_defn} \\
    \nonumber \\
    \ket{\text{cat}}_\alpha: \quad &W_\text{in}(\bi r) = \exp(-\textbf{r}\cdot\textbf{r}/2) \bigg[2\pi \bigg(1\pm e^{-2|\alpha|^2}\bigg) \bigg]^{-1} \nonumber \\
    &\hskip2.5cm \cdot \bigg[ e^{-2|\alpha|^2} \cosh(2\textbf{r}\cdot \alpha) \pm \cos(2\textbf{r} \cdot \varpi\alpha) \bigg], \label{eq:wigner_cat_defn}
\end{eqnarray}
where \(\mathcal{L}_n(x) = (n!)e^x \frac{\rm{d}^n}{\rm{d} x^n} (x^n e^{-x})\) is the \(n^\text{th}\)  Laguerre polynomial \cite{bennett_PhD_2017}. Cat states are either even (\(+\)) or odd (\(-\)) superpositions of coherent states with amplitude \(\alpha\) (\(\ket{\text{cat}}_\alpha \propto \ket{\alpha}\pm \ket{-\alpha}\)). We consider only an even cat state. In the main text we use the mode operator convention \(\hat a = (\hat X+\rm{i} \hat Y)/\sqrt{2}\) so that \(W_\text{in}(\bi r) \mapsto 2 W_\text{in}(\sqrt{2} \bi r )\) when applying Eqs. \eqref{eq:wigner_fock_defn}, \eqref{eq:wigner_cat_defn}. 

\section{Insensitivity of the transfer to feedback detection losses}
\label{sec:apx:insensitivity_feedback_loss}

\begin{figure}[t]
    \centering
    \includegraphics{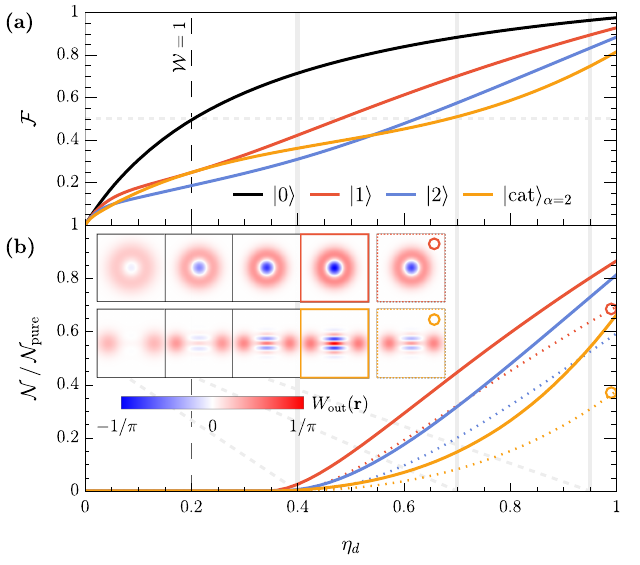}
    \caption{\textbf{Insensitivity of the transfer to feedback detection losses}. (a) Transfer fidelity \(\mathcal{F}\) (Eq.~\eqref{eq:fidelity_def}) as a function of the homodyne detection loss, \(\eta_d\). Transducer performance exceeding that of an LOCC protocol (\(\qtw>1\), Eq.~\eqref{eq:def_quantum_transfer_witness}) is achievable for \(\eta_d > 1/5\) (see vertical black dashed line). (b) Wigner-negativity \(\mathcal{N}\) (Eq.~\eqref{eq:negativity_def}) of the transduced state relative to that of the pure input, \(\mathcal{N}_\text{pure}\). Insets: output Wigner function \(W_\text{out}(\mathbf{r})\) (see Eq.~\eqref{eq:def_W_out}) in phase space \(\mathbf{r}=(X,Y)^T\) generated at the detection losses identified by grey vertical lines (left-to-right: \(\eta_d = 0.4, \, 0.7, \, 0.95\)). The pure input state is identified with a coloured border. The transferred state without detection loss (\(\eta_d=1\)) and with \(\eta_L,\eta_M=0.98\) is identified with a dotted border. For both (a) and (b), \(\CL / \nth = 10\). Solid curves indicate unit cavity in-coupling efficiencies \(\eta_L, \, \eta_M = 1\) and dotted curves indicate the degradation from small cavity in-coupling losses \(\eta_L, \, \eta_M = 0.98\).}
    \label{fig:detection_loss_analysis_SI}
\end{figure}

Here, we comment on how the transfer fidelity and preserved Wigner negativity change as a function of feedback detection losses, \(\eta_d\). Feedback detection losses only contribute noise \(V_\text{det} \propto (1-\eta_d)/\eta_d\) without degrading the transfer efficiency (see vertical contour in Fig.~\ref{fig:TV+W}a). For this reason, the fidelity remains an appropriate figure of merit (with \(\eta_L, \, \eta_M=1\), \(\CL/\nth = 10\)). 

Fig. \ref{fig:detection_loss_analysis_SI}a demonstrates that transfer of a vacuum state (black solid line) can attain the classical fidelity limit for coherent states \(\mathcal{F} =1/2\) with 80\% detection loss (beyond which \(V_\text{det} > 1\)). The classical fidelity bound also coincides with \(\qtw=1\) (see vertical black dashed line), which is consistent with the quantum transfer witness' interpretation as a threshold for beyond LOCC performance. The no-cloning bound for coherent states \(\mathcal{F} = 2/3\) is also exceeded for \(\eta_d > 1/3\), beyond which detection losses add less than vacuum noise \(V_\text{det} < 1/2\). 

The non-classical states considered in Fig.~\ref{fig:detection_loss_analysis_SI}a (single photon (red), two photon (blue), cat state (yellow)) all attain a transfer fidelity greater than \(\mathcal{F} =1/2\) for \(\eta_d \gtrsim 0.7\) (with \(\eta_L, \, \eta_M=1\), \(\CL/\nth = 10\)). Furthermore, Fig.~\ref{fig:detection_loss_analysis_SI}b and the insets therein demonstrate that these states maintain Wigner-negativity after the transfer until \(\eta_d \lesssim 0.4\) (see solid curves). As acknowledged in the main text, the transfer quality is far more sensitive to the optical cavity extraction efficiency. We illustrate this in Fig.~\ref{fig:detection_loss_analysis_SI}b by including the transferred negativity with additional small cavity in-coupling losses \(\eta_L, \eta_M = 0.98\) (dotted curves). In the case of the single photon state without detection loss (\(\eta_d=1\), solid red curve), the transferred negativity is reduced by a further \(\delta(\mathcal{N}/\mathcal{N}_\text{pure})= 0.17\) when 2\(\%\) extraction losses are included (dotted red curve). The same magnitude in negativity degradation without cavity extraction losses (solid red curve) requires detection losses exceeding \(13\%\).

Indeed, similar to the results reported in Ref.~\cite{Warwick_Paper}, we conclude that feedback-enabled transduction is highly robust to losses incurred during the continuous homodyne measurement.

\section{The quantum transfer witness \(\qtw\) as a transducer figure of merit}

Here, we relate the \textit{quantum transfer witness} \(\qtw\) to bounds on classical transducer performance and entanglement metrics. Through which, we unlock further utility of \(\qtw\) as a single-valued figure of merit to compare transducer designs. 

The quantum transfer witness \(\qtw\) was formerly introduced as the \textit{gain-normalised conditional variance product}, \(\mathcal{M}\) (\(\equiv \qtw^2\)), in Ref. \cite{bowen_M_2003}. In which, the authors consider a frequency-domain channel of the form
\begin{eqnarray}
    X_\text{out}^\pm &= g^\pm X_\text{in}^\pm + X_\text{noise}^\pm \,, \label{eq:general_channel}
\end{eqnarray}
where for brevity we instead adopt the notation \(X^\pm\) to denote the quadrature operators (\(X\rightarrow X^+\), \(Y \rightarrow X^-\)). All quadratures \(X^\pm\) in this section should be understood as the fluctuations on top of any coherent field amplitude (i.e. they are zero-mean \(\langle X^\pm\rangle = 0\)). Eq. \eqref{eq:general_channel} describes the degradation of some input mode \(X_\text{in}^\pm\) by a transfer gain \(g^\pm\) and added noise \(X_\text{noise}^\pm\). For such a channel, the authors define the quantity
\begin{eqnarray}
    \mathcal{M} &= \frac{V_q}{(|g^+ g^-| + 1)^2}, \label{eq:M_def}
\end{eqnarray}
where
\begin{eqnarray}
    V_q &= \underbrace{\bigg(V[X_\text{out}^+]-\frac{|\langle X_\text{in}^+ X_\text{out}^+\rangle|^2}{V[X_\text{in}^+]}\bigg)}_{V_\text{cv}^+} \cdot
    \underbrace{\bigg(V[X_\text{out}^-]-\frac{|\langle X_\text{in}^- X_\text{out}^-\rangle|^2}{V[X_\text{in}^-]}\bigg)}_{ V_\text{cv}^-} \label{eq:conditional_variance_defn}
\end{eqnarray}
is the product of the amplitude and phase conditional variances \(V_\text{cv}^\pm\) which measure the similarity between the input and output states \cite{ralph_teleportation_1998}. Here, the quadrature normalisation convention \([X^+, X^-]=2 \rm{i}\) is used and we denote the variance of an operator \(\mathcal{O}\) as \(V[\mathcal{O}]\). The definition of \(\mathcal{M}\) in Eq.~\eqref{eq:M_def} is motivated by the authors' recognition that, in the context of continuous variable quantum teleportation, if the contributions to \(X_\text{noise}^\pm\) incurred in reconstructing the input state \(X_\text{in}^\pm\) are separable, then necessarily \(V_q \geq (|g^+ g^-|+1)^2\). Therefore, \(\mathcal{M} < 1\) is only achievable using entangled noise sources. Correspondingly, the value of \(\mathcal{M}\) benchmarks a teleportation protocol against the best performance achievable without quantum resources, and is a useful single-valued figure of merit for comparing quantum teleportation protocols in non-unity gain regimes \cite{bowen_M_2003}. 

In the next section, we define \(\qtw\) by recasting the construction of \(\mathcal{M}\) in a quantum state transfer context. From which, we see that the condition \(\qtw<1\) also naturally marks beyond classical transducer performance. In the final section, we unlock further significance of \(\qtw\) by demonstrating that it precisely quantifies the degree to which the Simon separability criterion~\cite{simon_separability_2000} is violated after one mode of a perfectly entangled two-mode Gaussian state encounters degradation of the channel form in Eq.~\eqref{eq:general_channel}. Therefore, we conclude that \(\qtw<1\) is jointly a necessary and sufficient condition for a transducer to preserve Gaussian entanglement.

\subsection{\(\qtw<1\) as a condition for beyond-LOCC performance}
\label{sec:apx:witness_beyond_LOCC}

We begin by considering a general electro-optic feed-forward setup for performing continuous variable state transfer using only local operations and classical communication (LOCC). We will follow closely the arguments in both Refs. \cite{grangier_grosshans_2000, bowen_M_2003}. From hereon, we instead use the quadrature normalisation \([X^+, X^-] = \rm{i}\), which is consistent with the main text.

The LOCC state-transfer protocol begins with a joint measurement of the optical input state quadratures \(X_\text{in}^\pm\) via idealised homodyne detection, yielding measurement results
\begin{eqnarray}
    M^\pm &= \Lambda^\pm X_\text{in}^\pm  + N_\text{meas}^\pm \, , \label{eq:LOCC_meas_outcomes}
\end{eqnarray}
where \(\Lambda^\pm\) are the measurement gains and \(N_\text{meas}^\pm\) quantifies any measurement noise incurred. Since \(M^\pm\) are detected photocurrents, the operators commute \([M^+, M^-] = 0\) \cite{grangier_grosshans_2000, bowen_CV_entanglement_2003}. Assuming the measurement noise is uncorrelated with the input, it follows from applying Eq.~\eqref{eq:LOCC_meas_outcomes} to \([M^+, M^-] = 0\) that the measurement noise variances necessarily satisfy
\begin{eqnarray}
    V[N_\text{meas}^+]V[N_\text{meas}^-] \geq |\Lambda^+ \Lambda^-|^2/4. \label{eq:V_meas_inequality}
\end{eqnarray}
The second component of the LOCC state-transfer protocol is to reconstruct the input state onto an independent microwave field using the imperfect measurement results \(M^\pm\) and amplitude/phase modulators. Assuming idealised modulation, we may model the output microwave field quadratures as
\begin{eqnarray}
    X_\text{out}^\pm &= \Upsilon^\pm M^\pm + N_\text{rec}^\pm \, ,
\end{eqnarray}
where \(\Upsilon^\pm\) are the electronic feed-forward gains and \(N_\text{rec}^\pm\) describes the noise fluctuations on the reciever beam, which necessarily satisfy \cite{ralph_teleportation_1998}
\begin{eqnarray}
    V[N_\text{rec}^+] \,V[N_\text{rec}^-]\geq 1/4. \label{eq:V_rec_inequality}
\end{eqnarray}
After defining \(g^\pm = \Upsilon^\pm \Lambda^\pm\) and \(X_\text{noise}^\pm = \Upsilon^\pm N_\text{meas}^\pm + N_\text{rec}^\pm\), the classical detection and reconstruction scheme is equivalent to a channel of the form in Eq.~\eqref{eq:general_channel}. 

Now, we can characterise the noise performance of the LOCC state-transfer by computing the conditional variance product in Eq.~\eqref{eq:conditional_variance_defn}. Provided the input signal and noise are stationary and uncorrelated, it follows that
\begin{eqnarray}
    V_\text{cv}^\pm &= V[X_\text{noise}^\pm].
\end{eqnarray}
The conditional variance product is then simply the product of the added-noise variances~\(V_q = V[X_\text{noise}^+] V[X_\text{noise}^-]\).

Now, we are interested in deriving a lower bound on \(V_q\) by considering the internal structure of the noise \(X_\text{noise}^\pm = \Upsilon^\pm N_\text{meas}^\pm + N_\text{rec}^\pm\) granted by the classical detection and reconstruction scheme. Since the noise sources are independent and uncorrelated, we have that \(V[X_\text{noise}^\pm] = |\Upsilon^\pm|^2 V[N_\text{meas}^\pm] + V[N_\text{rec}^\pm]\). Therefore,
\begin{eqnarray}
    V_q &= \bigg( |\Upsilon^+|^2 V[N_\text{meas}^+] + V[N_\text{rec}^+] \bigg) \bigg( |\Upsilon^-|^2 V[N_\text{meas}^-] + V[N_\text{rec}^-] \bigg) \\
    &\geq (|g^+ g^-|^2 + 1)/4 + |\Upsilon^+|^2 V[N_\text{meas}^+] V[N_\text{rec}^-] \nonumber \\
    &\hspace{3.4cm} + |\Upsilon^-|^2 V[N_\text{meas}^-] V[N_\text{rec}^+], 
\end{eqnarray}
where we have invoked the inequalities in Eqs.~\eqref{eq:V_meas_inequality}, \eqref{eq:V_rec_inequality} and \(g^\pm = \Upsilon^\pm \Lambda^\pm\). We may similarly invoke the same inequalities once more to retrieve
\begin{eqnarray}
    V_q &\geq (|g^+ g^-|^2 + 1)/4 + |\Upsilon^+|^2 V[N_\text{meas}^+] V[N_\text{rec}^-] \nonumber \\
    &\hspace{3.4cm} + \frac{|\Upsilon^-|^2 |\Lambda^+ \Lambda^-|^2}{16}\bigg(\frac{1}{V[N_\text{meas}^+] V[N_\text{rec}^-]}\bigg).
\end{eqnarray}
By minimising the right hand side of the above inequality with respect to \(V[N_\text{meas}^+] V[N_\text{rec}^-]\), one finds that
\begin{eqnarray}
    V[N_\text{meas}^+] V[N_\text{rec}^-] &= \frac{|\Upsilon^-||\Lambda^+ \Lambda^-|}{4|\Upsilon^+|} 
\end{eqnarray}
grants the minimal conditional variance product attainable when using idealised electro-optic feed-forward to transduce a stationary input state
\begin{eqnarray}
    V_q \geq (|g^+ g^-| + 1)^2/4. \label{eq:Vq_LOCC_min}
\end{eqnarray}
The ratio of the conditional variance product to the lower bound in Eq. \eqref{eq:Vq_LOCC_min} defines the \textit{quantum transfer witness}
\begin{eqnarray}
    \qtw^2 &= \frac{4 V[X_\text{noise}^+]V[X_\text{noise}^-]}{(|g^+ g^-| + 1)^2}, \label{eq:W_squared_SI}
\end{eqnarray}
as per the main text. The choice \(\qtw^2\) on the left hand side is motivated by the results of the next section. Strictly, the conditional variance product \(V_q = V[X_\text{noise}^+]V[X_\text{noise}^-]\) and its lower bound in Eq.~\eqref{eq:Vq_LOCC_min} used in this construction are restricted to the case of input states with stationary statistics. Despite this, \(\qtw<1\) serves as a useful heuristic for identifying advantage beyond LOCC state-transfer. Indeed, for stationary input states, a transducer satisfying \(\qtw<1\) necessarily imprints less noise on the output than a classical detection and reconstruction scheme. 

The conditional variance product is also constrained by the Heisenberg uncertainty requirements \([X_\text{in/out}^+, X_\text{in/out}^-]=\mathrm{i}\) \cite{bowen_M_2003}. Using these conditions and Eq.~\eqref{eq:general_channel}, it follows that \([X_\text{noise}^+, X_\text{noise}^-] = \mathrm{i} (1-g^+g^-)\), provided the input and noise are uncorrelated. The conditional variance product therefore admits the fundamental lower bound
\begin{eqnarray}
    V_q \geq |1-g^+ g^-|^2/4,
\end{eqnarray}
from which we can identify a forbidden region \(\qtw<\qtw^\text{min}\) where
\begin{eqnarray}
    \qtw^\text{min} &= \frac{|g^+g^- - 1|^2}{(|g^+ g^-|+1)^2}.
\end{eqnarray}
The forbidden region is shaded grey in Fig. \ref{fig:TV+W}a in the main text.

\subsection{\(\qtw<1\) as a condition for entanglement preservation}
\label{sec:apx:witness_entanglement}

Here, we show via the Simon separability criterion \cite{simon_separability_2000} that \(\qtw\) jointly serves to quantify a transducer's ability to preserve entanglement. Precisely, \(\qtw\) quantifies the degree of inseparability preserved after one mode of a perfectly entangled two-mode input state is transduced. In this section, we also adopt the quadrature notation \(X\rightarrow X^+\), \(Y \rightarrow X^-\), unlike the main text, and denote the variance of an operator \(\mathcal{O}\) as \(V[\mathcal{O}]\).

We begin by introducing the Simon separability criterion. A bipartite quantum state is \textit{separable} if its density matrix may be written as \(\rho = \sum_j p_j  \, \rho_j^A \otimes \rho_j^B\), where the coefficients \(p_j\) are non-negative and \(\rho_j^{A/B}\) describe the density matrices of subsystems A/B for each product state \(j\) in the mixture~\cite{simon_separability_2000}. One method for interrogating whether a state is separable is to inspect whether the \textit{partial transpose} of the density matrix -- obtained via transposition with respect to one subsystem only -- is positive semidefinite (i.e. all eigenvalues are non-negative). Indeed, for \((1\times N)\)-mode Gaussian states, Positivity of the Partial Transpose (PPT) is both a necessary and sufficient condition for separability~\cite{adesso_entanglement_61pg_2007}. For a two-mode Gaussian state characterised by quadratures \(\bi R = ( \hat X_1^+, \hat X_1^-, \hat X_2^+, \hat X_2^-)^T\) and covariance matrix \(\mathbf{V}_{ij} = \langle \bi R_i \bi R_j + \bi R_j \bi R_i \rangle - \langle \bi R_i \rangle \langle \bi R_j \rangle \), the PPT criterion admits the simple form~\cite{adesso_entanglement_61pg_2007}
\begin{eqnarray}
    \tilde{\mathbf{V}} + \frac{\mathrm{i}}{2} m^2 \mathbf{\Omega} \geq 0, \label{eq:simon_symplectic_ineq}
\end{eqnarray}
where \(\tilde{\mathbf{V}}\) is the partially transposed covariance matrix, \(m^2\) determines the quadrature normalisation \([\hat X^+, \hat X^-]=m^2 \rm{i}\), and \(\mathbf{\Omega}\) is a symplectic form given by
\begin{eqnarray}
    \mathbf{\Omega} = 
    \bigg(\begin{array}{cc}
        \underline{\omega} & \underline{0} \\
        \underline{0} & \underline{\omega}
    \end{array}\bigg), 
    \quad 
    \underline{\omega} = \bigg(\begin{array}{cc}
        0 & 1 \\ -1 & 0
    \end{array}\bigg), \quad \mathbf{\Omega}^{-1} = \mathbf{\Omega}^T = -\mathbf{\Omega}.
\end{eqnarray}
Eq. \eqref{eq:simon_symplectic_ineq} is both a compact statement of the uncertainty principle and a constraint that ensures positivity of the partially transposed density matrix \cite{simon_separability_2000,adesso_entanglement_61pg_2007}. The inequality \eqref{eq:simon_symplectic_ineq} may be conveniently recast into a condition on the minimum symplectic eigenvalue \(\tilde{\nu}_-\) of the 
partially transposed covariance matrix \(\tilde{\mathbf{V}}\)~\cite{adesso_entanglement_61pg_2007}
\begin{eqnarray}
    \tilde{\nu}_- \geq m^2/2, \label{eq:simon_criterion}
\end{eqnarray}
which is computed via \cite{Warwick_Textbook}
\begin{eqnarray}
    \tilde{\nu}_- &= \bigg[\frac{1}{2}\bigg( \tilde{\Delta} - \sqrt{\tilde{\Delta}^2 -4 \det(\mathbf{V})  } \bigg) \bigg]^{1/2}, \label{eq:symplectic_eval_def}
\end{eqnarray}
with the \textit{seralian} \(\tilde{\Delta}\) defined as
\begin{eqnarray}
    \tilde{\Delta} = \det(\mathbf{A}) + \det(\mathbf{B}) - 2 \det(\mathbf{C}), \quad \mathbf{V} = \bigg(\begin{array}{cc}
         \mathbf{A} & \mathbf{C} \\
         \mathbf{C}^T & \mathbf{B}
    \end{array}\bigg). \label{eq:seralian_def}
\end{eqnarray}
We may then define the \textit{degree of inseparability}
\begin{eqnarray}
    \mathcal{I} = 2 \tilde{\nu}_- / m^2 \label{eq:degree_of_inseparability}
\end{eqnarray}
to quantify the extent to which the inequality in Eq. \eqref{eq:simon_criterion} is violated. Therefore, \(\mathcal{I} < 1\) is a necessary and sufficient condition for two-mode Gaussian entanglement. \\

Now, we turn to demonstrate that the quantum transfer witness measures the degree of inseparability \(\qtw = \mathcal{I}\) remaining after one mode of a perfectly entangled two-mode Gaussian state is transduced according to a channel of the form in Eq.~\eqref{eq:general_channel}. We will proceed by using a two-mode squeezed state as a proxy for modelling a perfectly entangled EPR state in the continuous variable domain. We will then construct the corresponding covariance matrix in the case where one of the entangled modes is degraded according to a general channel of the form in Eq. \eqref{eq:general_channel}. In the context of quantum state transfer, this will represent one mode incurring a noise penalty upon frequency conversion. Finally, we will compute the degree of inseparability in Eq.~\eqref{eq:degree_of_inseparability} after the transfer in the limit of perfect input squeezing, from which we find \(\qtw = \mathcal{I}\).

We begin with the covariance matrix of a two-mode squeezed state corresponding to operators \(\bi R_\text{in} = ( \hat X_\text{in,1}^+, \hat X_\text{in,1}^-, \hat X_\text{in,2}^+, \hat X_\text{in,2}^-)^T\) with normalisation convention \([X_\text{in,1/2}^+, X_\text{in,1/2}^-]=\rm{i}\) (\(m^2 = 1\)) and real squeezing parameter \(r\) \cite{adesso_entanglement_61pg_2007}
\begin{eqnarray}
    \mathbf{V} &= \frac{1}{2} \left( 
    \begin{array}{cccc}
         \cosh(2r) & 0 & \sinh(2r) & 0 \\
        0 & \cosh(2r) & 0 & -\sinh(2r) \\
        \sinh(2r) & 0 & \cosh(2r) & 0 \\
        0 & -\sinh(2r) & 0 & \cosh(2r)
    \end{array} \right) \label{eq:squeezed_cov_mat} = \bigg(\begin{array}{cc}
         \mathbf{A} & \mathbf{C} \\
         \mathbf{C}^T & \mathbf{B}
    \end{array}\bigg).
\end{eqnarray}
The quadratures \(\bi R_\text{in}\) exhibit the correlation properties
\begin{eqnarray}
    \langle (\hat X_\text{in,1}^+  - \hat X_\text{in,2}^+)^2\rangle = \langle (\hat X_\text{in,1}^-  + \hat X_\text{in,2}^-)^2\rangle = e^{-2r}.
\end{eqnarray}
In the limit of perfect squeezing (\(r \rightarrow \infty\)), the quadrature sum and difference channels between the beams become perfectly correlated -- permitting an EPR paradox \cite{bowen_CV_entanglement_2004}. For \(\mathbf{V}\) given in Eq.~\eqref{eq:squeezed_cov_mat}, it also follows that \(\mathcal{I} = \exp(-2r)\) so that, in the same limit, the state becomes `perfectly' inseparable with respect to the PPT criterion. Such a state may, for instance, be generated by mixing two identically squeezed vacuum states with a \(\pi/2\) phase difference on a 50-50 beam splitter \cite{bowen_CV_entanglement_2004}. 

Now, we consider degrading one of the entangled modes in the two-mode squeezed state according to a general channel of the form in Eq.~\eqref{eq:general_channel}. We take 
\(X_\text{in,1}^\pm \rightarrow g^\pm X_\text{in,1}^\pm + X_\text{noise}^\pm\), where \(g^\pm\) are necessarily real-valued in the time domain. Assuming 
the noise is uncorrelated with the input state, the covariance matrix in Eq.~\eqref{eq:squeezed_cov_mat} transforms as
\begin{eqnarray}
    \mathbf{A} &\mapsto  \text{diag}\bigg((g^+)^2, \, (g^-)^2 \bigg) \, \mathbf{A} + \text{diag}\bigg(V[X_\text{noise}^+], \, V[X_\text{noise}^-]\bigg), \\
    \mathbf{C} &\mapsto \text{diag}\bigg(g^+, \, g^-\bigg) \, \mathbf{C}. 
\end{eqnarray}
For \(r \gg 1\), one may readily verify that
\begin{eqnarray}
    \text{det}(\mathbf{V}) &\rightarrow \frac{1}{4}V[X_\text{noise}^+]V[X_\text{noise}^-] \cosh^2(2r), \\
    \tilde \Delta &\rightarrow \frac{1}{4}\bigg((g^+ g^-) +1\bigg)^2 \cosh^2(2r),
\end{eqnarray}
so that
\begin{eqnarray}
    \frac{\text{det}(\mathbf{V})}{\tilde \Delta} \rightarrow \frac{\qtw^2}{4}.
\end{eqnarray}
Now, we may equivalently write Eq.~\eqref{eq:symplectic_eval_def} as
\begin{eqnarray}
    \tilde{\nu}_- &= \bigg[\frac{\tilde{\Delta}}{2}\bigg(1 - \sqrt{1 -\frac{4 \det(\mathbf{V})}{\tilde{\Delta}^2}  } \bigg) \bigg]^{1/2}.
\end{eqnarray}
Since \(\text{det}(\mathbf{V})/\tilde{\Delta}^2 \propto 1/\cosh^2(2r)\rightarrow 0\) for \(r\gg 1\), it follows after a binomial approximation that
\begin{eqnarray}
    \lim_{r\rightarrow \infty}\tilde{\nu}_- &= 
    \lim_{r\rightarrow \infty} \bigg[\frac{\text{det}(\mathbf{V})}{\tilde{\Delta}}\bigg]^{1/2} = \frac{\qtw}{2}.
\end{eqnarray}
Therefore, the degree of inseparability defined in Eq.~\eqref{eq:degree_of_inseparability} admits the limiting value
\begin{eqnarray}
    \lim_{r\rightarrow \infty} \mathcal{I} &= \qtw.
\end{eqnarray}
We have shown that for a perfectly entangled two-mode Gaussian state (\(\mathcal{I}=0\)), the degree of inseparability remaining after one mode is transduced according to a channel of the form in Eq.~\eqref{eq:general_channel} is precisely the quantum transfer witness \(\mathcal{I}=\qtw\). Since \(\mathcal{I}<1\) is a necessary and sufficient condition for the output Gaussian state to be entangled, \(\qtw<1\) jointly marks a bound on transducer loss and noise parameters for which Gaussian entanglement is preserved upon performing state-transfer of one sub-system. \\

If instead both sub-systems of the perfectly entangled input are transduced with identical losses and tranfer gain, it can be shown from both the Simon criterion and the product form of the Duan criterion \cite{bowen_CV_entanglement_2004, duan_inseparability_2000} that
\begin{eqnarray}
    \mathcal{I} &= 2\sqrt{V[X_\text{noise}^+]V[X_\text{noise}^-]} \,.
\end{eqnarray}
In this case, entanglement preservation \(\mathcal{I}<1\) necessitates that the geometric mean of the added noise on each quadrature is less than vacuum. For quadrature-symmetric losses, this simply corresponds with adding less than vacuum noise to each quadrature -- independent of the transfer gain. Indeed, this is a stronger criterion than that afforded by \(\mathcal{I} = \qtw < 1\) in the single-transfer case.

\section{References}
\printbibliography[heading=none]
\end{refsection}

\end{document}